\documentclass[aps,showpacs,nofootinbib,superscriptaddressn,twocolumn]{revtex4}
\usepackage[mathscr]{eucal}
\usepackage{color}
\usepackage{graphicx}
\usepackage{epsf}
\usepackage{bm}
\usepackage{epsfig}
\usepackage{feynmf}
\usepackage{amsmath}
\usepackage[figuresleft]{rotating}
\usepackage{young}
\usepackage{amsfonts}
\usepackage{amssymb}

\usepackage{booktabs}
\usepackage{bm}

\newcommand{\pion}{{\pi}}
\newcommand{\kaon}{{K}}
\newcommand{\kaonb}{{\bar{K}}}
\newcommand{\kaonS}{{K^*}}
\newcommand{\kaonSb}{{\bar{K}^*}}
\newcommand{\D}{{D}}

\newcommand{\Ds}{{D_s}}

\newcommand{\DS}{{D^*}}

\newcommand{\DsS}{{D_s^*}}

\newcommand{\etap}{{\eta^\prime}}

\newcommand{\Lambdac}{{\Lambda_c}}
\newcommand{\SigmaS}{{\Sigma^*}}
\newcommand{\Sigmac}{{\Sigma_c}}
\newcommand{\SigmacS}{{\Sigma_c^*}}
\newcommand{\Cascada}{{\Xi}}
\newcommand{\CascadaS}{{\Xi^*}}
\newcommand{\Cascadac}{{\Xi_c}}
\newcommand{\CascadacS}{{\Xi_c^*}}
\newcommand{\Cascadacc}{{\Xi_{cc}}}
\newcommand{\CascadaccS}{{\Xi_{cc}^*}}
\newcommand{\Cascadacp}{{\Xi_c^\prime}}
\newcommand{\Omegac}{{\Omega_c}}
\newcommand{\Omegacc}{{\Omega_{cc}}}
\newcommand{\Omegaccc}{{\Omega_{ccc}}}
\newcommand{\OmegacS}{{\Omega_c^*}}
\newcommand{\OmegaccS}{{\Omega_{cc}^*}}

\newcommand{\SU}{{\rm SU}}
\newcommand{\U}{{\rm U}}
\newcommand{\vJ}{{\bm J}}
\newcommand{\bc}{{\bar c}}
\newcommand{\MeV}{{\rm MeV}}
\newcommand{\GeV}{{\rm GeV}}

\newcommand{\ignore}[1]{} 

\begin{document}

\title{Charmed and strange baryon resonances with heavy-quark spin symmetry}

\author{O.~Romanets$^1$, L.~Tolos$^{2,1}$, C.~Garc\'ia-Recio$^3$, J.~Nieves$^4$,
  L.~L.~Salcedo$^3$, R.~G.~E.~Timmermans$^1$}
\affiliation{$^1$KVI,
  University~of~Groningen, Zernikelaan~25,~9747AA~Groningen, The~Netherlands
  \\
$^2$ Institut~de~Ci\`encies~de~l'Espai~(IEEC/CSIC),
  Campus~Universitat~Aut\`onoma~de~Barcelona, Facultat~de~Ci\`encies,
  Torre~C5,~E-08193~Bellaterra,~Spain \\
$^3$Departamento~de~F\'isica~At\'omica, Molecular~y~Nuclear, and Instituto
  Carlos I de F{\'\i}sica Te\'orica y Computacional, Universidad~de~Granada,
  E-18071~Granada, Spain\\
$^4$ Instituto~de~F\'isica~Corpuscular~(centro~mixto~CSIC-UV),
  Institutos~de~Investigaci\'on~de~Paterna, Aptdo.~22085,~46071,~Valencia,
  Spain}

\date{\today}


\begin{abstract}
We study charmed and strange baryon resonances that are generated dynamically
by a unitary baryon-meson coupled-channel model which incorporates
heavy-quark spin symmetry. This is accomplished by extending the SU(3)
Weinberg-Tomozawa chiral Lagrangian to SU(8) spin-flavor symmetry 
plus a suitable symmetry breaking.  The model produces resonances with
negative parity from $s$-wave interaction of pseudoscalar and vector mesons
with $1/2^+$ and $3/2^+$ baryons.  Resonances in all the isospin, spin, and
strange sectors with one, two, and three charm units are studied.
Our results are compared with experimental data from several facilities, such
as the CLEO, Belle or BaBar Collaborations, as well as with other theoretical
models. Some of our dynamically generated states can be readily assigned to
resonances found experimentally, while others do not have a straightforward
identification and require the compilation of more data and also a refinement
of the model. In particular, we identify the $\Xi_c(2790)$ and $\Xi_c(2815)$
resonances as possible candidates for a heavy-quark spin symmetry doublet.
\end{abstract}

\maketitle
\tableofcontents

\section{Introduction}

In the last decades there has been a growing interest in the properties of charmed
hadrons in connection with many experiments such as CLEO, Belle, BaBar, and
others
\cite{Aubert:2003fg,Briere:2006ff,Krokovny:2003zq,Abe:2003jk,Choi:2003ue,%
  Acosta:2003zx,Abazov:2004kp,Aubert:2004ns,Abe:2007jn,Abe:2007sya,Abe:2004zs,%
  Aubert:2007vj,Uehara:2005qd,Albrecht:1997qa,Frabetti:1995sb,Aubert:2006sp,%
  Abe:2006rz, Artuso:2000xy,Albrecht:1993pt,Frabetti:1993hg,Edwards:1994ar,%
  Ammar:2000uh,Brandenburg:1996jc, Ammosov:1993pi, Aubert:2008if,%
  Mizuk:2004yu,Lesiak:2008wz, Frabetti:1998zt, Gibbons:1996yv, Avery:1995ps,%
  Csorna:2000hw, Alexander:1999ud, Aubert:2007dt, Chistov:2006zj,%
  Jessop:1998wt, Aubert:2006je}. Also, the planned experiments such as PANDA
and CBM at the FAIR facility at GSI \cite{cbm,panda}, which involve the studies of
charm physics, open the possibility for observation of more states with exotic
quantum numbers of charm and strangeness in the near future. The observation
of new states and the plausible explanation of their nature is a very active
topic of research. The ultimate goal is to understand whether those states can
be described with the usual three-quark baryon or quark-antiquark meson
interpretation or, alternatively, qualify better as hadron molecules.

Recent approaches based on coupled-channels dynamics have proven to be very
successful in describing the existing experimental data. In particular,
unitarized coupled-channel methods have been applied in the baryon-meson
sector with the charm degree of freedom
\cite{Tolos:2004yg,Lutz:2003jw,Lutz:2005ip,Hofmann:2005sw,Hofmann:2006qx,%
  Mizutani:2006vq,JimenezTejero:2009vq}, partially motivated by the
parallelism between the $\Lambda(1405)$ and the $\Lambda_c(2595)$. In those
references, the baryon-meson interaction in the charm sector is constructed
using the $t$-channel exchange of vector mesons between pseudoscalar mesons
and baryons. Other existing coupled-channel approaches are based on the
J\"ulich meson-exchange model
\cite{Haidenbauer:2007jq,Haidenbauer:2008ff,Haidenbauer:2010ch} or on the
hidden gauge formalism \cite{Wu:2010jy}.

However, those previous models are not consistent with heavy-quark spin
symmetry~\cite{Isgur:1989vq,Neubert:1993mb,Manohar:2000dt}, which is a proper
QCD symmetry that appears when the quark masses, such as the charm mass,
become larger than the typical confinement scale. Aiming to incorporate
heavy-quark symmetry, an SU(8) spin-flavor symmetric model has recently been
developed \cite{GarciaRecio:2008dp,Gamermann:2010zz}, which includes vector
mesons similarly to the SU(6) approach developed in the light sector
Refs.~\cite{GarciaRecio:2005hy,Toki:2007ab}. Following this scheme, baryon
resonances in the charm sector have been studied, such as the $s$-wave states
in the charm $C=1$ and strangeness $S=0$ sector \cite{GarciaRecio:2008dp}, as
well as $C=-1$ sectors \cite{Gamermann:2010zz}, which are necessarily exotic.

The objective of this work is to continue those studies on dynamically
generated baryon resonances using heavy-quark spin symmetry constraints. We
will focus on charm $C=1$ and strangeness $S=-3,-2$ and $-1$, as well as on
sectors with $C=2$ and $3$. We therefore use the model of
Ref.~\cite{GarciaRecio:2008dp}, and as novelty we pay here special attention
to the pattern of spin-flavor symmetry breaking.
Flavor $\SU(4)$ is not a good symmetry in the limit of a heavy charm quark,
for this reason, instead of the breaking pattern $\SU(8)\supset\SU(4)$, in
this work we consider the pattern $\SU(8)\supset\SU(6)$, since the light
spin-flavor group ($\SU(6)$) is decoupled from heavy-quark transformations.  This
allows us to implement heavy-quark spin symmetry in the analysis and to
unambiguously identify the corresponding multiplets among the resonances
generated dynamically.
At the same time, we are also able to assign approximate heavy
(SU(8)) and light (SU(6)) spin-flavor multiplet labels to the states.

We would like to devote a few words here to critically discuss the extension  
of the  Weinberg-Tomozawa (WT) interaction to vector mesons and to flavor SU(4) assumed in this
work. Because at present we lack a robust scheme to
systematically construct an effective field theory approach to study four
flavor physics, one has to rely on models, including as many as possible known
features of QCD. Most of the available models in the literature, assume SU(4) flavor
symmetry in one way or another and this is probably the weakest point of the
whole approach, as well as some arbitrary pattern of symmetry breaking.

As usually understood, chiral symmetry refers only to the light-quark sector,
i.e., at best to flavor SU(3), and it could very well be the case that no
trace of it survives in sectors involving charm. However, spectroscopic data
indicate that the large charm-quark mass acts mainly in an additive way on the
hadron masses. This still leaves the possibility that the effect of the
large quark mass introduces only moderate changes in other hadron
properties. Chiral symmetry breaking fixes the strength of the lowest order
interaction between Goldstone bosons and other hadrons (here baryons) in a
model independent way, this is the WT interaction, but chiral symmetry does
not fix the interaction between vector-mesons and baryons. On the other hand,
heavy quark spin symmetry (HQSS) connects vector and pseudoscalar mesons
containing charmed quarks. It does not tell anything about mesons made out of
light quarks. Nevertheless, there exist several predictions (relative
closeness of baryon octet and decuplet masses, the axial current coefficient
ratio F/D = 2/3, the magnetic moment ratio $\mu_p/\mu_n=-3/2$) which are
remarkably well satisfied in nature~\cite{Lebed:1994ga}, which suggests that
spin symmetry could be a good approximate symmetry in the light sector. This
is the spin-flavor SU(6) symmetry of the quark model already identified in the
hadronic properties before the advent of QCD (see, for instance
\cite{Hey:1982aj}). Moreover, in the large $N_c$ limit (being $N_c$ the number
of colors) there exists an exact spin-flavor symmetry for ground state
baryons~\cite{Dashen:1993jt}. In the meson sector, an underlying static chiral
$U(6)\times U(6)$ symmetry has been advocated by Caldi and
Pagels~\cite{Caldi:1975tx,Caldi:1976gz}, in which vector mesons would be
Goldstone bosons acquiring mass through relativistic
corrections. This scheme solves a number of theoretical problems in the
classification of mesons and also makes predictions which are in remarkable
agreement with the experiment.

The number of couplings between low-lying hadrons with four flavors is very
large, and the amount of available spectroscopic data is still reduced.  It is
clearly appealing to have a predictive model for four flavors including all
basic hadrons (pseudoscalar and vector mesons, and $1/2^+$ and $3/2^+$
baryons) which reduces to the WT interaction in the sector where Goldstone
bosons are involved and which incorporates heavy-quark symmetry in the sector
where charm quarks participate. The model assumption in the present extension
does not appear to be easy to justify directly from QCD, but, on one hand, it is worth
trying it in view of the reasonable semi-qualitative outcome of the SU(6)
extension in the three-flavor sector \cite{Gamermann:2011mq}. On the other hand, there is a formal plausibleness on how the SU(4) WT interaction in the charmed pseudoscalar meson-baryon sector did come out in the vector-meson exchange
picture, as discussed in the t-channel vector model exchange (TVME) approach \cite{Hofmann:2005sw}. We just want to do a simple match making of the two, with the bonus that we improve on previous models since we incorporate spin symmetry HQSS in the charm sector.

The paper has been organized in the following way. In the next section a
description of the theoretical model is given. The third section presents the
results of our calculation, and in the last section we summarize the
conclusions of this work. In Appendix \ref{app:C-exchange} we show results
incorporating a suppression factor for the charm-exchange transitions.  The
tables of the interaction matrices for the different baryon-meson channels are
collected in Appendix \ref{app:tables}.

\section{Theoretical Framework}

\subsection{Spin-flavor and heavy-quark structure of the baryon-meson 
interaction}

For the baryon-meson interaction we use the model of
\cite{GarciaRecio:2008dp,Gamermann:2010zz}. As mentioned, this model obeys
SU(8) spin-flavor symmetry and also HQSS in the
sectors studied in this work.  The SU(6) version of the model has been also
applied to the study of mesonic \cite{GarciaRecio:2010ki} and baryonic
\cite{Gamermann:2011mq} light resonances.

The model is based on an extension of the WT SU(3) chiral
Lagrangian to implement spin-flavor symmetry
\cite{GarciaRecio:2005hy,GarciaRecio:2006wb}. The channel space is augmented
with vector mesons and $3/2^+$ baryons, in addition to pseudoscalar mesons and
$1/2^+$ baryons. The interaction includes only $s$-wave, which is appropriate
for the description of low-lying odd parity baryon-meson resonances.

In the SU(8) spin-flavor scheme, the mesons, $M$, fall in the irrep ${\bf
  63}$-plet (adjoint representation) plus a singlet, while the baryons, $B$,
are placed in the ${\bf 120}$-plet, which is fully symmetric. This refers to
the lowest-lying hadrons with all quarks in relative $s$-wave. The extension
of the WT Lagrangian is a contact interaction between baryon and meson
modeling the zero-range $t$-channel exchange of mesons in the adjoint
representation. Schematically,
\begin{equation}
\label{symbolic}
{\mathcal L}^{\SU (8)}_{ \rm WT} 
=
 \frac{1}{f^2} [[M^{\dagger} \otimes M]_{\bf 63_{a}}
  \otimes [B^{\dagger} \otimes B]_{ \bf 63} ]_{ \bf 1} 
.
\end{equation}

In the $s$-channel, the baryon-meson space reduces into the following SU(8)
irreps:
\begin{equation}
{\bf 63} \otimes {\bf 120} = {\bf 120} \oplus {\bf 168} \oplus {\bf 2520}
\oplus {\bf 4752}
,
\end{equation}
therefore the single {\bf 63}-like coupling in the $t$-channel
(see Eq.~(\ref{symbolic})) corresponds to four $s$-channel couplings. These are
proportional to the following eigenvalues:
\begin{eqnarray}
\lambda_{\bf 120}=-16 , ~~\lambda_{\bf 168}=-22 ,~~\lambda_{\bf
  2520}=6,~~\lambda_{\bf 4752}=-2. \
\end{eqnarray}

In our convention for the potential, a negative sign in the eigenvalues
implies an attractive interaction. Then, from the eigenvalues, we find that
the multiplets ${\bf 120}$ and ${\bf 168}$ are the most attractive ones while
the ${\bf 4752}$-plet is weakly attractive and the ${\bf 2520}$-plet is
repulsive. As a consequence, dynamically-generated baryon resonances are most
likely to occur within the ${\bf 120}$ and ${\bf 168}$ sectors. The other
SU(8) irrep are necessarily exotic, as they cannot be obtained from $qqq$
states, that is, from ${\bf 4}\otimes{\bf 4}\otimes{\bf 4}$ in flavor
SU(4).\footnote{The states in the ${\bf 168}$ cannot be obtained from ${\bf
    8}\otimes{\bf 8}\otimes{\bf 8}$ of spin-flavor if a relative $s$-wave is
  assumed but they appear by allowing $p$-wave states, so exotic is better
  defined with regards to flavor.} It is also instructive to draw the
attention here to some of the findings of Ref.~\cite{GarciaRecio:2005hy} when
the number of colors $N_C $ departs from 3 \cite{GarciaRecio:2006wb}. There it
is shown that, in the {\bf 168} SU(8) irreducible space, the SU(8) extension
of the WT $s$-wave baryon-meson interaction scales as ${\cal O}(1)$.
Note that SU(3) WT counterpart in some channels also scale as ${\cal
O} (1)$   because the coupling strength for some channels behaves as 
${\cal O}(N_C)$, which compensates ${\cal O}(1/N_C)$ coming from the square of the meson decay constant~\cite{Hyodo:2007np}. However, the WT interaction behaves as ${\cal O}(1/N_C)$ within the {\bf 120}
and {\bf 4752} baryon-meson spaces. This presumably implies that {\bf 4752}
states do not appear in the large $N_C$ QCD spectrum, since both excitation
energies and widths grow with an approximate $\sqrt{N_C}$ rate.

To take into account the breaking of flavor symmetry introduced by the heavy
charmed quark, we consider the reduction
\begin{equation}
\SU(8) \supset \SU(6)\times \SU_C(2) \times \U_C(1)
\label{eq:4}
\end{equation}
where $\SU(6)$ is the spin-flavor group for three flavors. $\SU_C(2)$ is the
rotation group of quarks with charm. We consider only $s$-wave interactions so
$\vJ_C$ is just the spin carried by the charmed quarks or antiquarks.  Finally
$U_C(1)$ is the group generated by the charm quantum number $C$.

 The two main SU(8) multiplets have the following
reductions
\begin{eqnarray}
{\bf 120} &=&{\bf 56}_{\bf 1,0} \oplus {\bf 21}_{\bf 2,1} 
\oplus {\bf 6}_{\bf 3,2} \oplus {\bf 1}_{\bf 4,3} 
,
\nonumber \\
{\bf 168} &=& {\bf 70}_{\bf 1,0} \oplus {\bf 21}_{\bf 2,1} 
\oplus {\bf 15}_{\bf 2,1} \oplus {\bf 6}_{\bf 1,2} \oplus 
{\bf 6}_{\bf 3,2} \oplus {\bf 1}_{\bf 2,3}
.
\label{eq:5}
\end{eqnarray}
For the r.h.s. we use the notation ${\bm R}_{2J_C+1,C}$, where ${\bm R}$ is
the SU(6) irrep label (for which we use the dimension), $J_C$ is the spin
carried by the quarks with charm, and $C$ is the charm. Therefore, with $C=1$
there are two ${\bf 21}_{\bf 2,1}$, one from ${\bf 120}$ and another from
${\bf 168}$, and one ${\bf 15}_{\bf 2,1}$ only from ${\bf 168}$. With $C=2$
there are two ${\bf 6}_{\bf 3,2}$, one from each SU(8) irrep, and one ${\bf
  6}_{\bf 1,2}$ from ${\bf 168}$. Finally, there are two representations with
$C=3$, ${\bf 1}_{\bf 4,3}$ and ${\bf 1}_{\bf 2,3}$.

The SU(6) multiplets can be reduced under $\SU(3)\times \SU_l(2)$. The factor
$\SU_l(2)$ refers to the spin of the light quarks (i.e., with flavors $u$, $d$,
and $s$). In order to connect with the labeling $(C,S,I,J)$ based on isospin
multiplets ($S$ is the strangeness, $I$ the isospin, $J$ the spin), we further
reduce $\SU_l(2) \times \SU_C(2)\supset\SU(2)$ where $\SU(2)$ refers to the
total spin $J$, that is, we couple the spins of light and charmed quarks to
form $\SU(3)$ multiplets with well-defined $J$. So, for instance, the
multiplet ${\bf 21}_{\bf 2,1}$ reduces as ${\bf 6}_{\bf 2}\oplus{\bf 3}^*_{\bf
  2}\oplus{\bf 6}_{\bf 4}$, where we use the notation ${\bm r}_{2J+1}$ and
${\bm r}$ stands for the SU(3) irrep.\footnote{The ${\bf 21}_{\bf 2,1}$ irrep
  can be realized by a baryon with quark structure $llc$ with the two light
  quarks in a symmetric spin-flavor state. In the light sector, and from the
  point of view of SU(3), this is $({\bf 3}_{\bf 2}\otimes {\bf 3}_{\bf 2})_s=
  {\bf 6}_3\oplus{\bf 3}^*_{\bf 1}$, the subindex being $2J_l+1$. The coupling
  of $J_l=0,1$ with $J_C=1/2$ gives the decomposition quoted in the text. The
  ${\bf 15}_{\bf 2,1}$ reduction follows similarly, but the pair $ll$ is
  antisymmetric.}  The charmed $\SU(6)$ multiplets reduce as follows
\begin{eqnarray}
{\bf 21}_{\bf 2,1} &=& {\bf 6}_{\bf 2}\oplus{\bf 3}^*_{\bf 2}
\oplus{\bf 6}_{\bf 4}
,
\nonumber \\
{\bf 15}_{\bf 2,1} &=& {\bf 6}_{\bf 2}\oplus{\bf 3}^*_{\bf 2}
\oplus{\bf 3}^*_{\bf 4}
,
\nonumber \\
{\bf 6}_{\bf 3,2} &=& {\bf 3}_{\bf 2}\oplus{\bf 3}_{\bf 4}
,
\nonumber \\
{\bf 6}_{\bf 1,2} &=& {\bf 3}_{\bf 2}
,
\nonumber \\
{\bf 1}_{\bf 2,3} &=& {\bf 1}_{\bf 2}
,
\nonumber \\
{\bf 1}_{\bf 4,3} &=& {\bf 1}_{\bf 4}
.
\label{eq:6}
\end{eqnarray}

\begin{figure*}
\begin{center}
\includegraphics[scale=1.]{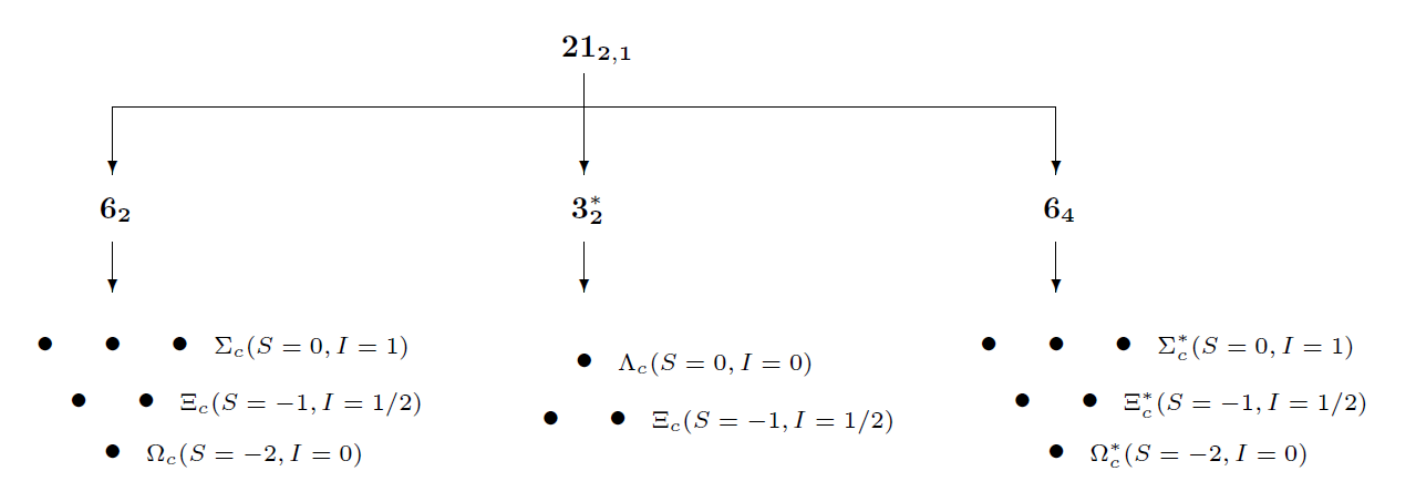}
\end{center}
\caption{$\SU(3)\times\SU(2)$ reduction of the ${\bf 21_{2,1}}$ multiplet
of $\SU(6)\times\SU_C(2)\times\U_C(1)$.}
\label{fig:21rep}
\end{figure*}

\begin{figure*}
\begin{center}
\includegraphics[scale=1.]{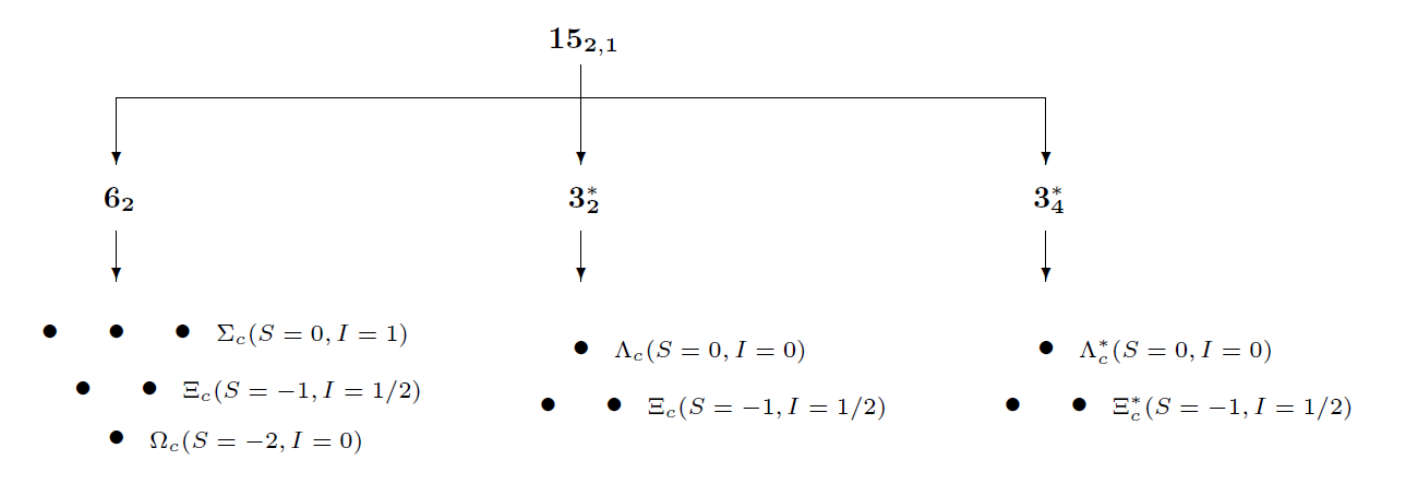}
\end{center}
\caption{$\SU(3)\times\SU(2)$ reduction of the ${\bf 15_{2,1}}$ multiplet
of $\SU(6)\times\SU_C(2)\times\U_C(1)$.}
\label{fig:15rep}
\end{figure*}

\begin{figure*}
\begin{center}
\includegraphics[scale=1.]{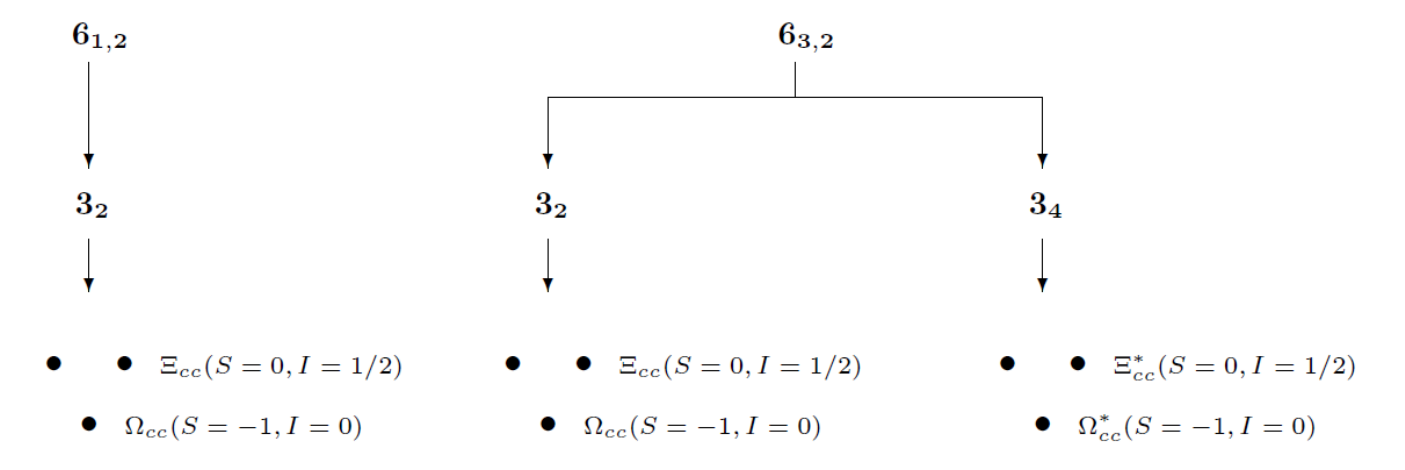}
\end{center}
\caption{$\SU(3)\times\SU(2)$ reduction of the multiplets ${\bf 6_{3,2}}$ and
${\bf 6_{1,2}}$ of $\SU(6)\times\SU_C(2)\times\U_C(1)$.}
\label{fig:6rep}
\end{figure*}

The decomposition of the $\SU(6)\times \SU_C(2) \times \U_C(1)$ multiplets
under $\SU(3)\times \SU(2)$ is shown in
Figs.~\ref{fig:21rep}, \ref{fig:15rep}, \ref{fig:6rep} for the multiplets in
Eq.~(\ref{eq:5}) with $C=1,2,3$ (except the singlets). The further reduction
into $(C,S,I,J)$ multiplets is also displayed.

Collecting the various $CSIJ$ multiplets in the strongly attractive
representations ${\bf 120}$ and ${\bf 168}$, we can estimate the expected
number of dynamically generated baryon-meson resonances. These expected
numbers of states are shown in Table~\ref{tbl:expected}.  In the next section we
find that none of these states for the sectors with charm goes to a wrong Riemann
sheet in the complex $s$-plane, and so they can be identified with physical
states.\footnote{This is not always the case, for instance in
  \cite{GarciaRecio:2003ks,Gamermann:2011mq}, some resonances move to
  unphysical regions of the Riemann surface after breaking of the symmetry.}

\begin{table}[h]
\begin{center}
\begin{tabular}{| c | r | c | c | c | c |}

\hline 
 &  &  &  &  \multicolumn{2}{|c|}{ $J^P$ }   \\ 

$C$ & $S$ & $I$ & state &  \multicolumn{2}{|c|}{ $ \frac{1}{2}^{-} $   $ \frac{3}{2}^{-} $ }  \\ 

\hline

\cline{2-6}
1& 0 & 0 & $\Lambda_c$ & 3 & 1 \\
   
\cline{3-6}
    & & 1 & $\Sigma_c$ & 3 & 2 \\
   
\cline{2-6}
   & $-1$ & $1/2$ & $\Xi_c$ & 6 & 3 \\

 & $-2$ & 0 & $\Omega_c$ & 3 & 2 \\

\hline

\cline{2-6}
2 & 0 & 1/2 & $\Xi_{cc}$ & 3 & 2 \\

& $-1$ & 0 & $\Omega_{cc}$ & 3 & 2 \\

\hline
3 & 0 & 0 & $\Omega_{ccc}$ & 1 & 1 \\

\hline
\end{tabular}
\end{center}
\caption{Expected number of baryonic resonances for the various $CSIJ$
  sectors.}
\label{tbl:expected}
\end{table}

Heavy-quark spin operators are suppressed by the inverse mass of the heavy
quark, therefore HQSS is a fairly good approximate symmetry of QCD
\cite{Isgur:1989vq,Neubert:1993mb} and it is mandatory to implement
it in any hadronic model involving charmed quarks. HQSS implies conservation
of the number of charmed quarks, $N_c$, and of the number of charmed
antiquarks, $N_\bc$, with corresponding symmetry group
$\U_c(1)\times\U_\bc(1)$. In addition, there is invariance under the group
$\SU(2)\times \SU(2)$ of separate rotations of spin of $c$ and $\bc$.
Although such invariance is not automatically ensured by requiring spin-flavor
symmetry,\footnote{Spin-flavor symmetry ensures invariance under equal spin
  rotations of $c$ and $\bc$.} spin-flavor does imply HQSS whenever only heavy
quarks are present (but not heavy antiquarks), or only heavy antiquarks are
present (but not heavy quarks).  This observation suggests a simple
prescription to enforce HQSS in the interaction for the charmed sectors
considered in this work, namely, to drop meson-baryon channels containing
$c\bc$ pairs. Specifically, we consider the sectors
$(N_c,N_\bc)=(1,0),(2,0),(3,0)$, for which the SU(8)-extended WT interaction
fulfills HQSS.  It should be noted that SU(8) is no longer an exact symmetry
in the truncated coupled-channel space. Nevertheless, for the low-lying
resonances, the omitted channels are kinematically suppressed anyway, due to
their large mass. Charmless sectors with hidden charm are necessarily exotic. The study of
these sectors is deferred for future work.

As a final comment, it should be noted that the $\SU(6)$ irrep ${\bf 56_{1,0}}$ 
in Eq.~(\ref{eq:5}) does not exactly coincide with the
usual ${\bf 56}$ irrep that one finds in spin-flavor with only $u$, $d$ and $s$
flavors. The latter is completely charmless, while the states in ${\bf
56_{1,0}}$ contain hidden charm components in general. Actually, in the
$\SU(8)$ case, there are further ${\bf 56_{1,0}}$ irreps (in ${\bf 2520}$ or
${\bf 4752}$). Using a suitable angle mixing (similar to the ideal mixing in
$\SU(3)$) one can recover the purely charmless ${\bf 56_{1,0}}$ and construct another
${\bf 56_{1,0}}$ of the form $|lll\rangle|c\bar{c}\rangle$ ($l$ standing for
light quarks). When the hidden charm components are dropped, one ${\bf
56_{1,0}}$ combination remains while the other one disappears. These
considerations can be extended to the other irreps in Eq.~(\ref{eq:5}). This explains why, when dropping the hidden charm components, we
still get the same number of expected states quoted in Table \ref{tbl:expected}, even if the total dimension of the full meson-baryon space is reduced.


\subsection{Unitarization in coupled channels}
\label{sec:2.B}

The tree-level baryon-meson interaction of the SU(8)-extended WT interaction,
reads
\begin{equation}
V_{ij}(s)= D_{ij}
\frac{2\sqrt{s}-M_i-M_j}{4\,f_i f_j} \sqrt{\frac{E_i+M_i}{2M_i}}
\sqrt{\frac{E_j+M_j}{2M_j}} 
\,.
\label{eq:vsu8break}
\end{equation}
Here, $i$ and $j$ are the outgoing and incoming baryon-meson channels,
respectively.
$M_i$, $E_i$, and $f_i$ stand, respectively, for the mass and the center of
mass energy of the baryon and the meson decay constant in the $i$ channel.
 $D_{ij}$ are the matrix elements for the various
$CSIJ$ sectors considered in this work. They  are displayed in Appendix
\ref{app:tables}. The matrix elements can be evaluated using the method described in the
Appendix A of Ref.~\cite{GarciaRecio:2008dp}, or using the Clebsch-Gordan
coefficients computed in Ref.~\cite{GarciaRecio:2010vf}. 

In order to calculate the scattering amplitudes, $T_{ij}$, we solve the
on-shell Bethe-Salpeter equation in coupled channels using the interaction
matrix $V$ as kernel:
\begin{equation}
\label{LS}
 T(s) =\frac{1}{1-V(s)G(s)} V(s).
\end{equation}
$G(s)$ is a diagonal matrix containing the baryon-meson propagator for each
channel.  $D$, $T$, $V$, and $G$ are matrices in coupled-channel space. All
these matrices are symmetric and block diagonal in $CSIJ$ sectors, producing
the corresponding symmetric submatrices $D^{CSIJ}$, $T^{CSIJ}$, $V^{CSIJ}$,
and $G^{CSIJ}$.

The bare loop function $G^{0}_{ii}(s)$ is logarithmically ultraviolet
divergent and needs to be renormalized. This can be done by one-subtraction at
a subtraction point $\sqrt{s}=\mu_i$,
\begin{equation}
G_{ii}(s)= G^{0}_{ii}(s)-G^{0}_{ii}(\mu_i^2)
.
\label{eq:subs}
\end{equation}
Here we adopt the prescription of \cite{Hofmann:2005sw}, namely, $\mu_i$
depends only on $CSI$ and equals $\sqrt{m_{\rm th}^2+M_{\rm th}^2}$, where
$m_{\rm th}$ and $M_{\rm th}$ are, respectively, the masses of the meson and
the baryon of the channel with the lowest threshold in the given $CSI$ sector.
$G_{ii}$ is determined (see Eq.~(14) of Ref.~\cite{Gamermann:2011mq}) by the loop
function ${\bar J}_0$ defined in the Appendix of Ref.~\cite{Nieves:2001wt} for the
different relevant Riemann sheets.

An enlightening discussion on the dependence on the subtraction point has
been presented in \cite{Hyodo:2008xr}. There, a ``natural'' value is
introduced for the subtraction point, namely, the mass of the lightest baryon
in the given coupled-channel sector (see \cite{Oller:2000fj} for an
alternative definition of natural value). As argued in \cite{Hyodo:2008xr},
the natural value needs not coincide with the phenomenological one, and a
comparison between both provides valuable information on the nature of the
resonances generated dynamically, to wit, quark vs molecular structures. In
the present exploratory work the phenomenological values, obtained from
reproducing experimental data on the position of the resonances, are not
available in general. With regard to the
prescription of Ref.~\cite{Hofmann:2005sw}, this choice was justified in  \cite{Hofmann:2005sw} by guaranteeing an approximate crossing symmetry
although, as noted in \cite{GarciaRecio:2008dp} such a claim appears somewhat dubious because
crossing symmetry involves isospin mixtures. Thus choosing an alternative
subtraction point might lead to yet another reasonable result. This
prescription for the subtraction point was indeed used in the SU(6)
model~\cite{Gamermann:2011mq}. The SU(6)
approach recovered previous results for the lowest-lying $1/2^-$ and $3/2^-$ baryon
resonances appearing in the scattering of the octet of Goldstone bosons off
the lowest baryon octet and decuplet given in Refs.~\cite{Sarkar:2009kx,Oset:2009vf}, and lead to new predictions for
higher energy resonances, giving a phenomenological confirmation of its
plausibility.

In Ref.~\cite{GarciaRecio:2008dp} the value of the subtraction point was slightly modified to
obtain the position of the $\Lambda_c(2595)$ resonance. In the present work the
value has not been modified because, on one hand, results for the resonances
in $C=1,S=0$ sector did not change substantially by varying the value of the
subtraction point and, on the other hand, there is scarce experimental
information about resonances in the other strange and charm sectors beyond
$C=1,S=0$.

The dynamically-generated baryon resonances can be obtained as poles of the
scattering amplitudes for given $CSIJ$ quantum numbers.  One has to check both
first and second Riemann sheets of the variable $\sqrt{s}$. The poles
of the scattering amplitude on the first Riemann sheet (FRS) on the real axis
that appear below threshold are interpreted as bound states. The poles that
are found on the second Riemann sheet (SRS) below the real axis and above
threshold are called resonances. Poles on the SRS on, or below the real axis
but below threshold will be called virtual states. Poles appearing in
different positions than the ones mentioned can not be associated with
physical states and are, therefore, artifacts. For any $CSIJ$ sector, there
are as many branching points as channels involved, which implies a complicated
geometry of the complex $s$-variable space \cite{Nieves:2001wt}.

The mass and the width of the resonance can be found from the position of the
pole on the complex energy plane. Close to the pole, the $T$-matrix behaves as
\begin{equation}
\label{Tfit}
 T_{ij} (s) \approx \frac{g_i g_j}{\sqrt{s}-\sqrt{s_R}}
\,.
\end{equation}
The mass and width of the resonance follow from
$\sqrt{s_R}=m_R-\frac{i}{2}\Gamma_R$, while the dimensionless constant $g_i$
is the coupling of the resonance to the $i$ channel. Since the dynamically-generated states may couple differently to their baryon-meson components, we will examine the $ij$-channel independent quantity
\begin{equation}
 \tilde{T}^{IJSC}(s) \equiv \max_j\ \sum_i\,
|T^{IJSC}_{ij}(s)|\ ,
\label{ttilde}
\end{equation}
which allows us to identify all the resonances within a given sector at once.

The matrix elements $D_{ij}$ display exact SU(8) invariance, but this symmetry
is severely broken in nature, so we implement various symmetry-breaking
mechanisms. As said we only keep channels without charmed antiquarks, to
comply with HQSS. This means to remove channels with extra $c\bc$ pairs. Such
channels are heavier than the basic ones so they would be kinematically
suppressed anyway. However, the suppression introduced by HQSS in the matrix
elements is more severe and we simply take the infinite $c$-quark mass limit
in those would-be couplings (but, of course, not in the charmed hadron
masses).

In addition, several soft symmetry-breaking mechanisms are
introduced.  In the present work  we use physical values for the masses of the hadrons and for the
decay constants of the mesons since we consider that meson-baryon states interact via a point-like interaction
given by the SU(8) model extension of the WT interaction. The values used in this work are quoted in
Table \ref{masses}. We have checked in a previous work~\cite{Gamermann:2011mq} that this approach leads to
a reasonable description of the odd parity light baryon resonances. Indeed, we
found that most of the low-lying three and four star odd parity baryon
resonances with spin 1/2 and 3/2 can be dynamically generated within this
scheme.

Also in Appendix
\ref{app:C-exchange} we discuss the effects induced by a possible reduction in
the matrix elements for which an exchange of charm between meson and baryon
takes place. The introduction of these quenching factors does not spoil
heavy-quark spin symmetry, however, it amounts to a further source of flavor
breaking.  In schemes formulated in terms of exchanges of vector mesons, this
reduction would be induced by the larger mass of the charmed vector meson
exchanged as compared to those of the vector mesons belonging to the $\rho$
nonet, which are  exchanged when there is not exchange of charm.



\begin{table*}

\begin{center}
\begin{tabular}{| cc c  r c c | cc r c c |}
\hline
Meson & mass & decay constant
& ~$\SU(6)$ & $\SU(3)$ & HQSS 
& Baryon & mass
& ~$\SU(6)$ & $\SU(3)$ & HQSS
\cr
\hline
$\pi$ & $138.03$ & $92.4$
& $\bm{35_{1,0}}$ & $\bm{8_1}$ & singlet 
& $N$ & $938.92$
& $\bm{56_{1,0}}$ & $\bm{8_2}$ & singlet
\\

$K$ & $495.68$ & $113.0$
& $\bm{35_{1,0}}$ & $\bm{8_1}$ & singlet 
& $\Lambda$ & $1115.68$
& $\bm{56_{1,0}}$ & $\bm{8_2}$ & singlet
\\

$\eta$ & $547.45$ & $111.0$
& $\bm{35_{1,0}}$ & $\bm{8_1}$ & singlet 
& $\Sigma$ & $1193.15$
& $\bm{56_{1,0}}$ & $\bm{8_2}$ & singlet
\\

$\rho$ & $775.49$ & $153.0$
& $\bm{35_{1,0}}$ & $\bm{8_3}$ & singlet
& $\Xi$ & $1318.11$ 
& $\bm{56_{1,0}}$ & $\bm{8_2}$ & singlet
\\

$K^*$ & $893.88$ & $153.0$
& $\bm{35_{1,0}}$ & $\bm{8_3}$ & singlet 
& $\Delta$ & $1210.00$
& $\bm{56_{1,0}}$ & $\bm{10_4}$ & singlet
\\

$\omega$ & $782.57$ & $138.0$
& $\bm{35_{1,0}}$ & ideal & singlet
& $\Sigma^*$ & $1384.57$
& $\bm{56_{1,0}}$ & $\bm{10_4}$ & singlet
\\

$\phi$ & $1019.46$ & $163.0$
& $\bm{35_{1,0}}$ & ideal & singlet
& $\Xi^*$ & $1533.40$
& $\bm{56_{1,0}}$ & $\bm{10_4}$ & singlet
\\

$\eta^\prime$ & $957.78$ & $111.0$ 
& $\bm{1_{1,0}}$ & $\bm{1_1}$ & singlet 
& $\Omega$ & $1672.45$
& $\bm{56_{1,0}}$ & $\bm{10_4}$ & singlet
\\

$D$  & $1867.23$ & $157.4$
& $\bm{6^*_{2,1}}$ & $\bm{3^*_1}$ & doublet
& $\Lambda_c$ & $2286.46$
& $\bm{21_{2,1}}$ & $\bm{3^*_2}$ & singlet
\\

$D^*$ & $2008.35$ & $157.4$
& $\bm{6^*_{2,1}}$ & $\bm{3^*_3}$ & doublet
& $\Xi_c$ & $2469.45$
& $\bm{21_{2,1}}$ & $\bm{3^*_2}$ & singlet
\\

$D_s$ & $1968.50$ & $193.7$
& $\bm{6^*_{2,1}}$ & $\bm{3^*_1}$ & doublet
& $\Sigma_c$ & $2453.56$
& $\bm{21_{2,1}}$ & $\bm{6_2}$ & doublet
\\

$D_s^*$ & $2112.30$ & $193.7$
& $\bm{6^*_{2,1}}$ & $\bm{3^*_3}$ & doublet
& $ \Sigma_c^*$ & $2517.97$
& $\bm{21_{2,1}}$ & $\bm{6_4}$ & doublet
\\

$\eta_c$ & $2979.70$ & $290.0$
& $\bm{1_{1,0}}$ & $\bm{1_1}$ & doublet
& $\Xi^\prime_c$ & $2576.85$
& $\bm{21_{2,1}}$ & $\bm{6_2}$ & doublet
\\

$J/\psi$ & $3096.87$ & $290.0$
& $\bm{1_{3,0}}$ & $\bm{1_3}$ & doublet
& $\Xi_c^*$ & $2646.35$
& $\bm{21_{2,1}}$ & $\bm{6_4}$ & doublet
\\

&  & & & &
& $\Omega_c$ & $2697.50$ 
& $\bm{21_{2,1}}$ & $\bm{6_2}$ & doublet
\\

& & & & &
& $\Omega_c^*$ & $2768.30$
& $\bm{21_{2,1}}$ & $\bm{6_4}$ & doublet
\\

& & & & &
& $\Xi_{cc}$ & $3519.00$
& $\bm{6_{3,2}}$ & $\bm{3_2}$ & doublet
\\

& & & & &
& $\Xi_{cc}^*$ & $3600.00$
& $\bm{6_{3,2}}$ & $\bm{3_4}$ & doublet
\\

& & & & &
& $\Omega_{cc}$ & $3712.00$
& $\bm{6_{3,2}}$ & $\bm{3_2}$ & doublet
\\

& & & & &
& $\Omega_{cc}^*$ & $3795.00$
& $\bm{6_{3,2}}$ & $\bm{3_4}$ & doublet
\\

& & & & &
& $\Omega_{ccc}$ & $4799.00$
& $\bm{1_{4,3}}$ & $\bm{1_4}$ & singlet
\\

\hline
\end{tabular}
\end{center}
\caption{Baryon masses, $M_i$, and meson masses, $m_i$, and decay constants
  $f_i$, (in MeV) used throughout this work.  The masses are taken from the
  PDG \cite{Nakamura:2010zzi}, except the masses for $\Xi_{cc}^*$,
  $\Omega_{cc}$, $\Omega_{cc}^*$, and $\Omega_{ccc}$.  While $\Xi_{cc}^*$ is
  obtained from $\Xi_{cc}$ summing $80\,\MeV$, similar to the
  $\Xi_{c}'-\Xi_{c}^*$ mass splitting, the masses for $\Omega_{cc}$,
  $\Omega_{cc}^*$ are given in Ref.~\cite{Albertus:2009ww} and for
  $\Omega_{ccc}$ in Ref.~\cite{Flynn:2011gf}. The decay constants $f_i$ are
  taken from Ref.~\cite{GarciaRecio:2008dp}, except for $f_{\eta_c}$ and $f_{J
    / \Psi}$. We take $f_{J / \Psi}$ from the width of the $J/ \Psi
  \rightarrow e^- e^+$ decay and we set $f_{\eta_c}$= $f_{J/\Psi}$, as
  predicted by HQSS and corroborated in the lattice evaluation of
  Ref.~\cite{Dudek:2006ej}.  The $\SU(6)\times\SU_C(2)\times\U_C(1)$ and
  $\SU(3)\times \SU(2)$ labels are also displayed. The last column indicates
  the HQSS multiplets. Members of a doublet are placed in consecutive rows.}
\label{masses}
\end{table*}


The symmetry breaking pattern, with regards to flavor, follows the chain
$\SU(8)\supset\SU(6)\supset\SU(3)\supset\SU(2)$, where the last group refers
to isospin. To tag the resonances with these quantum numbers, we start from
the $\SU(8)$-symmetric scenario, where hadrons in the same $\SU(8)$ multiplet
share common properties (mass and decay constants). This produces a single
resonance for the ${\bf 120}$-irrep and another for the ${\bf
  168}$-irrep. Subsequently, the $\SU(8)\supset\SU(6)$ breaking is introduced
by means of a deformation of the mass and decay constant
parameters. Specifically, we use
\begin{eqnarray}
m(x) &=& (1-x) m_{\SU(8)} + x \, m_{\SU(6)}
,
\nonumber\\
f(x)&=& (1-x) f_{\SU(8)} + x \, f_{\SU(6)}
.
\end{eqnarray}
The parameter $x$ runs from 0 (SU(8)-symmetric scenario) to $1$ (SU(6)
symmetric scenario). The symmetric masses and decay constants are assigned by
taking an average over the corresponding multiplet. The same procedure is
applied to the other breakings, with
\begin{eqnarray}
m(x^\prime) &=& (1-x^\prime) m_{\SU(6)} +  x^\prime\, m_{\SU(3)}
,
\nonumber\\
f(x^\prime)&=& (1-x^\prime) f_{\SU(6)} + x^\prime \, f_{\SU(3)}
,
\end{eqnarray}
and
\begin{eqnarray}
m(x^{\prime\prime}) &=& 
(1-x^{\prime\prime}) m_{\SU(3)} +  x^{\prime\prime}\, m_{\SU(2)}
,
\nonumber\\
f(x^{\prime\prime})&=& 
(1-x^{\prime\prime}) f_{\SU(3)} + x^{\prime\prime} \, f_{\SU(2)}
.
\end{eqnarray}

 It should be noted that SU(6) and SU(3), as well as HQSS, are broken only
  kinematically, through masses and meson decay constants. On the other hand,
  the breaking of SU(8) comes also from the interaction matrix elements, since
  we have truncated the SU(8) multiplets by removing channels with $c\bc$
  pairs, in order to enforce HQSS. Nevertheless, to have SU(8) assignations is
  important in our scheme to be able to isolate the dominant $\bm{168}$ and
  $\bm{120}$ SU(8) irreps, and get rid of the subdominant and exotic
  $\bm{4752}$. Therefore, instead of starting from an
  $\SU(6)\times\mbox{HQSS}$ symmetric scenario, we find it preferable to start
  from a SU(8) symmetric world, and let the charmed quarks to get heavier. In
  this way the offending channels with $c\bc$ pairs tend to decouple
  kinematically as we approach the physical point. At the end, we remove those
  channels and this introduces relatively small changes for the low-lying
  resonances that we are studying.  

The procedure just described allows us to
assign well-defined SU(8), SU(6) and SU(3) labels to the resonances.
Conceivably the labels could depend on the precise choice of symmetric points
or change if different paths in the parameter manifold were followed, but this
seems unlikely. At the same time, the HQSS multiplets form themselves at the physical
  point, since this symmetry is present in the interaction, and also,
  very approximately, in the properties of the basic hadrons. In order to
  unambiguously identify those multiplets, one simply has to adiabatically
  move to the HQSS point, by imposing exact HQSS in the masses and decay
  constants of the basic hadrons. The members of a multiplet get exactly
  degenerated under this test.

Because light spin-flavor and HQSS are independent symmetries, the members of
a HQSS multiplet always have equal SU(6), SU(3) and SU(2) labels. Quite often,
the SU(8) label is also shared by the members of a HQSS multiplet, but not
always, since this property is not ensured by construction.\footnote{Note that
  if HQSS were an exact symmetry of the basic hadrons, we could move from the
  physical point to the SU(6) symmetric point while preserving HQSS all the
  way. However, to reach the SU(8) symmetric point would require to restore
  channels with $c\bc$ pairs, breaking HQSS, and in the way members of a
  common HQSS can end up in different SU(8) irreps.}  

\ignore{
 The factor
$\SU_C(2) \times\U_C(1)$ implements HQSS in our case, since we only consider
sectors with $N_{\bar c}=0$. 
}


\section{ Dynamically generated charmed and strange baryon states }

\vspace{0.3cm}

In this section we show the dynamically generated states obtained in the
different charm and strange sectors. We have assigned to some of them a
tentative identification with known states from the PDG
\cite{Nakamura:2010zzi}. This identification is made by comparing the data
from the PDG on these states with the information we extract from the poles,
namely the mass, width and, most important, the couplings. The couplings give
us valuable information on the structure of the state and on the possible
decay channels and their relative strength. It should be stressed that there
will be mixings between states with the same $CSIJ^P$ quantum numbers but
belonging to different SU(8), SU(6) and/or SU(3) multiplets, since these
symmetries are broken both within our approach and in nature. Additional
breaking of SU(8) (and SU(6) and SU(3)) is expected to take place not only in
the kinematics but also in the interaction amplitudes. This will occur when
using more sophisticated models going beyond the (hopefully dominant) lowest
order retained here.

Masses, widths and main couplings of the resonances found are displayed in
Tables \ref{tab1001}-\ref{tab3001}. The tables are collected by the quantum
numbers $CSI$. States with equal $CSI$ and spin $J=1/2$ and $J=3/2$ have been
collected together in order to put HQSS multiplets members in consecutive
rows. As a rule, two states with $J=1/2$ and $J=3/2$ and equal SU(8), SU(6)
and SU(3) labels form a HQSS doublet (with some exceptions in the case of
  the SU(8) label). The other states are HQSS singlets.

In what follows, we occasionally use an asterisk in the symbol of the states
to emphasize that a resonance has spin $J=3/2$, for instance $\Lambda_c^*$
denotes a state with $CSIJ=(1,0,1/2,3/2)$. The symbol without asterisk may refer
to the generic case or to the $J=1/2$ case.



\subsection{${\bm \Lambda_c}$  states 
($\bm{  C = 1}$, $\bm{ S = 0} $, $\bm{ I=0}$ ) }

We present the poles obtained in the $C=1$, $S=0$ and $I=0$ sector coming from
the ${\bf 120}$ and ${\bf 168}$ SU(8) representations. Moreover, we determine
the coupling constants to the various baryon-meson channels through the
residues of the corresponding amplitudes, as in Eq.~(\ref{Tfit}). Results for
$C=1$ and $S=0$ were reported previously in
Ref.~\cite{GarciaRecio:2008dp}. However, the analysis of the dynamically
generated states in terms of the attractive $\SU(8) \supset \SU(6) \supset
\SU(3) \supset \SU(2)$ multiplets was not done in this previous reference.
Here we are able to assign SU(8), SU(6) and SU(3) labels to the resonances.
Simultaneously, we also classify the resonances into HQSS multiplets, in
practice doublets and singlets. This is of great interest as this symmetry is
less broken than spin-flavor, being of a quality comparable to flavor SU(3).


\subsubsection{Sector $J=1/2$}
In the sector $C=1$, $S=0$, $I=0$, $J=1/2$, there are 16 channels (the
threshold energies, in MeV, are shown below each channel):

\smallskip
\noindent
\begin{tabular}{cccccccc}
$\Sigma_c \pi $  & $ N D $  & $ \Lambda_c \eta $    & $ N D^* $    
& $ \Xi_c K $    & $ \Lambda_c \omega $     & $ \Xi'_c K $  & $ \Lambda D_s $   \\
 2591.6   &  2806.1  &  2833.9  &  2947.3  &  2965.1  &  3069.0  & 3072.5  &  3084.2  
\end{tabular}

\smallskip
\noindent
\begin{tabular}{cccccccc}
$ \Lambda D_s^*$ & $\Sigma_c \rho $  & $ \Lambda_c \eta' $ & $ \Sigma_c^* \rho $   & $ \Lambda_c \phi $    
& $ \Xi_c K^* $    & $ \Xi'_c K^* $     & $ \Xi_c^* K^* $    \\
   3228.0  & 3229.0  &  3244.2  &  3293.5  &  3305.9  &  3363.3  &  3470.7  & 3540.2 
\end{tabular}
\smallskip

\begin{figure}[h]
\begin{center}
\includegraphics[scale=0.65]{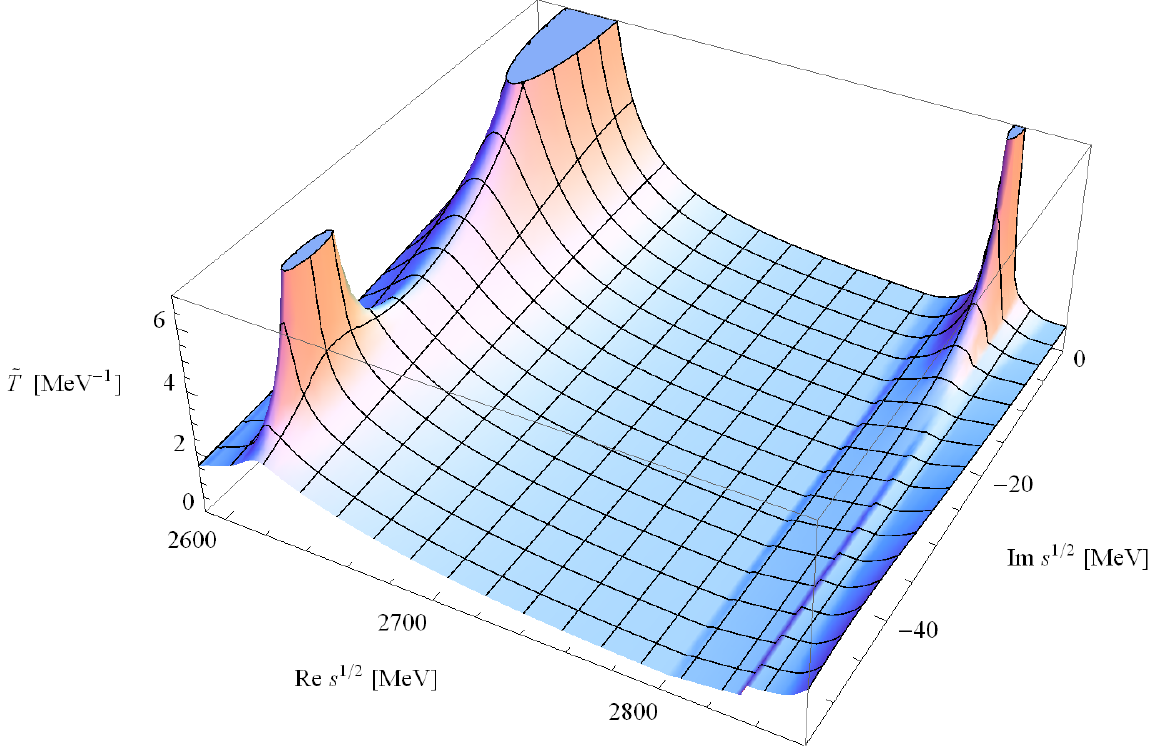}
\medskip
\vskip0.3cm 
\end{center}
\caption{(Color online) $\tilde T^{I=0,~J=\frac{1}{2},~S=0,~C=1}(s)$ amplitude  ($\Lambda_c$ resonances)}.
\label{fig1001}
\end{figure}

\begin{table*}[ht]
\begin{center}
\begin{tabular}{| c | c | c | c | c | c | c | c |}
\hline 

 SU(8)  & SU(6)  & SU(3) & & & Couplings  & &  \\
 irrep    & irrep   & irrep   & $M_{R}$ & $\Gamma_{R}$ &  to main channels &
 Status PDG & $J$  \\ 

\hline

$\bf 168$ & $\bf 15_{2,1}$ & $\bf 3_2^*$ &

2617.3 & 89.8 & $\bf{g_{\Sigma_c \pi}=2.3}$, $g_{N D}=1.6$, $g_{N D^*}=1.4$, &
& $1/2$ 
\\ & & & & & $g_{\Sigma_c \rho}=1.3$ & & \\

\hline

$\bf 168$ & $\bf 15_{2,1}$ & $\bf 3_4^*$ &

2666.6 & 53.7 & $\bf{g_{\Sigma_c^* \pi}=2.2}$, $g_{N D^*}=2.0$, $g_{\Sigma_c
  \rho}=0.8$, & $\Lambda_c(2625)$ & $3/2$ \\ & & & & & $g_{\Sigma_c^*
  \rho}=1.3$ & ***
& \\

\hline

$\bf 168$ & $\bf 21_{2,1}$ & $\bf 3_2^*$ &

2618.8 & 1.2 & $\bf{g_{\Sigma_c \pi }=0.3}$, $g_{N D}=3.5$, $g_{N D^*}=5.6$,
&$\Lambda_c(2595)$ & $1/2$ \\
 &  & & &    & $g_{\Lambda D_s}=1.4$, $g_{\Lambda D_s^*}=2.9$,
$g_{\Lambda_c \eta' }=0.9$ & *** & \\

\hline

$\bf 120$ & $\bf 21_{2,1}$ & $\bf 3_2^*$ &

2828.4 & 0.8  & $\bf{g_{N D}=0.3}$, $g_{\Lambda_c \eta}=1.1$, $g_{\Xi_c
  K}=1.6$,  & & $1/2$ \\
 &  & & &     & $g_{\Lambda D_s^*}=1.1$, $g_{\Sigma_c \rho}=1.1$,
$g_{\Sigma_c^* \rho}=1.0$,  & & \\
 &  & & &     & $g_{\Xi_c^* K^*}=0.8$ & & \\
\hline
\end{tabular}
\end{center}
\caption{$\Lambda_c$ ($J=1/2$) and $\Lambda_c^*$ ($J=3/2$) resonances
  predicted by our model in the {\bf 168} and {\bf 120} SU(8) irreps.  The
  first three columns contain the SU(8), SU(6) and SU(3) representations of
  the corresponding state. $M_R$ and $\Gamma_{R}$ stand for the mass and width
  of the state, in MeV. Next column displays the dominant couplings to the
  channels, ordered by their threshold energies.  In boldface we indicate the
  channels which are open for decay. The last column shows the spin of the
  resonance. Pairs of states with $J=1/2$ and $3/2$ and equal SU(8), SU(6) and
  SU(3) labels form HQSS doublets. They are displayed in consecutive rows.
  Tentative identifications with PDG resonances are shown when possible.}
\label{tab1001}
\end{table*}


The dynamically generated states are shown in Table~\ref{tab1001}.  We obtain
the three lowest-lying states of Ref.~\cite{GarciaRecio:2008dp} in this
sector. We display in Fig.~\ref{fig1001} the channel independent scattering
amplitude defined in Eq.~(\ref{ttilde}) in the SRS for this sector, where
these three poles clearly show up. However, those states appear with slightly
different masses as compared to Ref.~\cite{GarciaRecio:2008dp}. The reason is
that the subtraction point was slightly changed in this previous work in order
to reproduce the position of the $\Lambda_c(2595)$
\cite{Nakamura:2010zzi,Blechman:2003mq, Albrecht:1997qa,Frabetti:1995sb,
  Edwards:1994ar}. The same scaling factor of the subtraction point was
introduced in all the sectors in ~\cite{GarciaRecio:2008dp}.  Another
difference with ~\cite{GarciaRecio:2008dp} is that there the value
$f_{D^*_s}=f_{D^*}=157.4\,\MeV$ was used, whereas here we use the more correct
value $f_{D^*_s}=f_{D_s}=193.7\,\MeV$. These two modifications will affect the
comparison of other sectors too.  A permutation on the order of the two first
resonances as compared to Ref.~\cite{GarciaRecio:2008dp} is also observed.

The experimental $\Lambda_c(2595)$ resonance can be identified with the ${\bf
  21}_{2,1}$ pole that we found around $2618.8\,\MeV$, as similarly done in
Ref.~\cite{GarciaRecio:2008dp}. The width in our case is, however, bigger due
to the increase of the phase space available for decay. As indicated in
Ref.~\cite{GarciaRecio:2008dp}, we have not included the three-body decay
channel $\Lambda_c \pi \pi$, which already represents almost one third of the
decay events \cite{Nakamura:2010zzi}. Therefore, the experimental value of
$3.6^{+ 2.0}_{ -1.3}\,\MeV$ is still not reproduced. Our result for
$\Lambda_c(2595)$ agrees with previous works \ on $t-$channel vector-meson
exchange models
\cite{Tolos:2004yg,Hofmann:2005sw,Mizutani:2006vq,JimenezTejero:2009vq}, but
here as we first pointed out in Ref.~\cite{GarciaRecio:2008dp}, we claim a
large (dominant) $N D^*$ component in its structure. This is in sharp contrast
with the findings of the former references, where it was generated mostly as one
$N D$ bound state.
 
In Fig.~\ref{fig1001}, we also observe a second broad resonance at
$2617.3\,\MeV$ with a large coupling to the open channel $\Sigma_c \pi$, very
close to $\Lambda_c(2595)$. This is precisely the same two-pole pattern found
in the charmless $I = 0, S = -1$ sector for the $\Lambda(1405)$
\cite{Jido:2003cb,GarciaRecio:2003ks,GarciaRecio:2002td}.

As discussed in Ref.~\cite{GarciaRecio:2008dp}, the pole found at around
$2828\,\MeV$, and stemming from the {\bf 120} SU(8) irreducible
representation, is mainly originated by a strong attraction in the $\Xi_c K$
channel but it cannot be assigned to the $\Lambda _c(2880)$
\cite{Nakamura:2010zzi,Aubert:2006sp,Abe:2006rz, Artuso:2000xy} because of the
spin-parity determined by the Belle collaboration.

Some of the states found have coupling to channels with hadrons which are
themselves resonances, like $\Delta$, $\rho$ or $D^*$. Their widths can be
taken into account in the calculation by doing a convolution, as done for
instance in \cite{Roca:2006sz}. In practice the effect of introducing this
improvement is found to be negligible on the position of the dynamically
generated states. The reason is that in all cases the decay thresholds for
these channels are far above the pole, as compared to the widths involved. In
fact, the widths of the basic hadrons can be safely neglected in all sectors
for the low-lying states we obtain.


\subsubsection{Sector $J=3/2$}
For the $C=1$, $S=0$, $I=0$, $J=3/2$ sector, the channels and thresholds (in MeV) are:


\smallskip
\noindent
\begin{tabular}{cccccc}
$ \Sigma_c^* \pi $   & $ N D^*  $      & $ \Lambda_c \omega  $   & $ \Xi_c^* K  $    
& $ \Lambda D_s^*  $    & $ \Sigma_c \rho  $    \\
 2656.0  &  2947.3  &  3069.0  &  3142.0  &  3228.0  &  3229.1 
\end{tabular}

\smallskip
\noindent
\begin{tabular}{ccccc}
 $ \Sigma_c^* \rho  $  & $ \Lambda_c \phi  $   & $ \Xi_c K^*  $ 
& $ \Xi'_c K^*  $   & $  \Xi_c^* K^*  $  \\
  3293.5  &  3305.9  &  3363.3  & 3470.7   &  3540.2 
\end{tabular}
\smallskip

\begin{figure}[h]
\begin{center}
\includegraphics[scale=0.65]{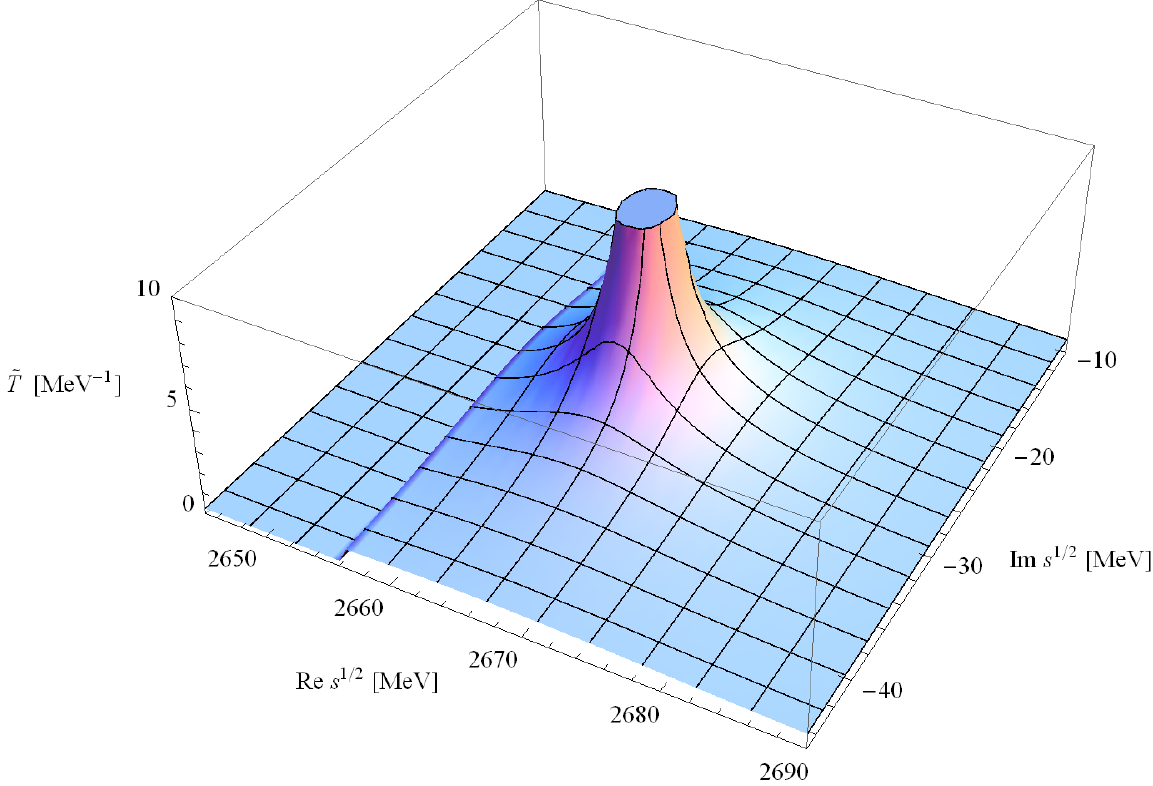}
\medskip
\vskip0.3cm 
\end{center}
\caption{ (Color online) $\tilde T^{I=0,~J=\frac{3}{2},~S=0,~C=1}(s)$  ($\Lambda_c^*$ resonance).}
\label{fig1002}
\end{figure}

We find one pole in this sector (see Fig.~\ref{fig1002} and
Table~\ref{tab1001}) located at $(2666.6 -i 26.7\,\MeV)$.
 
In Ref.~\cite{GarciaRecio:2008dp}, this structure had a Breit-Wigner shape
with a width of $38\,\MeV$ and coupled most strongly to $\Sigma_c^* \pi$. It
was assigned to the experimental $\Lambda_c(2625)$
\cite{Nakamura:2010zzi,Albrecht:1997qa,Edwards:1994ar,Frabetti:1993hg,Albrecht:1993pt}.
The $\Lambda_c(2625)$ has a very narrow width, $\Gamma < 1.9\,\MeV$, and
decays mostly to $\Lambda_c \pi \pi$. The reason for the assignment lies in
the fact that changes in the subtraction point could move the resonance closer
to the position of the experimental one, reducing its width significantly as it
will stay below its dominant $\Sigma_c^* \pi$ channel. In our present
calculation, we expect then a similar behavior.

A similar resonance was found at $2660\,\MeV$ in the $t$-channel vector-exchange model of Ref.~\cite{Hofmann:2006qx}. The novelty of our calculation
is that we obtain a non-negligible contribution from the baryon-vector meson
channels to the generation of this resonance, as already observed in
Ref.~\cite{GarciaRecio:2008dp}.


\subsection{${\bm \Sigma_c}$  states 
($\bm{  C = 1}$, $\bm{ S = 0} $, $\bm{ I=1}$)}



\subsubsection{Sector $J=1/2$}

The 22 channels and thresholds (in MeV) in this sector are:

\smallskip
\noindent
\begin{tabular}{cccccccc}
$\Lambda_c \pi $ & $ \Sigma_c \pi   $  & $  N D  $   & $ N D^*   $    
& $  \Xi_c K  $   & $\Sigma_c \eta $  & $ \Lambda_c \rho   $  & $  \Xi'_c K  $     \\
  2424.5  &  2591.6  &  2806.1   &   2947.3  &  2965.1   &  3001.0   &  3062.0   &  3072.5   
\end{tabular}

\smallskip
\noindent
\begin{tabular}{cccccccc}
$ \Sigma D_s  $   & $  \Delta D^*   $      & $  \Sigma_c \rho   $   & $ \Sigma_c \omega   $    
& $ \Sigma_c^* \rho  $    & $   \Sigma_c^* \omega  $     & $  \Sigma D_s^*   $  & $  \Xi_c K^*  $     \\
  3161.7 &  3218.3  & 3229.1   &  3236.1   &  3293.5   &   3300.5  &  3305.5   &   3363.3 
\end{tabular}

\smallskip
\noindent
\begin{tabular}{cccccc}
$ \Sigma_c \eta'  $   & $ \Xi'_c K^*   $      & $  \Sigma_c \phi   $   & $  \Sigma^* D_s^*  $    
& $ \Sigma_c^* \phi  $    & $  \Xi_c^* K^*   $   \\
 3411.3   &  3470.7  &  3473.0   &   3496.9  &   3537.4  & 3540.2  
\end{tabular}
\smallskip
%

\begin{figure}[h]
\begin{center}
\includegraphics[scale=0.65]{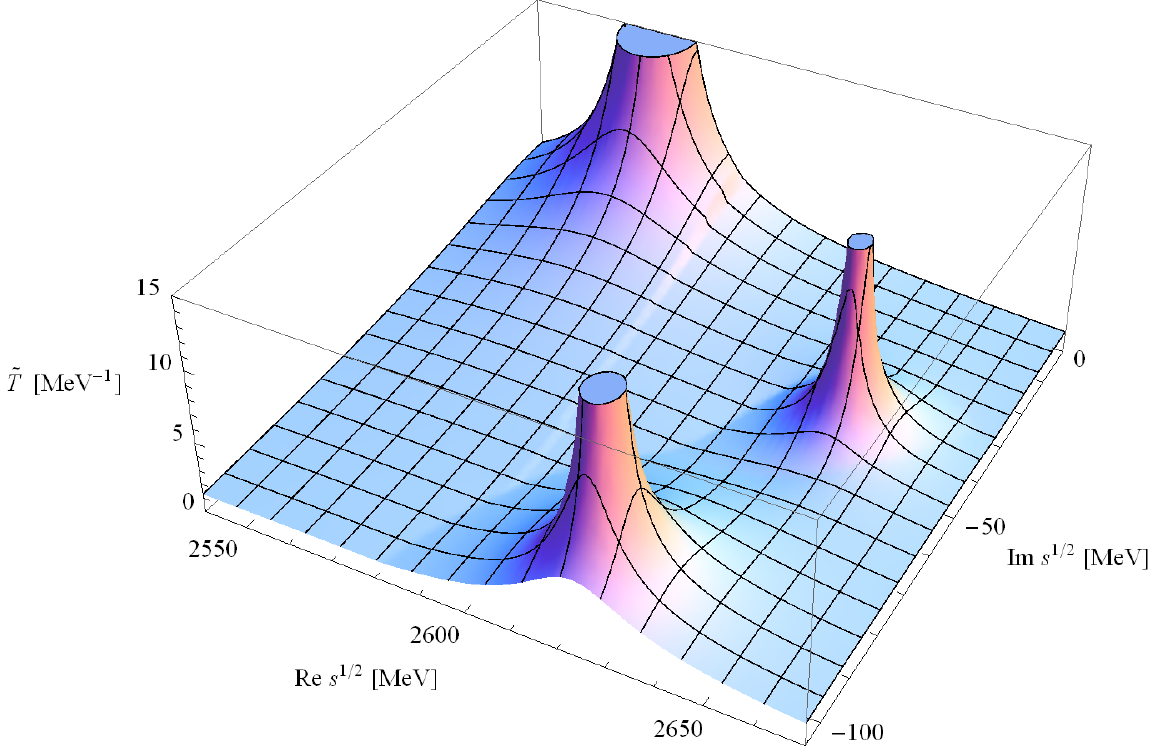}
\medskip
\vskip0.3cm 
\end{center}
\caption{ (Color online) $\tilde T^{I=1,~J=\frac{1}{2},~S=0,~C=1}(s)$ amplitude  ($\Sigma_c$ resonances)}
\label{fig1011}
\end{figure}

\begin{table*}[ht]
\begin{center}
\begin{tabular}{| c | c | c | c | c | c | c | c |}
\hline SU(8) & SU(6) & SU(3) & & & Couplings & \\ irrep & irrep & irrep &
$M_{R}$ & $\Gamma_{R}$ & to main channels & $J$  \\ \hline

$\bf 168$ & $\bf 21_{2,1}$ & $\bf 6_2$ &
2571.5 & 0.8 & $\bf{g_{\Lambda_c \pi}=0.1}$, $g_{N D}=2.2$, $g_{N D^*}=1.2$,
& $1/2$ \\
 &  & & &    & $g_{\Sigma D_s}=1.5$, $g_{ \Delta D^*}=6.6$, $g_{ \Sigma
  D_s^*}=1.1$, & \\
 &  & & &    & $g_{\Sigma^* D_s^*}=2.8$ & \\

\hline

$\bf 168$ & $\bf 21_{2,1}$ & $\bf 6_4$ &
2568.4 & 0.0 & $g_{ N D^*}=2.5$, $g_{\Delta D}=4.2$, $g_{\Delta D^*}=5.3$, &
$3/2$ \\
 &  & & &   & $g_{\Sigma D_s^*}=2.2$, $g_{\Sigma^* D_s}=1.5$, $g_{\Sigma^*
  D_s^*}=2.3$ & \\

\hline

$\bf 168$ & $\bf 15_{2,1}$ & $\bf 6_2$ &
2622.7 & 188.0 & $\bf{g_{\Lambda_c \pi}=1.9}$, $\bf{g_{\Sigma_c \pi}=0.2}$,
$g_{N D}=2.2$, & $1/2$ \\
 &  & & &     & $g_{N D^*}=3.8$, $g_{\Xi_c K}=0.8$, $g_{\Sigma_c
  \rho}=1.3$, & \\
 &  & & &     &  $g_{\Sigma_c^* \rho}=1.5$ & \\

\hline

$\bf 120$ & $\bf 21_{2,1}$ & $\bf 6_2$ &
2643.4 & 87.0 & $\bf{g_{\Lambda_c \pi}=0.2}$, $\bf{g_{\Sigma_c \pi}=2.0}$,
$g_{N D}=2.4$, & $1/2$ \\
 &  & & &     & $g_{N D^*}=1.7$, $g_{\Lambda_c \rho}=0.9$ $g_{\Delta
  D^*}=1.1$, & \\
 &  & & & & $g_{\Sigma_c \rho}=0.9$, $g_{\Sigma^* D_s^*}=1.3$ & \\

\hline

$\bf 120$ & $\bf 21_{2,1}$ & $\bf 6_4$ &
2692.9 & 67.0 & $\bf{g_{\Sigma_c^* \pi}=1.9}$, $g_{N D^* }=2.7$, $g_{\Lambda_c
  \rho}=1.0$, & $3/2$ \\
 &  & & & & $g_{\Sigma D_s^*}=1.0$, $g_{\Sigma^* D_s^*}=1.0$ & \\

\hline
\end{tabular}
\end{center}
\caption{As in Table \ref{tab1001}, for $\Sigma_c$ and $\Sigma_c^*$
  resonances.}
\label{tab1011}
\end{table*}


The three resonances obtained for $J=1/2$ (Table \ref{tab1011} and
Fig.~\ref{fig1011}) are predictions of our model, since no experimental data
have been observed in this energy region. Our predictions here nicely agree
with the three lowest lying resonances found in
Ref.~\cite{GarciaRecio:2008dp}. 

The model of Ref.~\cite{JimenezTejero:2009vq}, based on the full $t$-channel
vector exchange using baryon $1/2^+$ and pseudoscalar mesons as interacting
channels, predicts the existence of two resonances with
$I=1,~J=\frac{1}{2},~S=0,~C=1$. In this reference, the first one has a mass of
$2551\,\MeV$ with a width $0.15\,\MeV$. It couples strongly to the $\Sigma
D_s$ and $N D$ channels and, therefore, might be associated with the resonance
$\Sigma_c(2572)$ with $\Gamma=0.8\,\MeV$ of our model. Nevertheless, in our
model this resonance couples most strongly to the other channels which
incorporate vector mesons, such as $\Sigma^* D_s^*$ and $\Delta D^*$, as it is
shown in the Table~\ref{tab1011} and seen in Ref.~\cite{GarciaRecio:2008dp}.

The second resonance predicted in Ref.~\cite{JimenezTejero:2009vq} has a mass
of $2804\,\MeV$ and a width of $5\,\MeV$, and it cannot be compared to any of
our results because it is far from the energy region of our present {\nobreak
  calculations}. This resonance, though, was identified with the state found
in Ref.~\cite{Hofmann:2005sw} at a substantially lower energy, $2680\,\MeV$,
and in Ref.~\cite{Mizutani:2006vq} around $2750\,\MeV$.


\subsubsection{Sector $J=3/2$}

For the $\Sigma_c^*$ case, the 20 channels and thresholds (in MeV) are:

\smallskip
\noindent
\begin{tabular}{cccccccc}
$ \Sigma_c^* \pi $    & $  N D^*  $      & $ \Lambda_c \rho   $   & $  \Sigma_c^* \eta   $    
& $  \Delta D   $   & $  \Xi_c^* K  $     & $  \Delta D^*   $  & $  \Sigma_c \rho  $    \\
  2656.0  &  2947.3   &   3062.0   &   3065.4   &  3077.2   &   3142.0   &   3218.3   &   3229.1 
\end{tabular}

\smallskip
\noindent
\begin{tabular}{cccccccc}
  $  \Sigma_c \omega  $ & $ \Sigma_c^* \rho $  & $  \Sigma_c^* \omega  $      & $  \Sigma D_s^*  $   & $  \Sigma^* D_s   $    
& $  \Xi_c K^*   $    
& $ \Sigma_c^* \phi $   & $  \Xi_c^* K^*  $ \\
 3236.1  & 3293.5  & 3300.5   &   3305.5   &   3353.1   &  3363.3   &   3470.7   &   3473.0    
\end{tabular}

\smallskip
\noindent
\begin{tabular}{cccc}
   $  \Xi'_c K^*  $     & $  \Sigma_c \phi   $  & $  \Sigma_c^* \eta'  $ & $  \Sigma^* D_s^*  $   \\
   3475.8   &  3496.9 & 3537.4   &  3540.2   
\end{tabular}
\smallskip

\begin{figure}[h]
\begin{center}
\includegraphics[scale=0.65]{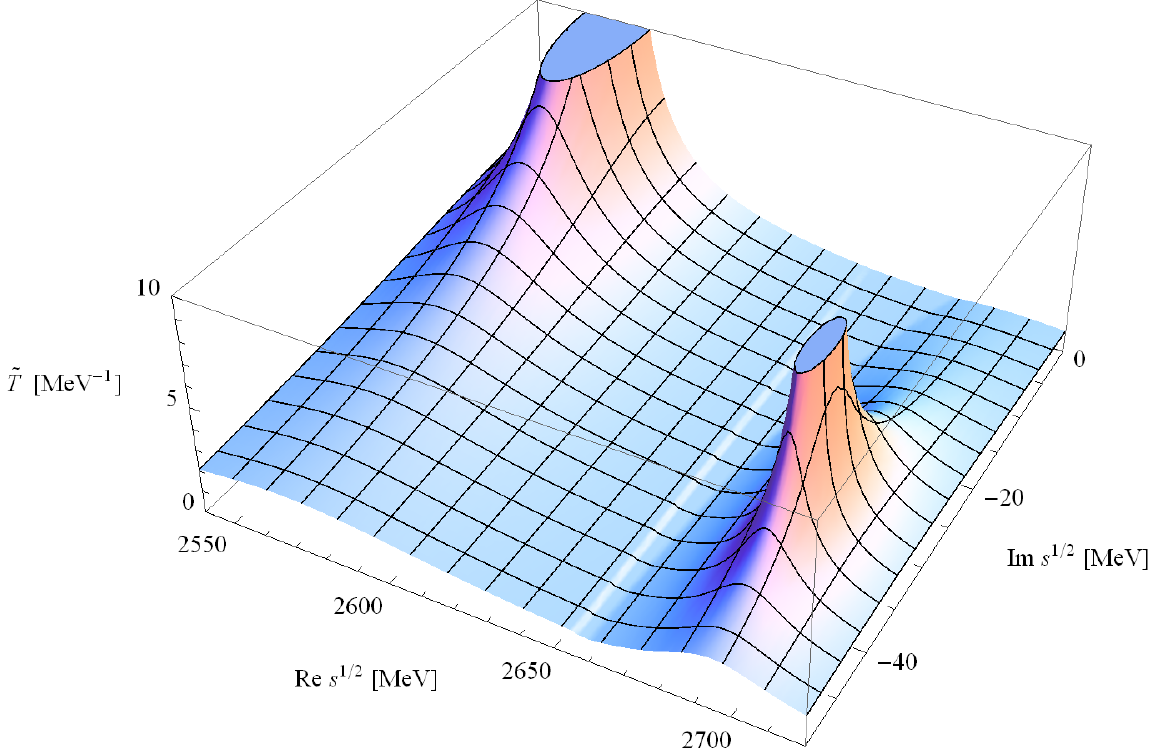}
\medskip
\vskip0.3cm 
\end{center}
\caption{ (Color online) $\tilde T^{I=1,~J=\frac{3}{2},~S=0,~C=1}(s) $ amplitude ($\Sigma_c^*$ resonances).}
\label{fig1012}
\end{figure}

The two predicted states are shown in Fig.~\ref{fig1012} and their properties
are collected in Table~\ref{tab1011}. A bound state at $2568.4\,\MeV$
($2550\,\MeV$ in Ref.~\cite{GarciaRecio:2008dp}), whose main baryon-meson
components contain a charmed meson, lies below the threshold of any possible
decay channel. This state is thought to be the charmed counterpart of the
hyperonic $\Sigma(1670)$ resonance. While the $\Sigma(1670)$ strongly couples
to $\Delta \bar{K}$ channel, this resonance is mainly generated by the
analogous $\Delta D$ and $\Delta D^*$ channels.

The second state at $2692.9\,\MeV$ has not a direct comparison with the
available experimental data, as discussed in Ref.~\cite{GarciaRecio:2008dp}.
In fact, the experimental $\Sigma_c(2520)$
\cite{Nakamura:2010zzi,Ammar:2000uh,Brandenburg:1996jc,Ammosov:1993pi} cannot
be assigned to any of these two states due to parity as well as because of the
dominant decay channel, $\Lambda_c^+ \pi$ ($d$-wave).

With regards to the experimental $\Sigma(2800)$
\cite{Nakamura:2010zzi,Aubert:2008if,Mizuk:2004yu}, there is also no
correspondence with any of our states due to its high mass and also the
empirically dominant $\Lambda_c \pi$ component. Heavier resonances were
produced in \cite{GarciaRecio:2008dp} but they come from the SU(8) irrep ${\bf
  4752}$ which we have disregarded here.

\subsection{${\bm \Xi_c}$  states 
($\bm{  C = 1}$, $\bm{ S = -1} $, $\bm{ I=1/2}$ ) }

We now study the $C=1$, $S=-1$, $I=1/2$ sector for different spin, $J=1/2$ and
$J=3/2$. None of the strange sectors, and in particular this one, were studied
in \cite{GarciaRecio:2008dp}. Those states are
labeled by $\Xi_c$ and and our model predicts the existence of nine states
stemming from the strongly attractive {\bf 120} and {\bf 168} SU(8)
irreducible representations.

\subsubsection{Sector $J=1/2$}

\begin{figure}[h]
\begin{center}
\includegraphics[scale=0.65]{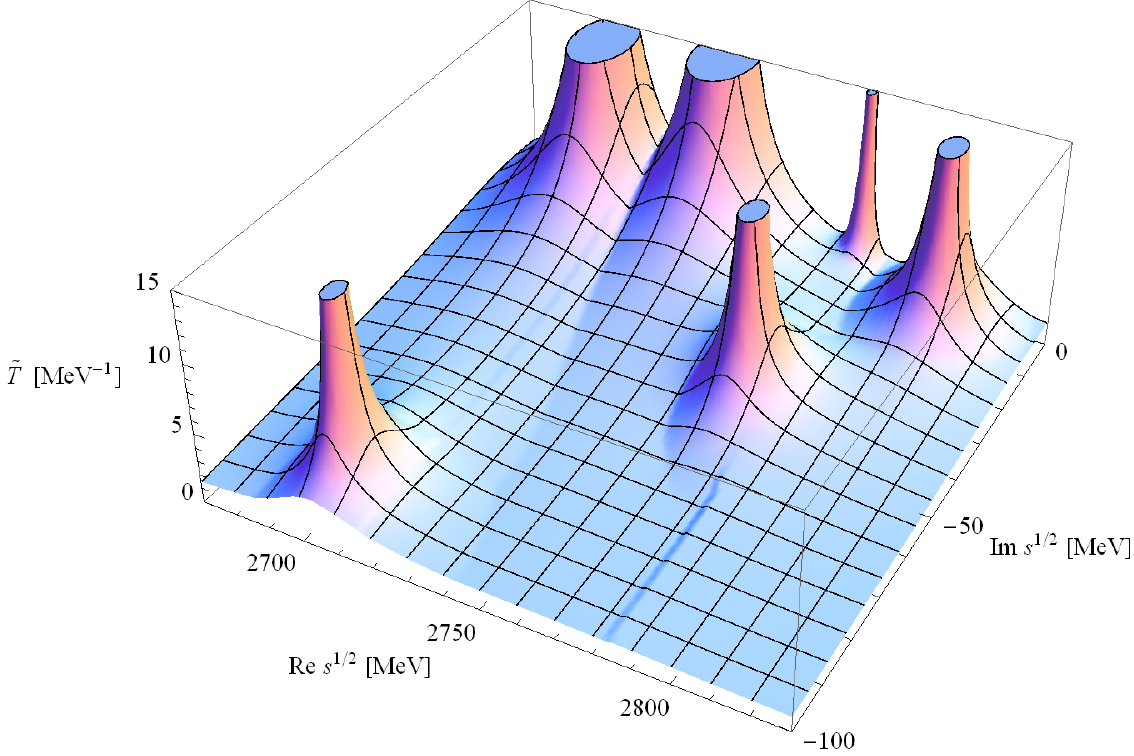}
\medskip
\vskip0.3cm 
\end{center}
\caption{ (Color online) $\tilde T^{I=\frac{1}{2},~J=\frac{1}{2},~S=-1,~C=1}(s) $ amplitude ($\Xi_c$ resonances)}
\label{fig1-111}
\end{figure}

The 31 channels and thresholds (in MeV) for this sector are:

\smallskip
\noindent
\begin{tabular}{cccccccc}
$\Xi_c \pi$   & $\Xi'_c \pi$      & $\Lambda_c \bar K$   & $\Sigma_c \bar K$    
& $\Lambda D$    & $\Xi_c  \eta$     & $\Sigma D$  & $\Lambda D^*$    \\
 2607.5  & 2714.9  &  2782.1  & 2949.2  & 2982.9  & 3016.9  & 3060.4 &  3124.0   
\end{tabular}

\smallskip
\noindent
\begin{tabular}{cccccccc}
  $\Xi'_c \eta$  &$\Lambda_c \bar K^*$     & $\Omega_c K$ & $\Sigma D^*$ & $\Xi_c \rho$     & $\Xi_c \omega$    
&  $\Xi D_s$   & $\Sigma_c \bar K^*$     \\
3124.3 & 3180.3  & 3193.2  & 3201.5  & 3244.9  &  3252.0 & 3286.6  & 3347.4 
\end{tabular}

\smallskip
\noindent
\begin{tabular}{cccccccc}
$\Xi'_c \rho$    & $\Xi'_c \omega$  &  $\Sigma^* D^*$    & $\Sigma_c^* \bar K^*$    & $\Xi_c^* \rho$    & $\Xi_c  \eta'$    & $\Xi_c^* \omega$ 
 & $\Xi D_s^*$    \\
   3352.3   & 3359.4 &  3392.9  &  3411.9 &  3421.8   & 3427.2   & 3428.9   & 3430.4   
\end{tabular}

\smallskip
\noindent
\begin{tabular}{ccccccc}
  $\Xi_c \phi$  & $\Xi'_c \eta'$   & $\Omega_c K^*$ & $\Xi'_c \phi$    & $\Xi^* D_s^*$     & $\Omega_c^* K^*$    &  $\Xi_c^* \phi$  \\
 3488.9   & 3534.6   & 3591.4 &  3596.3  & 3645.7  &  3662.2  &  3665.8 
\end{tabular}
\smallskip


\begin{table*}[ht]
\begin{center}
\begin{tabular}{| c | c | c | c | c | c | c | c |}
\hline 

 SU(8)  & SU(6)  & SU(3) &  &  &   Couplings & &  \\
 irrep    & irrep   & irrep   & $M_{R}$ & $\Gamma_{R}$ &  to main channels & Status PDG & $J$  \\ 
\hline

$\bf 168$ & $\bf 15_{2,1}$ & $\bf 6_2$ &

2702.8 & 177.8 & $\mathbf{g_{\Xi_c \pi}=2.4}$, $g_{\Lambda D}=1.2$, $g_{\Sigma
  D}=1.1$, & & $1/2$ \\
 &  & & &    & $g_{\Lambda D^*}=2.1$, $g_{\Sigma D^*}=1.7$, $g_{\Xi
  D_s^*}=1.1$ & & \\

\hline

$\bf 168$ & $\bf 21_{2,1}$ & $\bf 3_2^*$ &

2699.4 & 12.6 & $\bf{g_{\Xi_c \pi}=0.8}$, $ g_{\Lambda D}=1.2  $,     $
g_{\Sigma D}=3.4$,  &  & $1/2$  \\
 & & & &  &   $ g_{\Lambda D^*}=2.2  $,  $  g_{\Sigma D^*}=5.4 $, $  g_{\Xi D_s}=1.9$,  &  & \\
 & & & &  & $  g_{\Xi_c  \eta'}=1.0$, $   g_{\Xi D_s^*}=3.3$   & & \\

\hline

$\bf 168$ & $\bf 21_{2,1}$ & $\bf 6_2$ &

2733.0 & 2.2 & $\bf{g_{\Xi'_c \pi}=0.5}$, $g_{\Lambda D}=1.9$, $g_{\Sigma
  D}=1.8$, &  & $1/2$ \\
 & & & &    & $g_{\Lambda D^*}=0.9$, $g_{\Sigma D^*}=1.2$, $g_{\Xi
  D_s}=1.2$, & & \\
 & & & &  & $g_{\Sigma^* D^*}=5.8$, $g_{\Xi'_c \eta'}=0.9$, $g_{\Xi^*
  D_s^*}=3.3$ & & \\

\hline

$\bf 168$ & $\bf 21_{2,1}$ & $\bf 6_4$ &
2734.3 & 0.0 & $g_{\Lambda D^*}=2.2$, $g_{\Sigma D^*}=2.1$, $g_{\Sigma^*
  D}=3.6$, & & $3/2$ \\
 &  & & &   & $g_{\Sigma^* D^*}=4.6$, $g_{\Xi D_s^*}=1.3$, $g_{\Xi^*
  D_s}=2.1$, & & \\
 &  &&&   & $g_{\Xi^* D_s^*}=2.6$ & & \\

\hline

$\bf 120$  & $\bf 21_{2,1}$ & $\bf 3_2^*$ &

2775.4 & 0.6 & $\bf{g_{\Xi_c \pi}=0.1}$, $\bf{g_{\Xi'_c \pi}=0.1}$,
$g_{\Lambda_c \bar K}=1.4$, &  & $1/2$ \\
  & & & &     & $g_{\Xi_c  \eta}=0.9$, $g_{\Lambda D^*}=1.0$, $g_{\Sigma
  D^*}=1.4$, &   & \\
  & & & &     & $g_{\Sigma_c \bar K^*}=1.0$, $g_{\Sigma_c^* \bar K^*}=1.3$
& & \\

\hline
      
$\bf 168$ & $\bf 15_{2,1}$ & $\bf 3_2^*$ &

2772.9 & 83.7 & $\bf{g_{\Xi_c \pi}=0.1}$, $\bf{g_{\Xi'_c \pi}=2.3}$,
$g_{\Sigma_c \bar K}=1.2$, &  & $1/2$ \\
  & & & &     & $g_{\Lambda D}=2.1$, $g_{\Lambda D^*}=1.5$, $g_{\Omega_c
  K}= 0.9$, &   & \\
  & & & &     & $g_{\Sigma D^*}=0.9$, $g_{\Xi_c \rho}=1.0$, $g_{\Sigma_c
  \bar K^*}=0.9$, & & \\
  & & & &     & $g_{\Xi'_c \rho}=1.0$, $g_{\Sigma^* D^*}=1.4$, $g_{\Xi^*
  D_s^*}=1.1$ & & \\

\hline

$\bf 168$ & $\bf 15_{2,1}$ & $\bf 3_4^*$ &
2819.7 & 32.4 & $\bf{g_{\Xi_c^* \pi}=1.9}$, $g_{\Sigma_c^*  \bar K}=2.3$,
$g_{\Lambda D^*}=2.0$, &  & $3/2$ \\
 &  & & &  & $g_{\Lambda_c \bar K^*}=1.0$, $g_{\Xi_c^* \eta}=1.1$,
$g_{\Sigma D^*}=1.2$, &  &\\
 &  & & &  & $g_{\Xi_c \rho}=1.1$, $g_{\Sigma_c \bar K^*}=1.0$,
$g_{\Sigma_c^* \bar K^*}=2.0$ & & \\

\hline

$\bf 120$ & $\bf 21_{2,1}$ & $\bf 6_2$ &

2804.8 & 20.7 & $\bf{g_{\Xi'_c \pi}=1.1}$, $g_{\Sigma_c \bar K}=2.4$,
$g_{\Lambda D}=1.5$, & $\Xi_c(2790)$  & $1/2$  \\
  & & & &     & $g_{\Sigma D}=1.2$, $g_{\Xi'_c \eta}=1.3$, $g_{\Lambda_c
  \bar K^*}=1.2$, & *** & \\
  & & & &     & $g_{\Sigma D^*}=0.9$, $g_{\Sigma_c \bar K^*}=1.8$,
$g_{\Sigma^* D^*}=1.1$, &  & \\
  & & & &     & $g_{\Sigma_c^* \bar K^*}=1.0$, $g_{\Xi^* D_s^*}=1.2$ & & \\

\hline

$\bf 120$ & $\bf 21_{2,1}$ & $\bf 6_4$ &
2845.2 & 44.0 & $\bf{g_{\Xi_c^* \pi}=1.9}$, $g_{\Sigma_c^*  \bar K}=2.1$,
$g_{\Lambda D^*}=2.6$, &  $\Xi_c(2815)$ & $3/2$ \\
 &  & & &  & $g_{\Lambda_c \bar K^*}=1.4$, $g_{\Xi_c^* \eta}=1.2$,
$g_{\Sigma D^*}=1.2$, &  *** & \\
 &  &&&    & $g_{\Xi_c \rho}=0.9$, $g_{\Sigma_c \bar K^*}=0.9$,
$g_{\Sigma_c^* \bar K^*}=1.7$, & & \\
 &  &&&    & $g_{\Xi^* D_s}=0.9$, $g_{\Xi^* D_s^*}=1.1$ & & \\

\hline

\end{tabular}
\end{center}
\caption{As in Table \ref{tab1001}, for the $\Xi_c$ and $\Xi_c^*$ resonances.
}
\label{tab11-11}
\end{table*}

Six baryon resonances were expected (Table \ref{tbl:expected}) and found in
this sector.  The mass, width and couplings to the main channels are given in
Table~\ref{tab11-11} and Fig.~\ref{fig1-111}.  In the energy range where these six states predicted by
our model lie, three experimental resonances have been seen by the Belle, E687 and
CLEO Collaborations: $\Xi_c(2645)~ J^P=3/2^+$ \cite{Nakamura:2010zzi,
  Lesiak:2008wz,Frabetti:1998zt, Gibbons:1996yv,Avery:1995ps},
$\Xi_c(2790)~J^P=1/2^-$ \cite{Nakamura:2010zzi, Csorna:2000hw} and
$\Xi_c(2815)~J^P=3/2^-$
\cite{Nakamura:2010zzi,Lesiak:2008wz,Alexander:1999ud}. While $\Xi_c(2645)$
cannot be identified with any of our states for $J=1/2$ and $J=3/2$ due to
parity, the $\Xi_c(2790)$ might be assigned with one of the six resonances in
the $J=1/2$ sector. The experimental $J^P=3/2^-~\Xi_c(2815)~$ resonance will
be analyzed in the $J=3/2$ sector.

The state $\Xi_c(2790)$ has a width of $\Gamma<12-15\,\MeV$ and it decays to
$\Xi_c' \pi$, with $\Xi_c' \rightarrow \Xi_c \gamma$. We might associate it to
our $2733$, $2775.4$ or $2804.8$ states. Because the small coupling of
$2775.4$ to the $ \Xi'_c \pi$ channel, it seems unlikely that it might
correspond to the observed $\Xi_c(2790)$ state. In fact, the assignment to the
$2804.8$ state might be better since its larger $\Xi'_c\pi$ coupling and the
fact that a slight modification of the subtraction point can lower its
position to $2790\,\MeV$ and most probably reduce its width as it will get
closer to the $\Xi_c' \pi$ channel, the only channel open at those energies
that couples to this resonance. Moreover, this seems to be a reasonable
assumption in view of the fact that, in this manner, this $\Xi_c$ state is the
HQSS partner of the $2845$ $\Xi_c^*$ state, which we will identify with the
$\Xi_c(2815)$ resonance of the PDG. Nevertheless, it is also possible to
identify our pole at $2733\,\MeV$ from the {\bf 168} irreducible
representation with the experimental $\Xi_c(2790)$ state. In that case, one
might expect that if the resonance position gets closer to the physical mass
of $2790\,\MeV$, its width will increase and it will easily reach values of
the order of $10\,\MeV$.

In Ref.~\cite{JimenezTejero:2009vq} five baryon resonances were found in this
sector for a wide range of energies up to $2977\,\MeV$. As discussed in this
reference, none of these five states seemed to fit the experimental
$\Xi_c(2790)$ because of the small width observed. Higher-mass experimental
states, such as the $\Xi_c(2980)$ \cite{Nakamura:2010zzi,
  Aubert:2007dt,Lesiak:2008wz,Chistov:2006zj}, might correspond to one of the
two higher mass states in Ref.~\cite{JimenezTejero:2009vq}. In our
calculation, none of the states can be identified with such a heavy resonant
state.  In Ref.~\cite{Hofmann:2005sw} three resonances appear below $3\,\GeV$:
$2691\,\MeV$, $2793\,\MeV$, and $2806\,\MeV$, which mostly couple to $D
\Sigma$, $\bar K \Sigma_c$, and $D \Lambda$, respectively. Those states are
very similar in mass to some of those obtained in our calculations and we
might identify the first two states, 2691 and 2793, to our 2699.4 and 2804.8
states because of the dominant decay channel.



\subsubsection{Sector  $J=3/2$}

The 26 channels (thresholds in MeV are also given) in the $\Xi_c^*$ sector are:

\smallskip
\noindent
\begin{tabular}{ccccccc}
$ \Xi_c^* \pi $   & $ \Sigma_c^*  \bar K $      & $ \Lambda D^* $   & $ \Lambda_c \bar K^* $    
& $ \Xi_c^* \eta $    & $ \Sigma D^* $     & $ \Xi_c \rho $  \\
  2784.4  &  3013.6  &  3124.0  &  3180.3  &  3193.8  &  3201.5  & 3244.9   
\end{tabular}

\smallskip
\noindent
\begin{tabular}{ccccccc}
  $ \Sigma^* D $   & $ \Xi_c \omega $  & $ \Omega_c^* K $   & $ \Sigma_c \bar K^* $      & $ \Xi'_c \rho $  
    & $ \Xi'_c \omega $    
& $ \Sigma^* D^* $    \\
  3251.8  &  3252.0 & 3264.0  &  3347.4  &   3352.3  &  3359.4  &  3392.9  
\end{tabular}

\smallskip
\noindent
\begin{tabular}{ccccccc}
  $ \Sigma_c^* \bar K^* $     & $ \Xi_c^* \rho $  & $ \Xi_c^* \omega $   & $ \Xi D_s^* $  
&$ \Xi_c \phi $   & $ \Xi^* D_s $      & $ \Omega_c K^* $   \\
  3411.8  & 3421.8  &  3428.9  &  3430.4 & 3488.9   &  3501.9  &  3591.4  
\end{tabular}

\smallskip
\noindent
\begin{tabular}{ccccc}
  $ \Xi'_c \phi $    
& $ \Xi_c^* \eta' $    & $ \Xi^* D_s^* $     & $ \Omega_c^* K^* $  & $ \Xi_c^* \phi $ \\
  3596.3  &  3604.1  &  3645.7  &  3662.2 &  3665.8 
\end{tabular}
\smallskip

\begin{figure}[h]
\begin{center}
\includegraphics[scale=0.65]{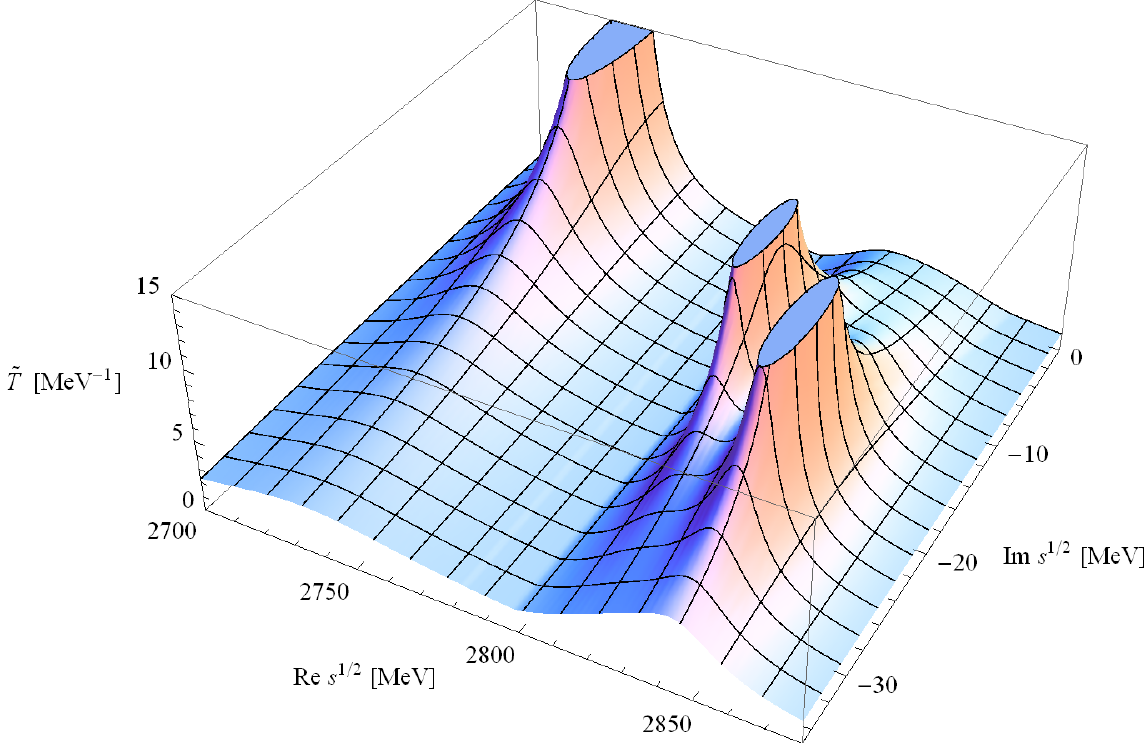}
\medskip
\vskip0.3cm 
\end{center}
\caption{ (Color online) $\tilde T^{I=\frac{1}{2},~J=\frac{3}{2},~S=-1,~C=1}(s) $ amplitude ($\Xi_c^*$ resonances).}
\label{fig1-112}
\end{figure}

The resonances predicted by the model and generated from the {\bf 120} and
{\bf 168} irreducible representations are compiled in Table \ref{tab11-11} and Fig.~\ref{fig1-112}.

The only experimental $J^P=3/2^-$ baryon resonance with a mass in the energy
region of interest is $\Xi_c(2815)$
\cite{Nakamura:2010zzi,Lesiak:2008wz,Alexander:1999ud}. The full width is
expected to be less than $3.5\,\MeV$ for $\Xi_c^+(2815)$ and less than
$6.5\,\MeV$ for $\Xi_c^0(2815)$, and the decay modes are $\Xi_{c+} \pi^+
\pi^-$, $\Xi_{c0} \pi^+ \pi^-$. We obtain two resonances at $2819.7\,\MeV$ and
$2845.2\,\MeV$, respectively, that couple strongly to $\Xi_c^* \pi$, with
$\Xi_c^* \rightarrow \Xi_c \pi$. Allowing for this possible indirect
three-body decay channel, we might identify one of them to the experimental
result. This assignment is, indeed, possible for the state at $2845.2\,\MeV$
if we slightly change the subtraction point. In this way, we will lower its
position and reduce its width as it gets closer to the open $\Xi_c^* \pi$
channel.

In Refs.~\cite{Hofmann:2005sw,Hofmann:2006qx} a resonance with a similar
energy of $2838\,\MeV$ and width of $16\,\MeV$ was identified with the $\Xi_c(2815)$.
It was suggested that its small width reflected its small coupling strength to
the $\Xi_c \pi$ channel.


\subsection{${\bm \Omega_c}$  states 
($\bm{  C = 1}$, $\bm{ S = -2} $, $\bm{ I=0}$ ) }

In this section we will discuss the $C=1$, $S=-2$ and $I=0$ resonant states
with $J=1/2$ and $J=3/2$ coming from the ${\bf 120}$ and ${\bf 168}$ SU(8)
representations. States with the $I=1$ and the $J= 5/2$ belong to the ${\bf
  4752}$-plet and are not discussed in this work.

\subsubsection{Sector $J=1/2$}

The 15 physical baryon-meson pairs that are incorporated in the $I=0$, $J=1/2$
sector are as follows:

\smallskip
\noindent
\begin{tabular}{cccccc}
$\Xi_{c} \bar K$   & $\Xi'_{c} \bar K$   & $\Xi D$      & $\Omega_{c} \eta$    
&$\Xi D^{*}$       &$\Xi_{c} \bar K^{*}$   \\
2965.1  & 3072.5  & 3185.3 & 3245.0 & 3326.5 & 3363.3 
\end{tabular}

\smallskip
\noindent
\begin{tabular}{cccccc}
  $\Xi'_{c} \bar K^{*}$ & $\Omega_{c} \omega$ &$\Xi_{c}^{*} \bar K^{*}$
  &$\Xi^{*} D^{*}$ &$\Omega_{c}^{*} \omega$ & $\Omega_{c} \eta'$ \\
 3470.7 & 3480.1 & 3540.2 & 3541.8 & 3550.9 & 3655.3
\end{tabular}

\smallskip
\noindent
\begin{tabular}{ccc}
   $\Omega_{c} \phi$      &$\Omega D_{s}^{*}$   &$\Omega_{c}^{*} \phi$ \\
 3717.0 & 3784.8 & 3787.8
\end{tabular}
\smallskip

\begin{figure}[h]
\begin{center}
\includegraphics[scale=0.65]{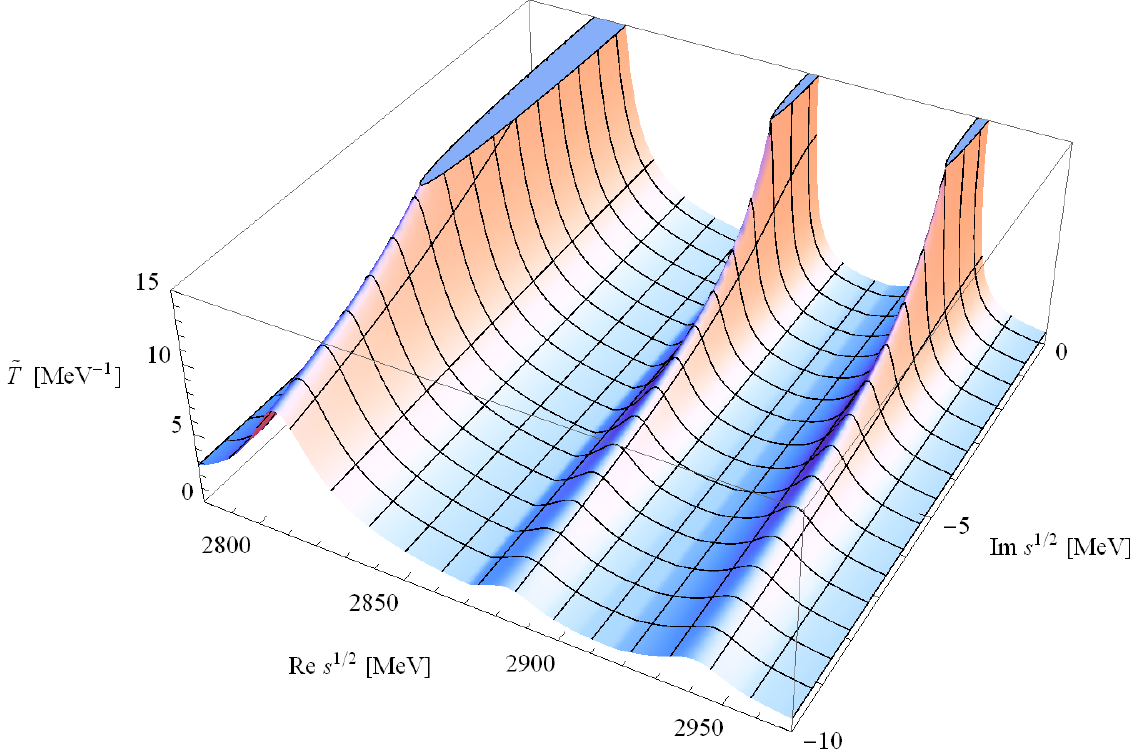}
\medskip
\vskip0.3cm 
\end{center}
\caption{ (Color online) $\tilde T^{I=0,~J=\frac{1}{2},~S=-2,~C=1}(s) $ amplitude ($\Omega_c$ resonances).}
\label{fig1-201}
\end{figure}

According to our analysis, there are three bound states which can be generated
dynamically as baryon-meson molecular entities resulting from the strongly
attractive representations of the SU(8) group. In Table~\ref{tab1-201} and
Fig.~\ref{fig1-201} we show the masses, widths, and the largest couplings of
those poles to the scattering channels.

\begin{table*}[ht]
\begin{center}
\begin{tabular}{| c | c | c | c | c | c | c |}
\hline 
 SU(8) & SU(6) & SU(3) & & & Couplings &  \\

irrep & irrep & irrep & $M_{R}$ & $\Gamma_{R}$ &  to main channels & $J$ \\ 

\hline

$\bf 168$ & $\bf 21_{2,1}$ & $\bf 6_2$ &
2810.9 & 0.0 & $g_{\Xi D} = 3.3$, $g_{\Xi D^{*}} = 1.7$, $g_{\Xi_{c} \bar
  K^{*}}=0.9$, & $1/2$ \\
 & & & & &  $g_{\Xi^{*} D^{*}} = 4.8$, $g_{\Omega_{c} \eta'}=0.9$, $g_{\Omega
  D_{s}^{*}} = 4.2 $ & \\

\hline

$\bf 168$ & $\bf 21_{2,1}$ & $\bf 6_4$ &
2814.3 & 0.0 & $g_{\Xi D^{*}}=3.7$, $g_{\Xi^{*} D}=3.1$, $g_{\Xi^{*}
  D^{*}}=3.8$, & $3/2$ \\
 & & & & & $g_{\Omega D_{s}}=2.7$, $g_{\Omega_{c}^{*} \eta'}=0.9$, $g_{\Omega
  D_{s}^{*}}=3.4$ & \\

\hline

$\bf 168$ & $\bf 15_{2,1}$ & $\bf 6_2$ &
2884.5 & 0.0 & $g_{\Xi_{c} \bar K} = 2.1$, $g_{\Xi D^{*}} = 1.7$, $g_{\Xi'_{c}
  \bar K^{*}} = 1.5$, & $1/2$ \\  
 & & & & & $g_{\Xi_{c}^{*} \bar K^{*}} = 1.8$, $g_{\Omega_{c} \phi}=0.9$,
$g_{\Omega_{c}^{*} \phi} = 1.1$ & \\

\hline

$\bf 120$ & $\bf 21_{2,1}$ & $\bf 6_2$ &
2941.6 & 0.0 & $g_{\Xi'_{c} \bar K} = 1.9$, $g_{\Xi D} = 1.5$, $g_{\Omega_{c}
  \eta} = 1.7$, & $1/2$ \\
 & & & & & $g_{\Xi_{c} \bar K^{*}} = 1.4$, $g_{\Xi'_{c} \bar K^{*}} = 1.1$,
$g_{\Omega_{c} \phi} = 1.0$, & \\
 & & & & & $g_{\Omega D_{s}^{*}}=0.9$ & \\

\hline

$\bf 120$ & $\bf 21_{2,1}$ & $\bf 6_4$ &
2980.0 & 0.0 & $g_{\Xi_{c}^{*} \bar K}=1.9$, $g_{\Omega_{c}^{*} \eta}=1.6$,
$g_{\Xi D^{*}}=1.4$, & $3/2$ \\
 & & & & & $g_{\Xi_{c} \bar K^{*}}=1.6$, $g_{\Xi_{c}^{*} \bar K^{*}}=1.3$,
$g_{\Omega_{c}^{*} \phi}=1.2$ & \\

\hline
\end{tabular}
\end{center}
\caption{$\Omega_c$ and $\Omega_c^*$ resonances.}
\label{tab1-201}
\end{table*}

There is no experimental information on those excited states. However, our
predictions can be compared to recent calculations of
Refs.~\cite{JimenezTejero:2009vq,Hofmann:2005sw}. In
Ref.~\cite{JimenezTejero:2009vq} three resonances were found, one with mass
$M_1=2959\,\MeV$ and width $\Gamma_{1}=0\,\MeV$, a second one with $M_2=2966\,\MeV$
and $\Gamma_{2}=1.1\,\MeV$, and the third one with $M_3=3117\,\MeV$ and
$\Gamma_{3}=16\,\MeV$.  The dominant baryon-meson channels are $\bar K \Xi'_c$,
$\bar K \Xi'_c$, and $D \Xi$, respectively. Three resonant states with lower
masses were also observed in Ref.~\cite{Hofmann:2005sw}, but with slightly
different dominant coupled channels.

In both previous references, vector baryon-meson channels were not considered,
breaking in this manner HQSS. In fact, it is worth noticing that the coupling
to vector baryon-meson states play an important role in the generation of the
baryon resonances in this sector. Furthermore, we have checked that other
states stemming to the ${\bf 4752}$-plet with the same quantum numbers might
be seen in this energy region and, therefore, a straightforward identification
of our states with the results of
Refs.~\cite{JimenezTejero:2009vq,Hofmann:2005sw} might not be possible.


\subsubsection{Sector  $J=3/2$}

In the $C=1$, $S=-2$, $I=0$, $J=3/2$ sector, there are 15 coupled channels:

\smallskip
\noindent
\begin{tabular}{cccccc}
$\Xi_{c}^{*} \bar K$ & $\Omega_{c}^{*} \eta$ & $\Xi D^{*}$ & $\Xi_{c} \bar K^{*}$
 &$\Xi^{*} D$ & $\Xi'_{c} \bar K^{*}$ \\
 3142.0  & 3315.8  &  3326.5  & 3363.3  & 3400.6  & 3470.7   
\end{tabular}

\smallskip
\noindent
\begin{tabular}{cccccc}
  $\Omega_{c} \omega$   &$\Xi_{c}^{*} \bar K^{*}$ & $\Xi^{*} D^{*}$  &$\Omega_{c}^{*} \omega$    &$\Omega D_{s}$ 
& $\Omega_{c} \phi$   \\
3480.1  & 3540.2 &3541.8  & 3550.9  & 3641.0  & 3717.0  
\end{tabular}

\smallskip
\noindent
\begin{tabular}{ccc}
    $\Omega_{c}^{*} \eta'$      &$\Omega D_{s}^{*}$   &$\Omega_{c}^{*} \phi$    \\
 3726.1  &  3784.8  & 3787.8
\end{tabular}
\smallskip

\begin{figure}[h]
\begin{center}
\includegraphics[scale=0.65]{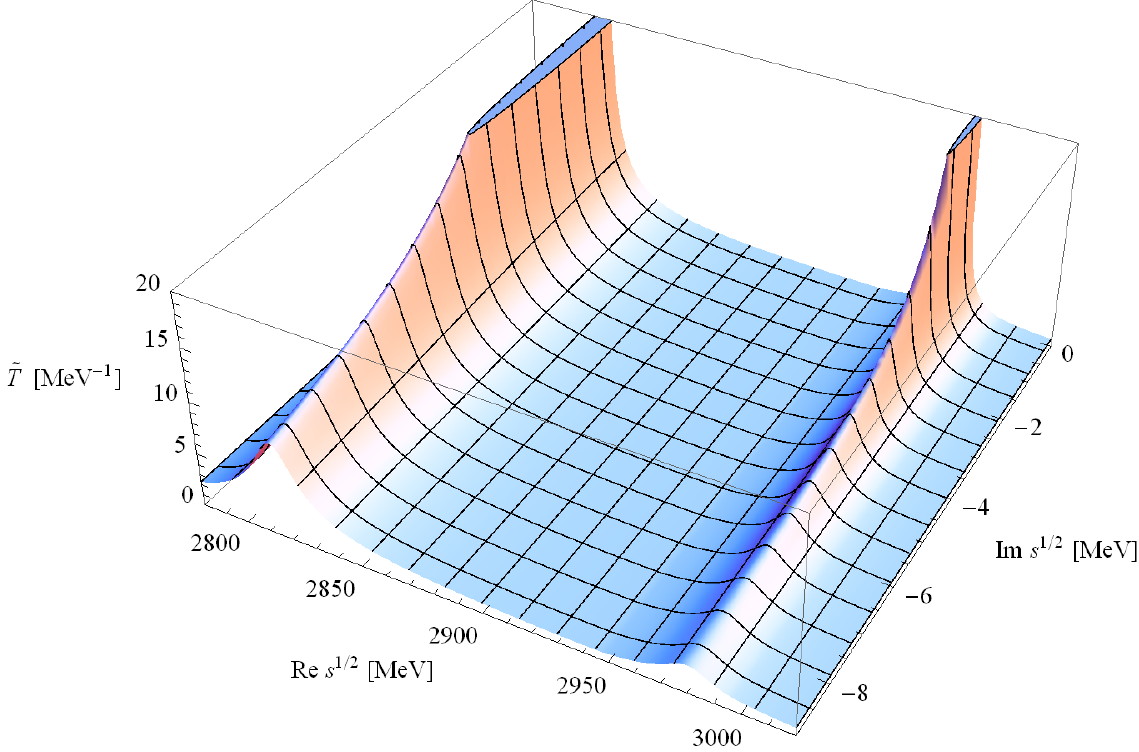}
\medskip
\vskip0.3cm 
\end{center}
\caption{ (Color online) $\tilde T^{I=0,~J=\frac{3}{2},~S=-2,~C=1}(s) $ amplitude ($\Omega_c^*$ resonances).}
\label{fig1-202}
\end{figure}

We obtain two bound $\Omega_c^*$ states (Table \ref{tab1-201} and
Fig.~\ref{fig1-202}), with masses $2814.3$, and
$2980.0$, which mainly couple to $\Xi D^*$ and $\Xi^* D^*$, and to $\Xi_c^* \bar K$,
respectively. As seen in the $J=1/2$ sector, no experimental information is
available. In Ref.~\cite{Hofmann:2006qx}, two states at $2843\,\MeV$ and $3008\,\MeV$
with zero width were found. Those states couple most strongly to $D \Xi$ and
$\bar K \Xi_c$, respectively, so an identification between the resonances in
both models is not possible.

\subsection{${\bm \Xi_{cc}}$  states 
($\bm{  C = 2}$, $\bm{ S = 0} $, $\bm{ I=1/2}$ ) }


In the $C=2$ sector no experimental information is available yet. Therefore,
all our results are merely predictions of the SU(8) WT model.

\subsubsection{Sector $J=1/2$}

The 22 channels (thresholds are also given in MeV) for $C=2$, $S=0$, $I=1/2$ and $J=1/2$, are as follows:

\smallskip
\noindent
\begin{tabular}{cccccc}
$ \Xi_{cc} \pi $      & $ \Xi_{cc} \eta  $   & $  \Lambda_c D  $    & $ \Omega_{cc} K$   
& $ \Xi_{cc} \rho  $    & $  \Lambda_c D^*  $   \\
 3657.0 & 4066.5  & 4153.7   &  4207.7 & 4294.5  &  4294.8  
\end{tabular}

\smallskip
\noindent
\begin{tabular}{cccccc}
  $ \Xi_{cc} \omega  $  & $ \Sigma_c D $ & $ \Xi_{cc}^* \rho  $    &  $ \Xi_{cc}^* \omega  $& $ \Xi_c D_s $   & $ \Sigma_c D^*  $    \\
   4301.6    & 4320.8  &   4375.5  & 4382.6  & 4438.0   &  4461.9 
\end{tabular}

\smallskip
\noindent
\begin{tabular}{cccccc}
 $ \Xi_{cc} \eta'$    $ \Sigma_c^* D^*  $     & $ \Xi_{cc} \phi $  & $  \Xi'_c D_s  $   & $\Xi_c D_s^*$ &$ \Omega_{cc} K^* $     \\
  4476.8   4526.3   &   4538.5   &  4545.4  &  4581.8   & 4605.9 
\end{tabular}

\smallskip
\noindent
\begin{tabular}{cccc}
 $ \Xi_{cc}^* \phi  $  & $ \Omega_{cc}^* K^* $   & $ \Xi'_c D_s^*  $   & $ \Xi_c^* D_s^*  $    \\
 4619.5   &   4688.9 & 4689.2  &  4758.7  
\end{tabular}
\smallskip

\begin{table*}[ht]
\begin{center}
\begin{tabular}{| c | c | c | c | c | c | c |}
\hline 
 SU(8) & SU(6)  & SU(3) &  & & Couplings & \\
 irrep    & irrep   & irrep   & $M_{R}$ & $\Gamma_{R}$ &  to main channels  &
 $J$ \\ 

\hline

$\bf 168$ & $\bf 6_{1,2}$ & $\bf 3_2$ &
3698.1 & 1.3 & $\bf{g_{\Xi_{cc} \pi}=0.3}$, $g_{\Lambda_c D^*}=2.1$,
$g_{\Sigma_c D}=3.2$, $g_{\Sigma_c D^*}=2.6$, & $1/2$ \\
 &  & & &   & $g_{\Sigma_c^* D^*}=4.1$, $g_{\Xi'_c D_s}=1.3$, $g_{ \Xi_c
  D_s^*}=1.4$, $g_{\Xi'_c D_s^*}=1.1$, & \\
 &  & & &   & $g_{\Xi_c^* D_s^*}=1.7$ & \\

\hline

$\bf 120$ & $\bf 6_{3,2}$ & $\bf 3_2$ &
3727.8 & 17.8 & $\bf{g_{\Xi_{cc} \pi}=1.0}$, $g_{\Lambda_c D}=2.0$,
$g_{\Sigma_c D}=1.1$, $g_{\Xi_c D_s}=1.5$, & $1/2$ \\
 &  & & &     & $g_{\Sigma_c D^*}=4.6$, $g_{\Xi_{cc} \eta'}=1.4$, 
$g_{\Xi_{cc}^* \rho}= 0.9$,
$g_{\Sigma_c^* D^*}=3.6$,& \\
 &  & & &     & $g_{\Xi'_c D_s^*}=2.0$,  $g_{\Xi_c^* D_s^* }=1.6$ & \\

\hline

$\bf 168$ & $\bf 6_{3,2}$ & $\bf 3_4$ &
3729.5 & 0.0 & $g_{ \Lambda_c D^*}=1.2$, $g_{ \Sigma_c^* D}=2.9$, $g_{\Sigma_c
  D^*}=1.8$, $g_{\Sigma_c^* D^*}=3.7$ & $3/2$ \\
 &  & & &   & $g_{ \Xi_c D_s^*}=1.3$, $g_{\Xi_c^* D_s}=1.2$,
$g_{\Xi_{cc}^* \eta'}=1.1$, $g_{ \Xi_c^* D_s^*}=1.5$ & \\

\hline

$\bf 168$ & $\bf 6_{3,2}$ & $\bf 3_2$ &
3727.4 & 120.2 & $\bf{g_{\Xi_{cc} \pi}=2.4}$, $g_{\Lambda_c D}=2.4$,
$g_{\Lambda_c D^*}=1.5$, $g_{\Sigma_c D^*}=2.3$, & $1/2$ \\
 &  & & &     & $g_{ \Sigma_c^* D^*}=1.4$, $g_{\Xi'_c D_s^*}=1.0$ & \\

\hline

$\bf 120$ & $\bf 6_{3,2}$ & $\bf 3_4$ &
3790.8 & 83.9 &  $\bf{g_{\Xi_{cc}^* \pi}=2.0}$, $g_{\Lambda_c D^*}=2.9$,
$g_{\Sigma_c^* D}=0.8$, $g_{\Sigma_c^* D^*}=1.1$, & $3/2$ \\
 &  & & &     & $g_{\Xi_c D_s^*}=0.8$, $g_{\Xi_c^* D_s}=0.8$,
$g_{\Xi_{cc}^* \eta'}=0.8$, $g_{\Xi_c^* D_s^*}=1.$ & \\


\hline

\end{tabular}
\end{center}
\caption{$\Xi_{cc}$ and $\Xi_{cc}^*$ resonances.  In this case, the HQSS
    classification differs from the SU(8) classification for the two HQSS
    doublets.}  
\label{tab2011}
\end{table*}

\begin{figure}[h]
\begin{center}
\includegraphics[scale=0.65]{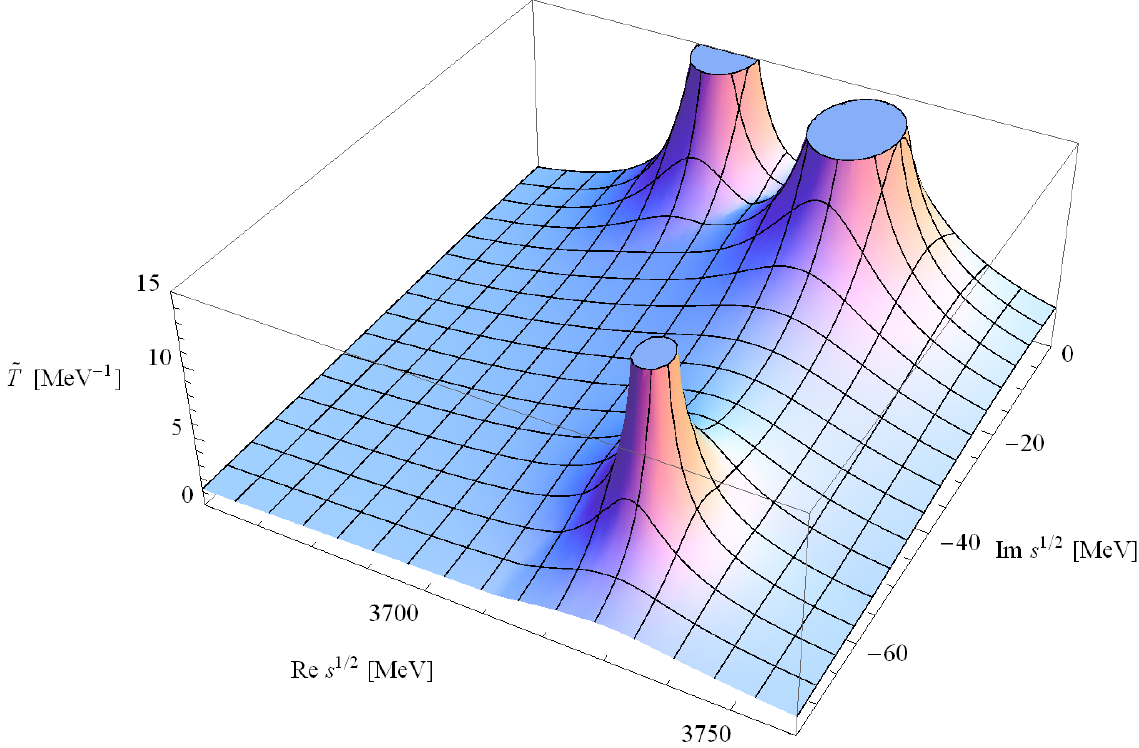}
\medskip
\vskip0.3cm 
\end{center}
\caption{ (Color online) $\tilde T^{I=\frac{1}{2},~J=\frac{1}{2},~S=0,~C=2}(s) $ amplitude ($\Xi_{cc}$ resonances).}
\label{fig2011}
\end{figure}

The three predicted poles in the $\Xi_{cc}$ sector can be seen in the
Table~\ref{tab2011} and Fig.~\ref{fig2011} together with the width and couplings to the main
channels. Their masses are $3698.1$, $3727.4$ and $3727.8\,\MeV$,
with widths $1.3$, $120.2$ and $17.8\,\MeV$, respectively. The dominant channels for the
generation of those states are, in order, $\Sigma_c^* D^*$, $\Xi_{cc} \pi$ and $\Lambda_c D$, and
$\Sigma_c D^*$.  In Ref.~\cite{Hofmann:2005sw} six states were found, two of
them coming from the weak interaction of the open-charm mesons and open-charm
baryons in the SU(4) anti-sextet and 15-plet. In this paper, we only consider
those states coming from the strongly attractive SU(8) {\bf 120}- and {\bf
  168}-plets. Therefore, only three states are expected in this sector.
Moreover, an identification among resonances in both models is complicated
because the strong coupling of our states to channels with vector mesons, not
considered in this previous reference.


\subsubsection{Sector $J=3/2$}

In the $\Xi_{cc}^*$ sector, the following 20 channels are coupled:

\smallskip
\noindent
\begin{tabular}{ccccccc}
$\Xi_{cc}^* \pi $   & $\Xi_{cc}^* \eta$      &   $  \Omega_{cc}^* K $   & $ \Xi_{cc} \rho $   & $  \Lambda_c D^*  $   
& $ \Xi_{cc} \omega $   &   $ \Xi_{cc}^* \rho $        \\
3738.0  & 4147.5 &    4290.7&  4294.5  & 4294.8   & 4301.6  & 4375.5  
\end{tabular}

\smallskip
\noindent
\begin{tabular}{ccccccc}
    $\Xi_{cc}^* \omega$ &   $  \Sigma_c^* D  $  & $\Sigma_c D^*  $   &  $ \Sigma_c^* D^*  $    & $ \Xi_{cc} \phi $   & $ \Xi_{cc}^* \eta'  $   &  $ \Xi_{c} D_s^* $   \\
   4382.6 &    4385.2     & 4461.9  & 4526.3     &  4538.5  &   4557.8 &  4581.8  
\end{tabular}

\smallskip
\noindent
\begin{tabular}{cccccc}
    $  \Omega_{cc} K^* $ & $\Xi_c^* D_s $  &   $ \Xi_{cc}^* \phi  $ &  $ \Omega_{cc}^* K^*  $  & $ \Xi'_c D_s^* $  & $ \Xi_c^* D_s^*  $    \\
   4605.9 &  4614.9   & 4619.5  &  4688.9  & 4689.2    &  4758.7    
\end{tabular}
\smallskip

\begin{figure}[h]
\begin{center}
\includegraphics[scale=0.65]{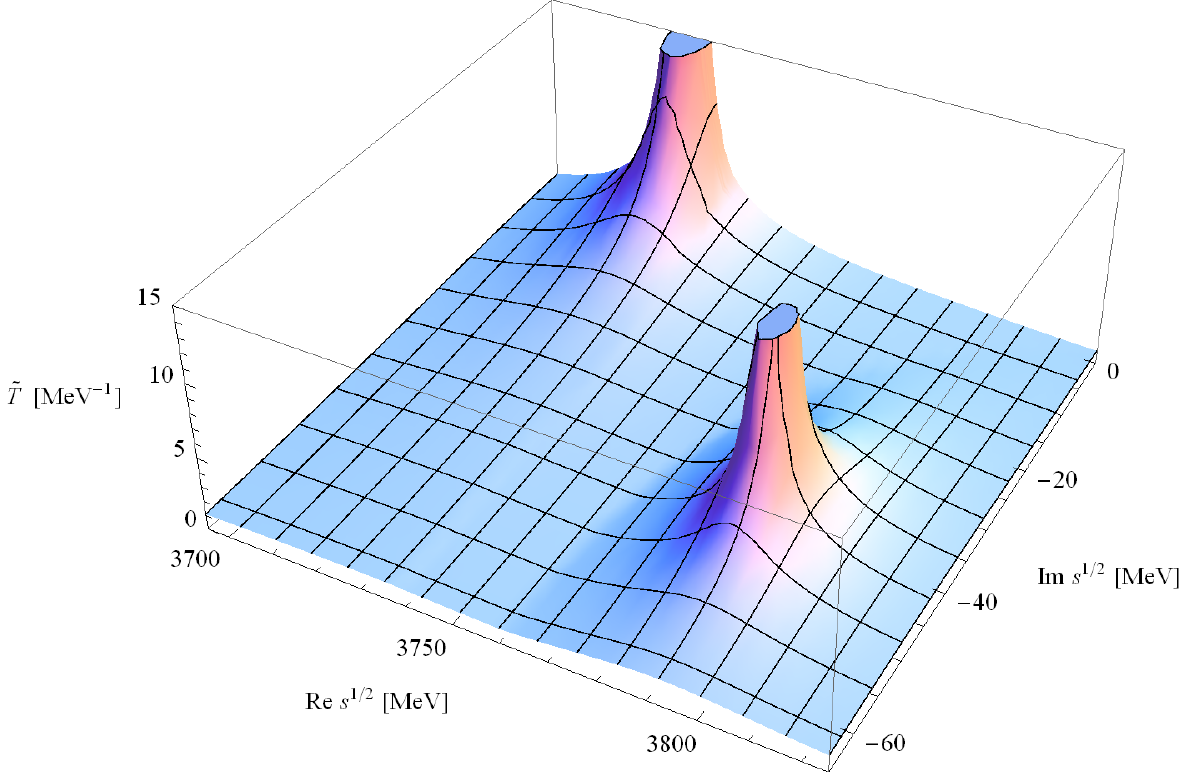}
\medskip
\vskip0.3cm 
\end{center}
\caption{ (Color online) $\tilde T^{I=\frac{1}{2},~J=\frac{3}{2},~S=0,~C=2}(s)
  $ amplitude ($\Xi_{cc}^*$ resonances).}
\label{fig2012}
\end{figure}
\
Two  states, with masses $3729.5$ and $3790.8\,\MeV$ have been obtained,
which couple mainly to $ \Sigma_c^* D$ and $ \Sigma_c^* D^*$, and to $\Xi_{cc}^* \pi$ and $\Lambda_c
D^*$, respectively (see Table~\ref{tab2011} and Fig.~\ref{fig2012}). 

In Ref.~\cite{Hofmann:2006qx}, 
two states were obtained at $3671\,\MeV$ and 
$3723\,\MeV$, with dominant coupling to the channels $ \Sigma_c D $ and $ \Xi_{cc} \pi$,
respectively. However, the analysis there was done on the basis that the
$\Xi_{cc}(3519)$ resonance found in  is, in fact, a $J^P=3/2^+$ state,
whereas, in our calculation this resonance is the ground state, $J^P=1/2^+$.
It is argued in~\cite{Hofmann:2006qx} that the second resonance should be more reliable in view of
the dominant coupling to a baryon-Goldstone boson. Moreover, it was
mentioned the necessity of implementing heavy-quark symmetry by incorporating
$0^-$ and $1^-$ charmed mesons as well as $1/2^+$ and $3/2^+$ baryon in the
coupled-channel basis. Therefore, in both models, the implementation of
heavy-quark symmetry seems to move to higher energies the expected resonant
states as well as to change their dominant molecular components.


\subsection{${\bm \Omega_{cc}}$ states 
($\bm{  C = 2}$, $\bm{ S = -1} $, $\bm{ I=0}$ ) }


\subsubsection{Sector $J=1/2$}

The 17 channels in this  $\Omega_{cc}$ sector are as follows:

\smallskip
\noindent
\begin{tabular}{ccccccc}
$\Xi_{cc} \bar K $   & $ \Omega_{cc} \eta  $   & $   \Xi_{c} D  $    & $  \Xi_{cc} \bar K^*  $    & $ \Xi'_c D   $   & $ \Xi_c D^*  $ &  $ \Xi_{cc}^* \bar K^*  $ \\
 4014.7  &  4259.5    &  4336.7  &  4412.9  &   4444.1  &   4477.8    &   4493.9 
\end{tabular}

\smallskip
\noindent
\begin{tabular}{ccccccc}
   $  \Omega_{cc} \omega  $       & $  \Omega_{cc}^* \omega   $ & $\Xi'_c D^* $  & $ \Xi_c^* D^* $      & $  \Omega_c D_s$ &   $ \Omega_{cc} \eta'  $&  $ \Omega_{cc} \phi  $    \\
  4494.6    &   4577.6 &  4585.2  & 4654.7   &  4666.0   & 4669.8&  4731.5 
\end{tabular}

\smallskip
\noindent
\begin{tabular}{ccc}
$ \Omega_c D_s^*  $ & $  \Omega_{cc}^* \phi  $     & $ \Omega_c^* D_s^*  $      \\
 4809.8 &  4814.5  &  4880.6    
\end{tabular}
\smallskip

\begin{figure}[h]
\begin{center}
\includegraphics[scale=0.65]{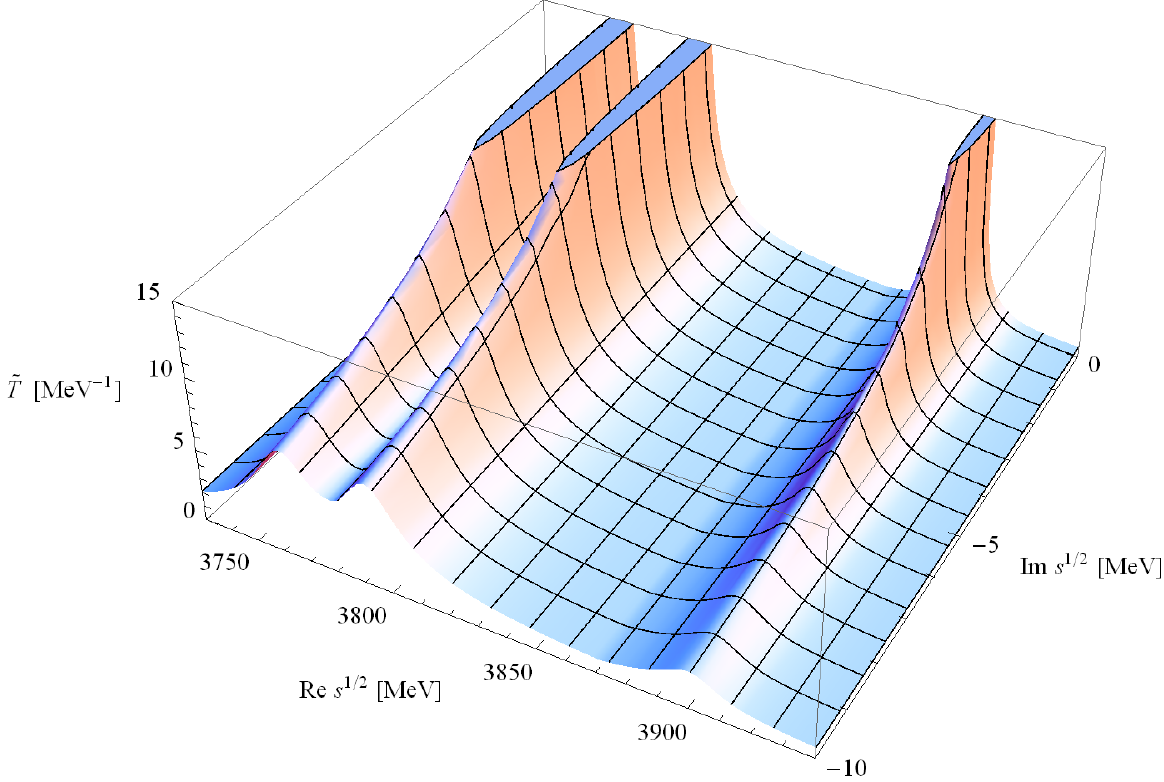}
\medskip
\vskip0.3cm 
\end{center}
\caption{ (Color online) $\tilde T^{I=0,~J=\frac{1}{2},~S=-1,~C=2}(s) $ amplitude ($\Omega_{cc}$ resonances).}
\label{fig2-101}
\end{figure}

The predicted bound states are three at $3761.8\,\MeV$, $3792.8\,\MeV$, and
$3900.2\,\MeV$, coupling strongly to $\Xi_c^* D^*$, $\Xi_c' D^*$ and $\Xi_{cc}
\bar K$, respectively. They are shown in Table~\ref{tab2m101} and
Fig.~\ref{fig2-101}. In Ref.~\cite{Hofmann:2005sw} four states were generated
from the SU(4) 3-plet at $3.71\,\GeV$, $3.74\,\GeV$ and $3.81\,\GeV$ and one
coming from the SU(4) 15-plet at $4.57 \,\MeV$. We might be tempted to
identify our three states with those coming from SU(4) 3-plet in
Ref.~\cite{Hofmann:2005sw} because the similar dominant channels if we do not
consider those including vector mesons and $3/2^+$ baryons. However, the only
clear identification that can be done is between our state at $3900.2\,\MeV$
and the one in Ref.~\cite{Hofmann:2005sw} at $3.81\,\GeV$ because in this case
the dominant channels coincide. For this state, channels with vector mesons
and/or $3/2^+$ baryons do not play a significant role.
 

\begin{table*}[ht]
\begin{center}
\begin{tabular}{| c | c | c | c | c | c | c |}
\hline 
 SU(8)  & SU(6)  & SU(3) &  &  & Couplings &   \\
 irrep    & irrep   & irrep   & $M_{R}$ & $\Gamma_{R}$ &  to main channels &
 $J$ \\ 

\hline

$\bf 168$ & $\bf 6_{1,2}$ & $\bf 3_2$ &
3761.8 & 0.0 & $g_{\Xi_{c} D}=1.2$, $g_{ \Xi'_c D }=2.7$, $g_{\Xi_c D^*}=2.9$,
& $1/2$ \\
 &  & & &    & $g_{\Xi'_c D^*}=2.0$, $g_{\Xi_c^* D^*}=3.6$, $g_{\Omega_c
  D_s}=1.9$, & \\
 &  & & &    & $g_{ \Omega_c D_s^*}=1.4$, $g_{ \Omega_c^* D_s^*}=2.5$ & \\

\hline

$\bf 168$ & $\bf 6_{3,2}$ & $\bf 3_2$ &
3792.8 & 0.0 & $g_{\Xi_{cc} \bar K}=0.9$,  $g_{\Xi_{c} D}=2.3$,
$g_{\Xi'_c D}=0.9$,  & $1/2$ \\
 &  & & &    & $g_{\Omega_{cc} \eta'}=1.2$, $g_{\Xi'_c D^*}=3.5$, 
$g_{\Xi_{cc}^* \bar K^*}=1.1$, & \\
 &  & & &    & $g_{\Xi_c^* D^*}=2.7$, $g_{\Omega_c
  D_s^*}=2.6$, $g_{ \Omega_c^* D_s^* }=2.0$ & \\

\hline

$\bf 168$ & $\bf 6_{3,2}$ & $\bf 3_4$ &
3802.9 & 0.0 & $g_{\Xi_c D^*}=2.5$, $g_{\Xi_c^* D}=2.6$, $g_{\Xi'_c D^*}=1.6$,
& $3/2$ \\
 &  & & &    & $g_{\Xi_{cc}^* \bar K^*}=0.9$,
 $g_{\Xi_c^* D^*}=3.3$, $g_{\Omega_c^* D_s}=2.0$,
 & \\
 &  & & &    & $g_{\Omega_c D_s^* }=1.2$, $g_{\Omega_{cc}^* \eta'}=1.1$,
 $g_{\Omega_c^* D_s^*}=2.5$ & \\

\hline

$\bf 120$ & $\bf 6_{3,2}$ & $\bf 3_2$ &
3900.2 & 0.0 & $g_{\Xi_{cc} \bar K}=2.1$, $g_{\Omega_{cc} \eta}=1.1$,
$g_{\Xi_{c} D}=1.6$, & $1/2$ \\
 &  & & &   &  $g_{\Xi_c D^*}=0.9$, $g_{ \Xi_{cc}^* \bar K^*}=1.3$, 
$g_{\Omega_c D_s^*}=1.$ & \\

\hline

$\bf 120$ & $\bf 6_{3,2}$ & $\bf 3_4$ &
3936.3 & 0.0 & $g_{\Xi_{cc}^* \bar K}=2.1$, $g_{\Xi_{cc} \bar K^*}=1.4$,
$g_{\Omega_{cc}^* \eta}=1.$, & $3/2$ \\
 &  & & &   & $g_{\Xi_c D^*}=1.6$, 
$g_{\Xi_{cc}^* \bar K^*}=1.3$, $g_{\Omega_c^* D_s^*}=0.9$ & \\
\hline

\end{tabular}
\end{center}
\caption{$\Omega_{cc}$ and $\Omega_{cc}^*$ resonances.}
\label{tab2m101}
\end{table*}


\subsubsection{Sector $J=3/2$}

The 16 channels in the $\Omega_{cc}^*$ sector are: 

\smallskip
\noindent
\begin{tabular}{ccccccc}
$\Xi_{cc}^* \bar K $   & $  \Omega_{cc}^* \eta   $&   $ \Xi_{cc} \bar K^*  $      & $ \Xi_c D^*  $ &   $ \Xi_{cc}^* \bar K^*  $ &
$ \Omega_{cc} \omega  $   & $  \Xi_c^* D  $        \\
 4095.7  &   4342.5   &   4412.9     & 4477.8  &  4493.9 & 4494.6 &  4513.6   
\end{tabular}

\smallskip
\noindent
\begin{tabular}{cccccccc}
$ \Omega_{cc}^* \omega  $  & $ \Xi'_c D^*  $    &$\Xi_c^* D^* $     &   $ \Omega_{cc} \phi  $ & $ \Omega_c^* D_s $   & $  \Omega_{cc}^* \eta'   $ & $  \Omega_c D_s^*   $         \\
  4577.6 &  4585.2    & 4654.7    &  4731.5  &  4736.8  &  4752.8  &  4809.8   
\end{tabular}

\smallskip
\noindent
\begin{tabular}{cc}
 $ \Omega_{cc}^* \phi  $ & $ \Omega_c^* D_s^*  $       \\
   4814.5 & 4880.6 
\end{tabular}

\smallskip

\begin{figure}[h]
\begin{center}
\includegraphics[scale=0.65]{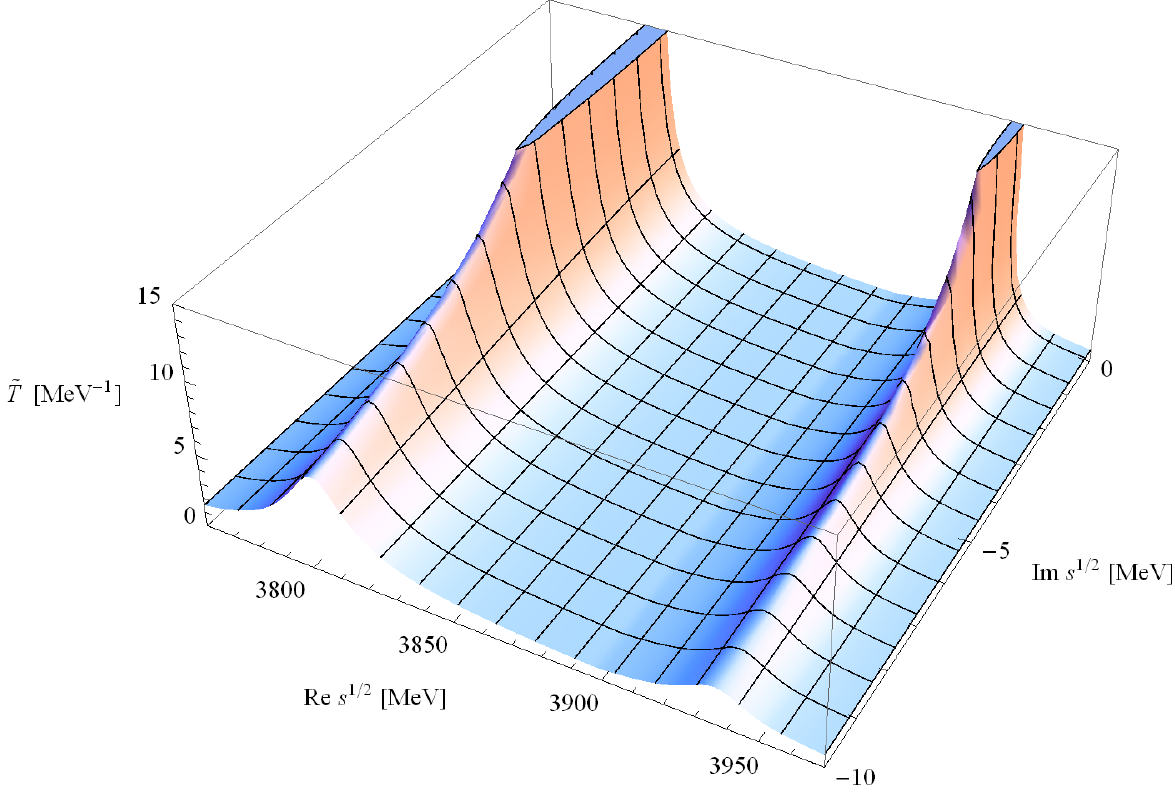}
\medskip
\vskip0.3cm 
\end{center}
\caption{ (Color online) $\tilde T^{I=0,~J=\frac{3}{2},~S=-1,~C=2}(s) $ amplitude ($\Omega_{cc}^*$ resonances).}
\label{fig2-102}
\end{figure}

Two bound states at $3802.9\,\MeV$ and $3936.3\,\MeV$ have been observed, which coupled
mostly to $\Xi_c^* D^*$ and $\Xi_{cc}^* \bar K$, respectively (see Table
\ref{tab2m101} and Fig.~\ref{fig2-102}). Compared to Ref.~\cite{Hofmann:2006qx}, we observe a similar
pattern as observed in the $C=2$, $S=0$, $I=1/2$, $J=3/2$ sector. The two expected
states are obtained with larger masses and the dominant molecular composition
incorporates a vector meson, or a vector meson and $3/2^+$ baryon state when
heavy-quark symmetry is implemented. As indicated also in
Ref.~\cite{Hofmann:2006qx}, the second resonance should be more reliable as
its main molecular contribution comes from the interaction of a baryon with a
Goldstone boson.


\subsection{${\bm \Omega_{ccc}}$  states 
($\bm{  C = 3}$, $\bm{ S = 0} $, $\bm{ I=0}$ ) }


We finally study baryon resonances with charm $C=3$ and strangeness $S=0$.

\subsubsection{Sector $J=1/2$}

The 8 coupled channels in the sector with $J=1/2$, are (thresholds in MeV are
also indicated):


\smallskip
\noindent
\begin{tabular}{cccc}
$ \Xi_{cc} D $   & $ \Xi_{cc} D^* $  &   $ \Omega_{ccc} \omega   $     &  $ \Xi_{cc}^* D^*  $          \\
5386.2  & 5527.3 & 5581.6   &5608.4    
\end{tabular}

\smallskip
\noindent
\begin{tabular}{cccc} $\Omega_{cc} D_s$&   $\Omega_{ccc} \phi $  & $  \Omega_{cc} D_s^*$ &  $ \Omega_{cc}^* D_s^*  $  \\
 5680.5 & 5818.5 & 5824.3 &    5907.3 
\end{tabular}
\smallskip

\begin{table*}[ht]
\begin{center}
\begin{tabular}{| c | c | c | c | c | c | c |}
\hline 
 SU(8)  & SU(6)  & SU(3) &  & & Couplings &   \\
 irrep    & irrep   & irrep   & $M_{R}$ & $\Gamma_{R}$ &  to main channels  &
 $J$ \\ 

\hline

$\bf 168$ & $\bf 1_{2,3}$ & $\bf 1_2$ &
4358.2 & 0.0 & $g_{ \Xi_{cc} D}=2.9$, $g_{\Omega_{cc} D_s}=1.3$, $g_{\Xi_{cc}
  D^*}=1.9$, $g_{ \Xi_{cc}^* D^*}=4.6$, & $1/2$ \\
 &  & & &   &  $g_{\Omega_{cc}^* D_s^*}=2.1$ & \\
\hline

$\bf 120$ & $\bf 1_{4,3}$ & $\bf 1_4$ &
4501.4 & 0.0 & $g_{\Xi_{cc} D^*}=2.9$,  $g_{\Xi_{cc}^* D}=2.4$,
$g_{\Omega_{cc} D_s^*}=1.8$,  $g_{\Xi_{cc}^* D^*}=2.9$,  & $3/2$  \\
 &  & & &   &  $g_{\Omega_{cc}^* D_s}=1.5$,  $g_{\Omega_{ccc} \eta'}=1.2$,
$g_{\Omega_{cc}^* D_s^*}=1.9$  & \\

\hline

\end{tabular}
\end{center}
\caption{$\Omega_{ccc}$ and $\Omega_{ccc}^*$ resonances. These two
  states are HQSS singlets.}
\label{tab3001}
\end{table*}


\begin{figure}[h]
\begin{center}
\includegraphics[scale=0.65]{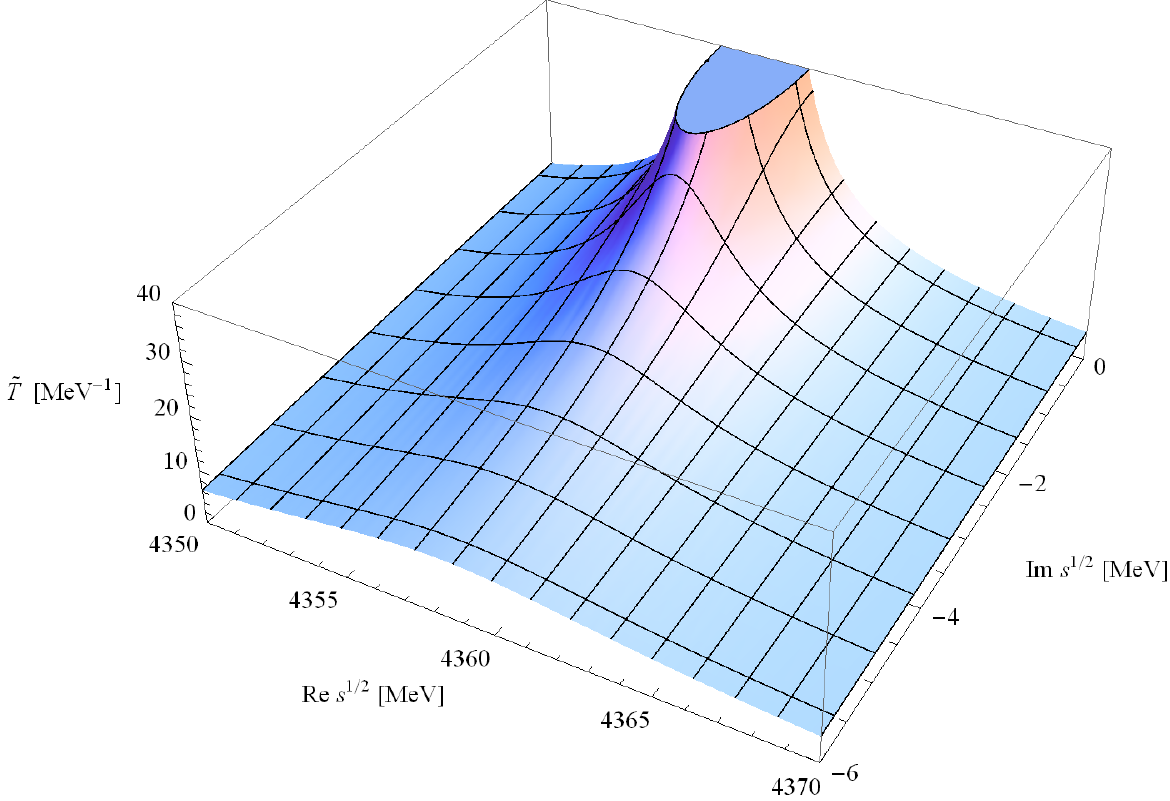}
\medskip
\vskip0.3cm 
\end{center}
\caption{ (Color online) $\tilde T^{I=0,~J=\frac{1}{2},~S=0,~C=3}(s) $ amplitude ($\Omega_{ccc}$ resonance).}
\label{fig3001}
\end{figure}

There is only one baryon state generated by the model in this sector. The mass
($4358.2\,\MeV$), width ($0\,\MeV$) and the couplings are shown in the
Table~\ref{tab3001} and Fig.~\ref{fig3001}. In Ref.~\cite{Hofmann:2005sw}, a
resonance at $4.31-4.33\,\GeV$ was also obtained. In both schemes, the
$\Omega_{ccc}$ resonance couples strongly to $\Xi_{cc} D$ and $\Omega_{cc}
D_s$ but in our SU(8) model, the dominant contribution comes from channels
with vector mesons and/or $3/2^+$ baryons.  Therefore, this result points to
the necessity of extending the coupled-channel basis to incorporate channels
with charmed vector mesons and $3/2^+$ baryons as required by heavy-quark
symmetry.


\subsubsection{ Sector $J=3/2$}

The 10 channels and thresholds (in MeV) in the sector $\Omega_{ccc}^*$  are as follows:

\smallskip
\noindent
\begin{tabular}{ccccc}
 $\Omega_{ccc} \eta$  & $\Xi_{cc}^* D $ & $\Xi_{cc} D^*$    &  $ \Omega_{ccc} \omega $    & $ \Xi_{cc}^* D^* $    \\
 5346.5 & 5467.2 & 5527.3 & 5581.6   &  5608.4 
\end{tabular}

\smallskip
\noindent
\begin{tabular}{ccccc}
 $\Omega_{ccc} \eta'$  &  $\Omega_{cc}^* D_s$   & $\Omega_{ccc} \phi $ &  $\Omega_{cc} D_s^* $&  $\Omega_{cc}^* D_s^* $   \\
  5756.8 &5763.5  &  5818.5 &  5824.3   & 5907.3 
\end{tabular}
\smallskip

\begin{figure}[h]
\begin{center}
\includegraphics[scale=0.65]{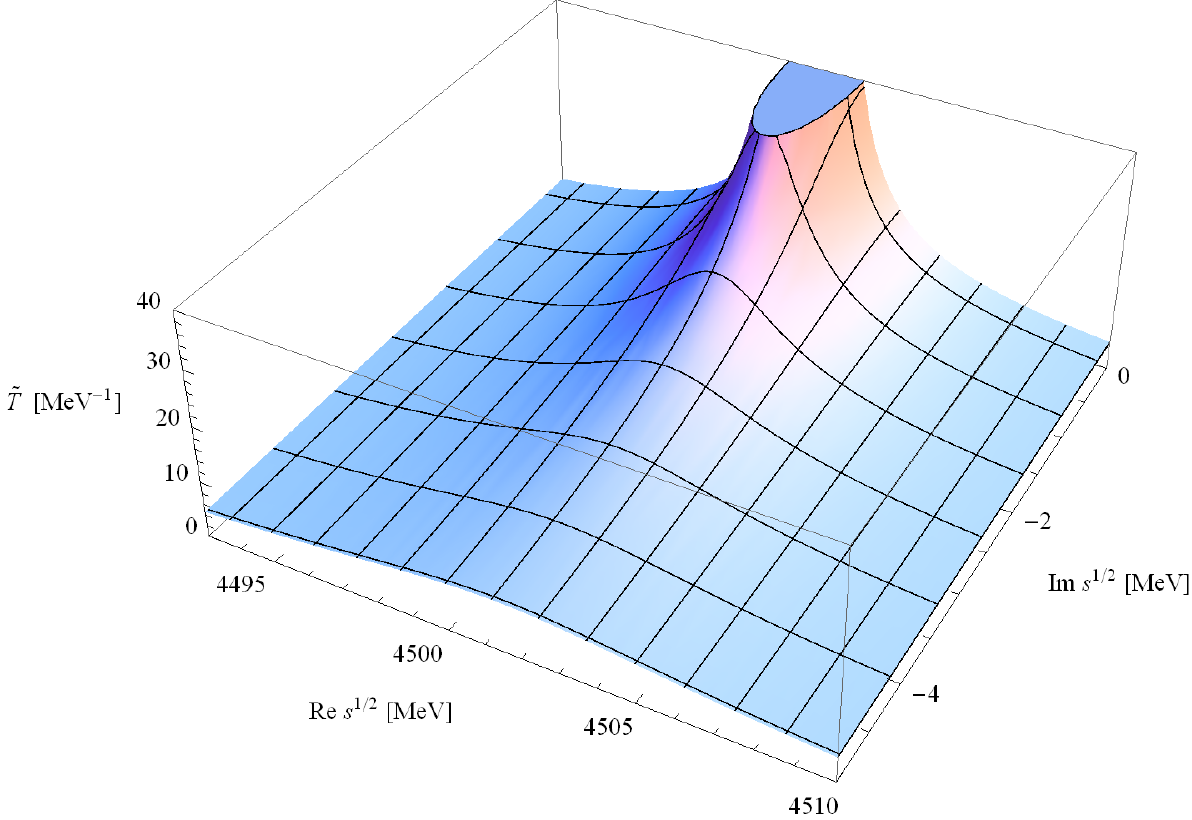}
\medskip
\vskip0.3cm 
\end{center}
\caption{ (Color online) $\tilde T^{I=0,~J=\frac{3}{2},~S=0,~C=3}(s) $
  amplitude ($\Omega_{ccc}^*$ resonance).}
\label{fig3002}
\end{figure}

The $\Omega_{ccc}^*$ resonance with $J=3/2$ has a mass approximately $1\,\GeV$
below the lowest baryon-meson threshold. This resonance stems from the ${\bf
  120}$ irrep of SU(8) and it is shown in Table~\ref{tab3001} and
Fig.~\ref{fig3002}. One resonance was also seen in Ref.~\cite{Hofmann:2006qx},
much below the first open threshold, coupling dominantly to $\Xi_{cc} D$. Our
results show that this bound state mainly couples to $\Xi_{cc} D^*$,
$\Xi_{cc}^* D^*$ and $\Xi_{cc}^* D$ states as we incorporate charmed vector
mesons and $3/2^+$ baryons according to heavy-quark symmetry. The large
  separation from the closest threshold suggests that interaction mechanisms
  involving $d$-wave could be relevant for this resonance. This remark applies
also to the $\Omega_{ccc}$ dynamically generated resonance with  $J=1/2$.

\subsection{HQSS in the results}
The factor $\SU_C(2)\times\U_C(1)$ in Eq.~(\ref{eq:4})
implements HQSS for the sectors studied in this work. The HQSS group acts by changing the coupling of spin of the
charmed quarks, relative to the spin of the block formed by light quarks. At
the level of basic hadrons, it reflects in the nearly degeneracy of $D$ and
$D^*$ mesons, which form a HQSS doublet.\footnote{We use ``doublet'' to
  indicate that only two multiplets with well-defined $CSIJ$ get mixed by the
  HQSS group. The space spanned by the eight $D$ or $D^*$ states reduces into
  two dimension four irreducible subspaces under HQSS, corresponding to the
  four spin states of $D$ and $D^*$ with given charge.} Other doublets are
$(\bar{D}_s,\bar{D}_s^*)$, and $(\eta_c,J/\psi)$ in mesons, and
$(\Sigma_c,\Sigma_c^*)$, $(\Xi_c^\prime,\Xi_c^*)$, $(\Omega_c,\Omega_c^*)$,
$(\Xi_{cc},\Xi_{cc}^*)$, $(\Omega_{cc},\Omega_{cc}^*)$, in baryons. On the
other hand, $\Lambda_c$, $\Xi_c$ and $\Omega_{ccc}$ are singlets, as are all
the other basic hadrons not containing charmed quarks. This information is
collected in Table \ref{masses}.\footnote{The classification of basic hadrons into HQSS multiplets can be
 obtained from the hadron wavefunctions in the Appendix A of
 \cite{GarciaRecio:2008dp}. For instance, for $\Sigma_c$ and $\Sigma_c^*$ the
 two light quarks are coupled to spin triplet (since they form an isospin
 triplet and color is antisymmetric) and this can give $J=1/2$ or $J=3/2$
 when coupled to the spin of the charmed quark. A systematic classification
 can be found in \cite{Falk:1991nq}.}

HQSS multiplets form also in the baryon-meson states. Specifically, in the
reduction in Eq.~(\ref{eq:6}) and Figs.~\ref{fig:21rep}, \ref{fig:15rep} and 
\ref{fig:6rep}, the pair $({\bf 6_2},{\bf 6_4})$ forms a HQSS
doublet in the reduction of ${\bf 21_{2,1}}$, while ${\bf 3^*_2}$ is a
singlet. Similarly, $({\bf 3^*_2},{\bf 3^*_4})$ in ${\bf 15_{2,1}}$, and
$({\bf 3_2},{\bf 3_4})$ in ${\bf 6_{3,2}}$, are doublets, whereas all other
$\SU(3)\times\SU(2)$ representations are HQSS singlets.

HQSS is much less broken than spin-flavor of light quarks, implemented by
$\SU(6)$, so HQSS is more visible in the results. If we imposed strict HQSS,
by setting equal masses and decay constants as required by the symmetry,
exactly degenerated HQSS multiplet would form, regardless of the amount of
breaking of SU(6). We break HQSS only through the use of physical masses and
decay constants\footnote{Most of the values in Table \ref{masses} are
    obtained from the experiment while some of them are guesses from models or
    from lattice calculations.},
but not in the interaction. Therefore we estimate that our breaking is no
larger than that present in full QCD. This suggests that the amount of
breaking we find is not an overestimation due to the model, on the contrary,
we expect to find more degeneracy than actually exists.

The approximate HQSS doublets can be observed in the results by comparing
states with equal $\SU(8)$ and $\SU(6)\times \SU_C(2)\times\U_C(1)$ labels
with $J=1/2$ and $J=3/2$. The only exception is for the $\Xi_{cc}$ states
  in Table~\ref{tab2011} where the SU(8) labels are mixed in the two
  doublets. As noted in Section \ref{sec:2.B} this reflects that exact SU(8)
  symmetry is broken in the interaction after dropping the channels with extra
  $c\bc$ pairs.

For convenience we have arranged the tables so that HQSS partners are in
consecutive rows. So, in Table \ref{tab1001}, the $\Lambda_c$ state
$2617.3\,\MeV$ with labels $({\bf 168,15_{21,},3^*_2})$ , matches the
$\Lambda_c^*$ state $2666.6\,\MeV$ with labels $({\bf
  168,15_{21,},3^*_4})$. The matching refers not only to the mass but also the
width and the couplings, taking into account that, e.g., $\Sigma_c$ in one
state corresponds to $\Sigma^*_c$ in the other. (Note that HQSS also implies
relations between couplings in the same resonance.) If the identifications in
Table \ref{tab1001} are correct, it would imply that $\Lambda_c(2595)$ is a
HQSS singlet whereas $\Lambda_c(2625)$ belongs to a doublet. Similarly, in
Table \ref{tab1011}, the $\Sigma_c$ state $2571.5\,\MeV$ is the HQSS partner
of the $\Sigma_c^*$ state $2568.4\,\MeV$ (both in $({\bf 168,21_{2,1})}$),
whereas the states $2643.4\,\MeV$ and $2692.9\,\MeV$ are partners in $({\bf
  120,21_{2,1}})$.  Of special interest is the case of $\Xi_c$ states. Here we
find that the two three star resonances $\Xi_c(2790)$ and $\Xi_c(2815)$ are
candidates to form a HQSS doublet.  Further doublets are predicted for
$\Omega_c$ and for the $C=2$ resonances, $\Xi_{cc}$ and $\Omega_{cc}$. On the
other hand, no doublet is present in the $\Omega_{ccc}$ sector. All these
considerations, follow unambiguously from the SU(8) structure if the
$\bm{168}$ and $\bm{120}$ irreps are dominant, as predicted by the extended WT
scheme.

\section{Summary}

In the present work, we have studied odd-parity charmed baryon resonances
within a coupled-channel unitary approach that implements the characteristic
features of HQSS. This implies, for instance, that $D$ and $D^*$ mesons
  have to be treated on an equal footing and that channels containing a
  different number of $c$ or $\bc$ quarks cannot be coupled. This is
accomplished by extending the $\SU(3)$ WT chiral interaction to $\SU(8)$
spin-flavor symmetry and implementing a strong flavor symmetry breaking. Thus,
our tree-level $s$-wave WT amplitudes are obtained not only by adopting the
physical hadron masses, but also by introducing the physical weak decay
constants of the mesons involved in the transitions.  Besides, and to deal
with the UV divergences that appear after summing the bubble chain implicit in
the Bethe-Salpeter equation, here we adopt the prescription of
Ref.~\cite{Hofmann:2005sw}. It amounts to force the renormalized loop function
to vanish at certain scale that depends only on $CSI$.  In this manner, we
have no free parameters. We have not refitted the subtraction points to
achieve better agreement in masses and widths of the few known states.

This scheme was first derived in Ref.~\cite{GarciaRecio:2008dp}, where results
for all non strange sectors with $C=1$ were already analyzed. Here, we have
discuss the predictions of the model for all $C=1$ strange sectors and have
also looked at the $C=2$ and $3$ predicted states. The $\SU(8)$ model
generates a great number of states, most of them stemming from the {\bf 4752}
representation. The interaction in this subspace, though attractive, is much
weaker than in the {\bf 168} and {\bf 120} ones ($-$2 vs $-$22 and
$-$16). Indeed, in the large $N_C$ limit, we expect the {\bf 4752} states will
disappear and only those related to the {\bf 168} representation will remain
\cite{GarciaRecio:2006wb}. Besides, being so weak the interaction in the {\bf
  4752} subspace, small corrections (higher orders in the expansion, $d$-wave
terms, etc) could strongly modify the properties of the states that arise from
this representation. For all this, we have restricted our study in this work
to the 288 states (counting multiplicities in spin and isospin) that stem from
the {\bf 168} and {\bf 120} representations, for which we believe the
predictions of the model are more robust.

A similar study for the light $\SU(6)$ spin-flavor sector was carried
out in Ref.~\cite{Gamermann:2011mq}. There, it was found that most of the low-lying
three- and four-star odd-parity baryon resonances with spin 1/2 and 3/2 can be
related to the {\bf 56} and {\bf 70} multiplets of the spin-flavor symmetry
group $\SU(6)$. These are precisely the charmless multiplets that appear in
the decomposition in Eq.~(\ref{eq:5}) of the {\bf 120} and {\bf 168} $\SU(8)$
representations. Thus, out of the 288 states mentioned above, we are left with
only 162 charmed states.\footnote{All exotic states, this is to say, those
  that cannot be accommodated within a simple $qqq$ picture of the baryon, lie
  in the {\bf 4752} space, and thus they have not been discussed here. Some of
  them ($C=-1$) were discussed in Ref.~\cite{Gamermann:2010zz}.}

To identify these states, we have adiabatically followed the trajectories of
the {\bf 168} and {\bf 120} poles, generated in a symmetric $\SU(8)$ world,
when the symmetry is broken down to $\SU(6) \times \SU_C(2)$ and later
$\SU(6)$ is broken down to $\SU(3) \times \SU(2)$. In this way, we have been
able to assign well-defined $\SU(8)$, $\SU(6)$ and $\SU(3)$ labels to the
resonances.  A first result of this work is that we have been able to identify
the {\bf 168} and {\bf 120} resonances among the plethora of resonances
predicted in Ref.~\cite{GarciaRecio:2008dp} for the different strangeless
$C=1$ sectors. As expected, they turn out to be the lowest lying ones, and we
are pretty confident about their existence. This appreciation is being
reinforced by our previous study of Ref.~\cite{Gamermann:2011mq} in the light
$\SU(6)$ sector. Thus, we interpret the $\Lambda_c(2595)$ and
$\Lambda_c(2625)$ as a members of the $\SU(8)$ {\bf 168}-plet, and in both
cases with a dynamics strongly influenced by the $ND^*$ channel, in sharp
contrast with previous studies inconsistent with HQSS. Moreover, the changes induced by a suppression factor
in the interaction when charm is exchanged do not modify 
the conclusions (see Appendix \ref{app:C-exchange}).  Second, we have
identified the HQSS multiplets in which the resonances are arranged.
Specifically, the $\Lambda_c$, $\Lambda^*_c$ sector arranges into two
singlets, the $\Lambda_c(2595)$ being one of them, an one doublet, which
contains the $\Lambda_c(2625)$. Similarly the $\Sigma_c$, $\Sigma_c^*$ sector
contains one singlet and two doublets. For the $\Xi_c$, $\Xi_c^*$ sector,
there are three doublets and three singlets. 
According to our tentative
identification, $\Xi_c(2790)$ and $\Xi_c(2815)$ form a HQSS
doublet.
Finally, $\Omega_c$, $\Omega_c^*$ states form two doublets and one singlet.
Third, we have worked out the predictions of the model of
Ref.~\cite{GarciaRecio:2008dp} for strange charmed and $C=2$ and $C=3$
resonances linked to the strongly attractive {\bf 168} and {\bf 120}
subspaces. To our knowledge, these are the first predictions in these sectors
deduced from a model fulfilling HQSS. The organization into HQSS multiplets
is also given in this case. There is scarce experimental information in these
sectors, and we have only identified the three-star $\Xi_c(2790)$ and
$\Xi_c(2815)$ resonances, but we believe that the rest of our predictions are
robust, and will find experimental confirmation in the future. Of particular
relevance to this respect will be the programme of PANDA at the future
facility FAIR.

\section*{Acknowledgments}

We thank Michael D\"oring for the fruitful discussion and introducing the
excellent method of calculating the couplings. This research was supported by
DGI and FEDER funds, under contracts FIS2011-28853-C02-02, FIS2008-01143/FIS,
FPA2010-16963 and the Spanish Consolider-Ingenio 2010 Programme CPAN
(CSD2007-00042), by Junta de Andaluc{\'\i}a grant FQM-225, by Generalitat
Valenciana under contract PROMETEO/2009/0090 and by the EU HadronPhysics2
project, grant agreement n. 227431. O.R. and L.T. wishes to acknowledge support from
the Rosalind Franklin Fellowship. L.T. acknowledges support from Ramon y Cajal
Research Programme, and from FP7-PEOPLE-2011-CIG under contract
PCIG09-GA-2011-291679.

\appendix
\section{Charm-exchange suppression}
\label{app:C-exchange}

In this section we discuss the effect of the inclusion of a suppression factor
in the interaction when charm-exchange is present. 

In our approach, the interactions are implemented by a contact term, and each
matrix element is affected by the decay constants of the mesons in the
external legs of the interaction vertex. In particular, the charm-exchange
terms always involve a $D \leftrightarrow \pi$-like transition, and thus they
carry a factor $1/(f_\pi f_D)$. This source of flavor symmetry breaking turns
out to enhance (suppress) these transitions with respect some others like
$ND\to ND$ ($\Sigma_c \pi \to \Sigma_c \pi$) where there is not charm
exchange, and that scale instead like $1/f_D^2$ ($1/f_\pi^2$).

On the other hand, only decay constants of light mesons are involved in the
$t$-channel vector-meson exchange models, as the one used in
\cite{Hofmann:2005sw,Mizutani:2006vq,Hofmann:2006qx}. Nevertheless, there is
another source of quenching for charm-exchange interactions, coming from the
larger mass of the charmed meson exchanged, as compared to those of the vector
mesons belonging to the $\rho$ nonet. Qualitatively, a factor $\kappa_c =1/4 \simeq m_{\rho}^2/m_{D^*}^2$ is applied in
the matrix elements involving charm exchange, whereas $\kappa_c = 1$ is kept in
the remaining matrix elements~\cite{Mizutani:2006vq}. The introduction of
these quenching factors does not spoil HQSS (note however, that neither the
scheme of Ref.~\cite{Hofmann:2005sw,Hofmann:2006qx} nor that of
Ref.~\cite{Mizutani:2006vq} are consistent with HQSS) but it is a new source
of flavor breaking.  In this Appendix, we study the effects of including this
suppression factor $\kappa_c$ within our scheme.  In this case, the potential
looks as follows:
\begin{equation}
\label{TVMEpotentiial}
V_{ij}(s)=\kappa_c D_{ij}
\frac{2\sqrt{s}-M_i-M_j}{4\,f_i f_j} \sqrt{\frac{E_i+M_i}{2M_i}}
\sqrt{\frac{E_j+M_j}{2M_j}} 
.
\end{equation}

In Tables \ref{tab1001suppr} and \ref{tab1011suppr} we show the results
including the $\kappa_c$ factor for the sectors with $C=1$, $S=0$. As it can
be seen, there are some small changes in the masses and the widths of the
resonances in comparison with the values shown in Tables~\ref{tab1001} and
\ref{tab1011}, while the values of the couplings also change in some cases.
However, in general, the changes induced by the inclusion of
this new source of flavor breaking are not dramatic, and they do not modify 
the main conclusions of this work.

\begin{table*}[ht]
\begin{center}
\begin{tabular}{| c | c | c | c | c | c | c | c | c |}
\hline 

$\SU(8)$  & $\SU(6)$  & $\SU(3)$ &  & & Couplings &   &  \\
 irrep    & irrep   & irrep  & $M_{R}$ & $\Gamma_{R}$ &  to main channels &
 Status PDG & $J$ \\ 

\hline

$\bf 168$ & $\bf 15_{2,1}$ & $\bf 3_2$ & 

2624.6 & 103.9 & $\bf{g_{\Sigma_c \pi}=2.3}$, $g_{N D}=0.4$, $g_{N D^*}=0.4$, &
& $1/2$ \\
 & & &  &  & $g_{\Sigma_c \rho}=1.6$ & & \\

\hline

$\bf 168$ & $\bf 15_{2,1}$ & $\bf 3_4$ & 

2675.1 & 65.7 & $\bf{g_{\Sigma_c^* \pi}=2.1}$, $g_{N D^*}=0.5$, $g_{\Sigma_c
  \rho}=0.9$, & $\Lambda_c(2625)$ & $3/2$ \\
 & & & & & $g_{\Sigma_c^* \rho}=1.6$ & *** & 
\\

\hline

$\bf 168$ & $\bf 21_{2,1}$ & $\bf 3_2$ &

2624.1 & 0.1 & $\bf{g_{\Sigma_c \pi }=0.1}$, $g_{ N D}=3.4$, $g_{N D^*}=5.7$,
&$\Lambda_c(2595)$ &  $1/2$ \\
 &  & & & & $g_{\Lambda D_s}=1.4$, $g_{\Lambda D_s^*}=3.0$, $g_{\Lambda_c
  \eta' }=0.2$ & *** & \\

\hline

$\bf 120$ & $\bf 21_{2,1}$ & $\bf 3_2$ &

2824.9 & 0.4  & $\bf{g_{N D}=0.1}$, $g_{\Lambda_c \eta}=1.1$, $g_{\Xi_c
  K}=1.9$,  & &  $1/2$\\
 &  & & & & $g_{\Lambda D_s^*}=1.1$, $g_{\Sigma_c \rho}=1.1$, $g_{\Sigma_c^*
  \rho}=1.4$,  & & \\
 &  & & & & $g_{\Xi_c^* K^*}=1.0$ & & \\
\hline
\end{tabular}
\end{center}
\caption{$\Lambda_c$ and $\Lambda_c^*$ resonances with inclusion of the
  suppression factor $\kappa_c$.}
\label{tab1001suppr}
\end{table*}

\begin{table*}[ht]
\begin{center}
\begin{tabular}{| c | c | c | c | c | c | c | c |}
\hline $\SU(8)$ & $\SU(6)$ & $\SU(3)$ & & & Couplings &  & \\ irrep & irrep &
irrep & $M_{R}$ & $\Gamma_{R}$ & to main channels & Status PDG & $J$  \\ \hline

$\bf 168$ & $\bf 21_{2,1}$ & $\bf 6_2$ &
2583.4 & 0.0 & $\bf{g_{\Lambda_c \pi}=0.03}$, $g_{N D}=2.4$, $g_{N D^*}=1.3$,
& & $1/2$ \\
 &  & & &    & $g_{\Sigma D_s}=1.7$, $g_{ \Delta D^*}=6.8$, $g_{ \Sigma
  D_s^*}=1.2$, & & \\
 &  & & &    & $g_{\Sigma^* D_s^*}=3.1$ & & \\

\hline

$\bf 168$ & $\bf 21_{2,1}$ & $\bf 6_4$ &
2577.8 & 0.0 & $g_{ N D^*}=2.7$, $g_{\Delta D}=4.3$, $g_{\Delta D^*}=5.4$, &  & $3/2$ \\
 &  & & &   & $g_{\Sigma D_s^*}=2.4$, $g_{\Sigma^* D_s}=1.6$, $g_{\Sigma^*
  D_s^*}=2.4$ & & \\

\hline

$\bf 168$ & $\bf 15_{2,1}$ & $\bf 6_2$ &
2691.3 & 137.6 & $\bf{g_{\Lambda_c \pi}=1.6}$, $\bf{g_{\Sigma_c \pi}=0.3}$, $g_{N D}=0.5$, &  & $1/2$ \\
 &  & & &     & $g_{N D^*}=0.7$, $g_{\Xi_c K}=1.1$, $g_{\Sigma_c
  \rho}=1.8$, & & \\
 &  & & &     &  $g_{\Sigma_c^* \rho}=2.5$, $g_{\Xi_c^* K^*}=0.9$ & & \\

\hline

$\bf 120$ & $\bf 21_{2,1}$ & $\bf 6_2$ &
2653.9 & 95.0 & $\bf{g_{\Lambda_c \pi}=0.2}$, $\bf{g_{\Sigma_c \pi}=2.0}$, $g_{N D}=0.6$, &  & $1/2$ \\
 &  & & &     & $g_{N D^*}=0.4$, $g_{\Lambda_c \rho}=1.7$ $g_{\Delta
  D^*}=0.4$, & & \\
 &  & & &     & $g_{\Sigma_c \rho}=1.3$, $g_{\Sigma^* D_s^*}=0.4$ & & \\

\hline

$\bf 120$ & $\bf 21_{2,1}$ & $\bf 6_4$ &
2697.2 & 65.8 & $\bf{g_{\Sigma_c^* \pi}=1.9}$, $g_{N D^* }=0.6$, $g_{\Lambda_c \rho}=1.7$, &  & $3/2$ \\
 &  & & &   & $g_{\Sigma_c^* \rho}=1.1$, $g_{\Sigma D_s^*}=0.3$,
$g_{\Sigma^* D_s^*}=0.3$ & & \\

\hline

\end{tabular}
\end{center}
\caption{$\Sigma_c$ and $\Sigma_c^*$ resonances with inclusion of the
  suppression factor $\kappa_c$.}
\label{tab1011suppr}
\end{table*}


\section{Baryon-meson matrix elements}
\label{app:tables}

In this Appendix the coefficients $D_{ij}$ appearing in
Eq.~(\ref{eq:vsu8break}) are shown for the various $CSIJ$ sectors studied in
this work (Tables \ref{tab:Xic11}-\ref{tab:OmegacccS}). The $D$-matrices for
the channels with $C=1,~S=0$ can be found in an Appendix B of
Ref.~\cite{GarciaRecio:2008dp}. The Table for $\Xi_c$ and $\Xi_c^*$ states has been divided
in three blocks.

\newpage
\begin{table*}
\centering
\caption{
  $ C=1$, $ S=-1$, $ I=1/2$, $ J= 1/2$.
}
\label{tab:Xic11}
\begin{ruledtabular}
\begin{tabular}{
rrrrrrrrrrrrrrrrr
}
 &  $ \Cascadac \pion $ 
 &  $ \Cascadacp \pion $ 
 &  $ \Lambdac \kaonb $ &  $ \Sigmac 
  \kaonb $ &  $ \Lambda \D $ 
 &  $ \Cascadac \eta $ &  $ \Sigma \D 
  $ &  $ \Lambda \DS $ 
 &  $ \Cascadacp \eta $ &  $ \Lambdac 
  \kaonSb $ &  $ \Omegac \kaon $ 
 &  $ \Sigma \DS $ &  $ \Cascadac \rho 
  $ &  $ \Cascadac \omega $ 
 &  $ \Cascada \Ds $ &  $ \Sigmac 
  \kaonSb $ 
\ignore{
&  $ \Cascadacp \rho $ 
 &  $ \Cascadacp \omega $ 
 &  $ \SigmaS \DS $ &  $ \SigmacS 
  \kaonSb $ &  $ \CascadacS \rho $ 
 &  $ \Cascadac \etap $ &  $ \CascadacS 
  \omega $ &  $ \Cascada \DsS $ 
 &  $ \Cascadac \phi $ &  $ \Cascadacp 
  \etap $ &  $ \Omegac \kaonS $ 
 &  $ \Cascadacp \phi $ &  $ \CascadaS 
  \DsS $ &  $ \OmegacS \kaonS $ 
 &  $ \CascadacS \phi $ 
}
\\
$ \Cascadac \pion $& $ -2 $ 
 & $ 0 $ 
 & $ \sqrt{\frac{ 3 }{ 2 }} 
   $ & $ 0 $ 
 & $ \sqrt{\frac{ 3 }{ 8 }} 
   $ & $ 0 $ 
 & $ \sqrt{\frac{ 3 }{ 8 }} 
   $ & $ \sqrt{\frac{ 9 }{ 8 
    }} $ & $ 0 $ & $ 0 $ 
 & $ 0 $ 
 & $ \sqrt{\frac{ 9 }{ 8 }} 
   $ & $ 0 $ & $ 0 $ 
 & $ 0 $ 
 & $ \sqrt{\frac{ 1 }{ 2 }} 
   $ 
\ignore{
& $ 2 $ & $ 0 $ 
 & $ 0 $ & $ 1 $ 
 & $ \sqrt{ 8 } $ 
 & $ 0 $ & $ 0 $ 
 & $ 0 $ & $ 0 $ 
 & $ 0 $ & $ \sqrt{ 3 } 
   $ & $ 0 $ & $ 0 $ 
 & $ \sqrt{ 6 } $ 
 & $ 0 $  
}
\\
$ \Cascadacp \pion $& $ 0 $ 
 & $ -2 $ & $ 0 $ 
 & $ \sqrt{\frac{ 1 }{ 2 }} 
   $ & $ \sqrt{\frac{ 9 }{ 8 
    }} $ & $ 0 $ 
 & $ -\sqrt{\frac{ 1 }{ 8 }} 
   $ & $ -\sqrt{\frac{ 3 }{ 8 
     }} $ & $ 0 $ 
 & $ \sqrt{\frac{ 3 }{ 2 }} 
   $ & $ -\sqrt{ 3 } $ 
 & $ \sqrt{\frac{ 1 }{ 24 }} 
   $ & $ 2 $ & $ 0 $ 
 & $ 0 $ 
 & $ \sqrt{\frac{ 2 }{ 3 }} 
   $ 
\ignore{
& $ -\sqrt{\frac{ 16 }{ 3 
     }} $ & $ 0 $ 
 & $ \sqrt{\frac{ 4 }{ 3 }} 
   $ & $ -\sqrt{\frac{ 1 }{ 3 
     }} $ & $ \sqrt{\frac{ 8 
    }{ 3 }} $ & $ 0 $ 
 & $ 0 $ & $ 0 $ 
 & $ 0 $ & $ 0 $ 
 & $ -2 $ & $ 0 $ 
 & $ 0 $ & $ \sqrt{ 2 } 
   $ & $ 0 $  
}
\\
\end{tabular}
\end{ruledtabular}
\end{table*}

\newpage
\begin{table*}
\centering
\caption{
  $ C=1$, $ S=-1$, $ I=1/2$, $ J= 1/2$ (cont.).
}
\label{tab:Xic12}
\begin{ruledtabular}
\begin{tabular}{
r
rrrrrrrrrrrrrrr
}
\ignore{
 &  $ \Cascadac \pion $ 
 &  $ \Cascadacp \pion $ 
 &  $ \Lambdac \kaonb $ &  $ \Sigmac 
  \kaonb $ &  $ \Lambda \D $ 
 &  $ \Cascadac \eta $ &  $ \Sigma \D 
  $ &  $ \Lambda \DS $ 
 &  $ \Cascadacp \eta $ &  $ \Lambdac 
  \kaonSb $ &  $ \Omegac \kaon $ 
 &  $ \Sigma \DS $ &  $ \Cascadac \rho 
  $ &  $ \Cascadac \omega $ 
 &  $ \Cascada \Ds $ &  $ \Sigmac 
  \kaonSb $ 
}
&  $ \Cascadacp \rho $ 
 &  $ \Cascadacp \omega $ 
 &  $ \SigmaS \DS $ &  $ \SigmacS 
  \kaonSb $ &  $ \CascadacS \rho $ 
 &  $ \Cascadac \etap $ &  $ \CascadacS 
  \omega $ &  $ \Cascada \DsS $ 
 &  $ \Cascadac \phi $ &  $ \Cascadacp 
  \etap $ &  $ \Omegac \kaonS $ 
 &  $ \Cascadacp \phi $ &  $ \CascadaS 
  \DsS $ &  $ \OmegacS \kaonS $ 
 &  $ \CascadacS \phi $ 
\\
$ \Cascadac \pion $
\ignore{
& $ -2 $ 
 & $ 0 $ 
 & $ \sqrt{\frac{ 3 }{ 2 }} 
   $ & $ 0 $ 
 & $ \sqrt{\frac{ 3 }{ 8 }} 
   $ & $ 0 $ 
 & $ \sqrt{\frac{ 3 }{ 8 }} 
   $ & $ \sqrt{\frac{ 9 }{ 8 
    }} $ & $ 0 $ & $ 0 $ 
 & $ 0 $ 
 & $ \sqrt{\frac{ 9 }{ 8 }} 
   $ & $ 0 $ & $ 0 $ 
 & $ 0 $ 
 & $ \sqrt{\frac{ 1 }{ 2 }} 
   $ 
}
& $ 2 $ & $ 0 $ 
 & $ 0 $ & $ 1 $ 
 & $ \sqrt{ 8 } $ 
 & $ 0 $ & $ 0 $ 
 & $ 0 $ & $ 0 $ 
 & $ 0 $ & $ \sqrt{ 3 } 
   $ & $ 0 $ & $ 0 $ 
 & $ \sqrt{ 6 } $ 
 & $ 0 $  
\\
$ \Cascadacp \pion $
\ignore{
& $ 0 $ 
 & $ -2 $ & $ 0 $ 
 & $ \sqrt{\frac{ 1 }{ 2 }} 
   $ & $ \sqrt{\frac{ 9 }{ 8 
    }} $ & $ 0 $ 
 & $ -\sqrt{\frac{ 1 }{ 8 }} 
   $ & $ -\sqrt{\frac{ 3 }{ 8 
     }} $ & $ 0 $ 
 & $ \sqrt{\frac{ 3 }{ 2 }} 
   $ & $ -\sqrt{ 3 } $ 
 & $ \sqrt{\frac{ 1 }{ 24 }} 
   $ & $ 2 $ & $ 0 $ 
 & $ 0 $ 
 & $ \sqrt{\frac{ 2 }{ 3 }} 
   $ 
}
& $ -\sqrt{\frac{ 16 }{ 3 
     }} $ & $ 0 $ 
 & $ \sqrt{\frac{ 4 }{ 3 }} 
   $ & $ -\sqrt{\frac{ 1 }{ 3 
     }} $ & $ \sqrt{\frac{ 8 
    }{ 3 }} $ & $ 0 $ 
 & $ 0 $ & $ 0 $ 
 & $ 0 $ & $ 0 $ 
 & $ -2 $ & $ 0 $ 
 & $ 0 $ & $ \sqrt{ 2 } 
   $ & $ 0 $  
\\
\end{tabular}
\end{ruledtabular}
\end{table*}

\newpage
\begin{table*}
\centering
\caption{
  $ C=1$, $ S=-1$, $ I=1/2$, $ J= 1/2$ (cont.).
}
\label{tab:Xic22}
\begin{ruledtabular}
\begin{tabular}{r
rrrrrrrrrrrrrrr
}
\ignore{
 &  $ \Cascadac \pion $ 
 &  $ \Cascadacp \pion $ 
 &  $ \Lambdac \kaonb $ &  $ \Sigmac 
  \kaonb $ &  $ \Lambda \D $ 
 &  $ \Cascadac \eta $ &  $ \Sigma \D 
  $ &  $ \Lambda \DS $ 
 &  $ \Cascadacp \eta $ &  $ \Lambdac 
  \kaonSb $ &  $ \Omegac \kaon $ 
 &  $ \Sigma \DS $ &  $ \Cascadac \rho 
  $ &  $ \Cascadac \omega $ 
 &  $ \Cascada \Ds $ &  $ \Sigmac 
  \kaonSb $ 
}
&  $ \Cascadacp \rho $ 
 &  $ \Cascadacp \omega $ 
 &  $ \SigmaS \DS $ &  $ \SigmacS 
  \kaonSb $ &  $ \CascadacS \rho $ 
 &  $ \Cascadac \etap $ &  $ \CascadacS 
  \omega $ &  $ \Cascada \DsS $ 
 &  $ \Cascadac \phi $ &  $ \Cascadacp 
  \etap $ &  $ \Omegac \kaonS $ 
 &  $ \Cascadacp \phi $ &  $ \CascadaS 
  \DsS $ &  $ \OmegacS \kaonS $ 
 &  $ \CascadacS \phi $ 
\\
$ \Cascadacp \rho $
\ignore{
& $ 2 $ 
 & $ -\sqrt{\frac{ 16 }{ 3 }} 
   $ & $ \sqrt{\frac{ 3 }{ 2 
    }} $ & $ \sqrt{\frac{ 2 
    }{ 3 }} $ 
 & $ -\sqrt{\frac{ 3 }{ 8 }} 
   $ & $ 0 $ 
 & $ \sqrt{\frac{ 1 }{ 24 }} 
   $ & $ \sqrt{\frac{ 25 }{ 8 
    }} $ & $ 0 $ 
 & $ -\sqrt{ 2 } $ 
 & $ -2 $ 
 & $ -\sqrt{\frac{ 25 }{ 72 }} 
   $ & $ \sqrt{\frac{ 4 }{ 3 
    }} $ & $ 2 $ & $ 0 $ 
 & $ -\sqrt{\frac{ 1 }{ 18 }} 
   $
}
 & $ -\frac{ 10 }{ 3  } 
   $ & $ -\sqrt{\frac{ 16 }{ 3 
     }} $ & $ \frac{ 2 }{ 3 
     } $ 
 & $ -\frac{ 1 }{ 3  } $ 
 & $ -\sqrt{\frac{ 2 }{ 9 }} 
   $ & $ 0 $ 
 & $ -\sqrt{\frac{ 2 }{ 3 }} 
   $ & $ 0 $ & $ 0 $ 
 & $ 0 $ 
 & $ -\sqrt{\frac{ 49 }{ 3 }} 
   $ & $ 0 $ & $ 0 $ 
 & $ -\sqrt{\frac{ 2 }{ 3 }} 
   $ & $ 0 $ 
\\
\end{tabular}
\end{ruledtabular}
\end{table*}

\newpage
\begin{table*}
\caption{
  $ C=1$, $ S=-1$, $ I=1/2$, $ J= 3/2$.
}
\label{tab:XicS11}
\begin{ruledtabular}
\begin{tabular}{rrrrrrrrrrrrrrrrrrrrrrrrrrrrrrrrrrrrrrrrrrrrrrrrrr}
 &  $ \CascadacS \pion $ 
 &  $ \SigmacS \kaonb $ &  $ \Lambda 
  \DS $ &  $ \Lambdac \kaonSb $ 
 &  $ \CascadacS \eta $ &  $ \Sigma \DS 
  $ &  $ \Cascadac \rho $ 
 &  $ \SigmaS \D $ &  $ \Cascadac 
  \omega $ &  $ \OmegacS \kaon $ 
 &  $ \Sigmac \kaonSb $ &  $ \Cascadacp 
  \rho $ &  $ \Cascadacp \omega $ 
\ignore{
 &  $ \SigmaS \DS $ &  $ \SigmacS 
  \kaonSb $ &  $ \CascadacS \rho $ 
 &  $ \CascadacS \omega $ 
 &  $ \Cascada \DsS $ &  $ \Cascadac 
  \phi $ &  $ \CascadaS \Ds $ 
 &  $ \Omegac \kaonS $ &  $ \Cascadacp 
  \phi $ &  $ \CascadacS \etap $ 
 &  $ \CascadaS \DsS $ &  $ \OmegacS 
  \kaonS $ &  $ \CascadacS \phi $ 
}
\\
$ \CascadacS \pion $& $ -2 $ 
 & $ \sqrt{\frac{ 1 }{ 2 }} 
   $ & $ -\sqrt{\frac{ 3 }{ 2 
     }} $ 
 & $ -\sqrt{\frac{ 3 }{ 2 }} 
   $ & $ 0 $ 
 & $ \sqrt{\frac{ 1 }{ 6 }} 
   $ & $ -2 $ 
 & $ \sqrt{\frac{ 1 }{ 2 }} 
   $ & $ 0 $ 
 & $ -\sqrt{ 3 } $ 
 & $ \sqrt{\frac{ 1 }{ 6 }} 
   $ & $ -\sqrt{\frac{ 4 }{ 3 
     }} $ & $ 0 $ 
\ignore{
 & $ \sqrt{\frac{ 5 }{ 6 }} 
   $ & $ \sqrt{\frac{ 5 }{ 6 
    }} $ 
 & $ -\sqrt{\frac{ 20 }{ 3 }} 
   $ & $ 0 $ & $ 0 $ 
 & $ 0 $ & $ 0 $ 
 & $ -1 $ & $ 0 $ 
 & $ 0 $ & $ 0 $ 
 & $ -\sqrt{ 5 } $ 
 & $ 0 $ 
}
\\
$ \SigmacS \kaonb $
 & $ \sqrt{\frac{ 1 }{ 2 }} 
   $ & $ -3 $ & $ 0 $ 
 & $ -\sqrt{ 3 } $ 
 & $ \sqrt{\frac{ 9 }{ 2 }} 
   $ & $ -\sqrt{\frac{ 4 }{ 3 
     }} $ 
 & $ -\sqrt{\frac{ 1 }{ 2 }} 
   $ & $ 1 $ 
 & $ -\sqrt{\frac{ 3 }{ 2 }} 
   $ & $ 0 $ 
 & $ -\sqrt{ 3 } $ 
 & $ \sqrt{\frac{ 1 }{ 6 }} 
   $ & $ \sqrt{\frac{ 1 }{ 2 
    }} $ 
\ignore{
& $ \sqrt{\frac{ 5 
    }{ 3 }} $ 
 & $ -\sqrt{ 15 } $ 
 & $ \sqrt{\frac{ 5 }{ 6 }} 
   $ & $ \sqrt{\frac{ 5 }{ 2 
    }} $ & $ 0 $ 
 & $ -\sqrt{ 3 } $ 
 & $ 0 $ & $ 0 $ 
 & $ 1 $ & $ 0 $ 
 & $ 0 $ & $ 0 $ 
 & $ \sqrt{ 5 } $ 
}
\\
\end{tabular}
\end{ruledtabular}
\end{table*}

\newpage
\begin{table*}
\caption{
  $ C=1$, $ S=-1$, $ I=1/2$, $ J= 3/2$ (cont.).
}
\label{tab:XicS12}
\begin{ruledtabular}
\begin{tabular}{rrrrrrrrrrrrrrrrrrrrrrrrrrrrrrrrrrrrrrrrrrrrrrrrrr}
\ignore{
 &  $ \CascadacS \pion $ 
 &  $ \SigmacS \kaonb $ &  $ \Lambda 
  \DS $ &  $ \Lambdac \kaonSb $ 
 &  $ \CascadacS \eta $ &  $ \Sigma \DS 
  $ &  $ \Cascadac \rho $ 
 &  $ \SigmaS \D $ &  $ \Cascadac 
  \omega $ &  $ \OmegacS \kaon $ 
 &  $ \Sigmac \kaonSb $ &  $ \Cascadacp 
  \rho $ &  $ \Cascadacp \omega $ 
}
 &  $ \SigmaS \DS $ &  $ \SigmacS 
  \kaonSb $ &  $ \CascadacS \rho $ 
 &  $ \CascadacS \omega $ 
 &  $ \Cascada \DsS $ &  $ \Cascadac 
  \phi $ &  $ \CascadaS \Ds $ 
 &  $ \Omegac \kaonS $ &  $ \Cascadacp 
  \phi $ &  $ \CascadacS \etap $ 
 &  $ \CascadaS \DsS $ &  $ \OmegacS 
  \kaonS $ &  $ \CascadacS \phi $ 
\\
$ \CascadacS \pion $
\ignore{
& $ -2 $ 
 & $ \sqrt{\frac{ 1 }{ 2 }} 
   $ & $ -\sqrt{\frac{ 3 }{ 2 
     }} $ 
 & $ -\sqrt{\frac{ 3 }{ 2 }} 
   $ & $ 0 $ 
 & $ \sqrt{\frac{ 1 }{ 6 }} 
   $ & $ -2 $ 
 & $ \sqrt{\frac{ 1 }{ 2 }} 
   $ & $ 0 $ 
 & $ -\sqrt{ 3 } $ 
 & $ \sqrt{\frac{ 1 }{ 6 }} 
   $ & $ -\sqrt{\frac{ 4 }{ 3 
     }} $ & $ 0 $ 
}
 & $ \sqrt{\frac{ 5 }{ 6 }} 
   $ & $ \sqrt{\frac{ 5 }{ 6 
    }} $ 
 & $ -\sqrt{\frac{ 20 }{ 3 }} 
   $ & $ 0 $ & $ 0 $ 
 & $ 0 $ & $ 0 $ 
 & $ -1 $ & $ 0 $ 
 & $ 0 $ & $ 0 $ 
 & $ -\sqrt{ 5 } $ 
 & $ 0 $ 
\\
$ \SigmacS \kaonb $
\ignore{
 & $ \sqrt{\frac{ 1 }{ 2 }} 
   $ & $ -3 $ & $ 0 $ 
 & $ -\sqrt{ 3 } $ 
 & $ \sqrt{\frac{ 9 }{ 2 }} 
   $ & $ -\sqrt{\frac{ 4 }{ 3 
     }} $ 
 & $ -\sqrt{\frac{ 1 }{ 2 }} 
   $ & $ 1 $ 
 & $ -\sqrt{\frac{ 3 }{ 2 }} 
   $ & $ 0 $ 
 & $ -\sqrt{ 3 } $ 
 & $ \sqrt{\frac{ 1 }{ 6 }} 
   $ & $ \sqrt{\frac{ 1 }{ 2 
    }} $ 
}
& $ \sqrt{\frac{ 5 
    }{ 3 }} $ 
 & $ -\sqrt{ 15 } $ 
 & $ \sqrt{\frac{ 5 }{ 6 }} 
   $ & $ \sqrt{\frac{ 5 }{ 2 
    }} $ & $ 0 $ 
 & $ -\sqrt{ 3 } $ 
 & $ 0 $ & $ 0 $ 
 & $ 1 $ & $ 0 $ 
 & $ 0 $ & $ 0 $ 
 & $ \sqrt{ 5 } $ 
\\
\end{tabular}
\end{ruledtabular}
\end{table*}

\newpage
\begin{table*}
\caption{
  $ C=1$, $ S=-1$, $ I=1/2$, $ J= 3/2$. (cont.).
}
\label{tab:XicS22}
\begin{ruledtabular}
\begin{tabular}{rrrrrrrrrrrrrrrrrrrrrrrrrrrrrrrrrrrrrrrrrrrrrrrrrr}
\ignore{
 &  $ \CascadacS \pion $ 
 &  $ \SigmacS \kaonb $ &  $ \Lambda 
  \DS $ &  $ \Lambdac \kaonSb $ 
 &  $ \CascadacS \eta $ &  $ \Sigma \DS 
  $ &  $ \Cascadac \rho $ 
 &  $ \SigmaS \D $ &  $ \Cascadac 
  \omega $ &  $ \OmegacS \kaon $ 
 &  $ \Sigmac \kaonSb $ &  $ \Cascadacp 
  \rho $ &  $ \Cascadacp \omega $ 
}
 &  $ \SigmaS \DS $ &  $ \SigmacS 
  \kaonSb $ &  $ \CascadacS \rho $ 
 &  $ \CascadacS \omega $ 
 &  $ \Cascada \DsS $ &  $ \Cascadac 
  \phi $ &  $ \CascadaS \Ds $ 
 &  $ \Omegac \kaonS $ &  $ \Cascadacp 
  \phi $ &  $ \CascadacS \etap $ 
 &  $ \CascadaS \DsS $ &  $ \OmegacS 
  \kaonS $ &  $ \CascadacS \phi $ 
\\
$ \SigmaS \DS $
\ignore{
 & $ \sqrt{\frac{ 5 }{ 6 }} 
   $ & $ \sqrt{\frac{ 5 }{ 3 
    }} $ & $ -\sqrt{ 5 } 
   $ & $ 0 $ 
 & $ \sqrt{\frac{ 5 }{ 6 }} 
   $ & $ -\sqrt{ 5 } $ 
 & $ 0 $ 
 & $ -\sqrt{ 15 } $ 
 & $ 0 $ & $ 0 $ 
 & $ \sqrt{\frac{ 20 }{ 9 }} 
   $ & $ \sqrt{\frac{ 10 }{ 9 
    }} $ & $ \sqrt{\frac{ 10 
    }{ 3 }} $ 
}
& $ -5 $ 
 & $ \frac{ 1 }{ 3  } $ 
 & $ \sqrt{\frac{ 1 }{ 18 }} 
   $ & $ \sqrt{\frac{ 1 }{ 6 
    }} $ & $ \sqrt{\frac{ 10 
    }{ 3 }} $ & $ 0 $ 
 & $ -\sqrt{ 10 } $ 
 & $ 0 $ & $ 0 $ 
 & $ \sqrt{\frac{ 5 }{ 3 }} 
   $ & $ -\sqrt{\frac{ 50 }{ 3 
     }} $ & $ 0 $ 
 & $ 0 $
\\
$ \SigmacS \kaonSb $
\ignore{
 & $ \sqrt{\frac{ 5 }{ 6 }} 
   $ & $ -\sqrt{ 15 } $ 
 & $ 0 $ 
 & $ -\sqrt{ 5 } $ 
 & $ \sqrt{\frac{ 15 }{ 2 }} 
   $ & $ \sqrt{\frac{ 20 }{ 9 
    }} $ & $ \sqrt{\frac{ 5 
    }{ 6 }} $
 & $ \sqrt{\frac{ 5 }{ 3 }} 
   $ & $ \sqrt{\frac{ 5 }{ 2 
    }} $ & $ 0 $ 
 & $ -\sqrt{ 5 } $ 
 & $ -\sqrt{\frac{ 5 }{ 18 }} 
   $ & $ -\sqrt{\frac{ 5 }{ 6 
     }} $ 
}
& $ \frac{ 1 }{ 3 
     } $ & $ -5 $ 
 & $ \sqrt{\frac{ 1 }{ 18 }} 
   $ & $ \sqrt{\frac{ 1 }{ 6 
    }} $ & $ 0 $ 
 & $ -\sqrt{ 5 } $ 
 & $ 0 $ & $ 0 $ 
 & $ \sqrt{\frac{ 5 }{ 3 }} 
   $ & $ 0 $ & $ 0 $ 
 & $ 0 $ 
 & $ \sqrt{\frac{ 25 }{ 3 }} 
   $ 
\\
\end{tabular}
\end{ruledtabular}
\end{table*}

\newpage
\begin{table*}
\centering
\caption{
  $ C=1$, $ S=-2$, $ I=0$, $ J= 1/2$.
}
\label{tab:Omegac}
\begin{ruledtabular}
\begin{tabular}{rrrrrrrrrrrrrrrrrrrrrrrrrrrrrrrrrrrrrrrrrrrrrrrrrr}
 &  $ \Cascadac \kaonb $ 
 &  $ \Cascadacp \kaonb $ 
 &  $ \Cascada \D $ &  $ \Omegac \eta 
  $ &  $ \Cascada \DS $ 
 &  $ \Cascadac \kaonSb $ 
 &  $ \Cascadacp \kaonSb $ 
 &  $ \Omegac \omega $ &  $ \CascadacS 
  \kaonSb $ &  $ \CascadaS \DS $ 
 &  $ \OmegacS \omega $ &  $ \Omegac 
  \etap $ &  $ \Omegac \phi $ 
 &  $ \Omega \DsS $ &  $ \OmegacS \phi 
  $ 
\\
$ \Cascadac \kaonb $& $ -1 $ 
 & $ 0 $ 
 & $ \sqrt{\frac{ 3 }{ 2 }} 
   $ & $ 0 $ 
 & $ \sqrt{\frac{ 9 }{ 2 }} 
   $ & $ 0 $ & $ 3 $ 
 & $ -\sqrt{ 2 } $ 
 & $ \sqrt{ 18 } $ 
 & $ 0 $ & $ -2 $ 
 & $ 0 $ & $ -2 $ 
 & $ 0 $ 
 & $ -\sqrt{ 8 } $ 
\\
$ \Cascadacp \kaonb $& $ 0 $ 
 & $ -1 $ 
 & $ \sqrt{\frac{ 1 }{ 2 }} 
   $ & $ \sqrt{ 6 } $ 
 & $ -\sqrt{\frac{ 1 }{ 6 }} 
   $ & $ 3 $ 
 & $ -\sqrt{\frac{ 4 }{ 3 }} 
   $ & $ \sqrt{\frac{ 8 }{ 3 
    }} $ & $ \sqrt{\frac{ 2 
    }{ 3 }} $ 
 & $ \sqrt{\frac{ 16 }{ 3 }} 
   $ & $ -\sqrt{\frac{ 4 }{ 3 
     }} $ & $ 0 $ 
 & $ \sqrt{\frac{ 16 }{ 3 }} 
   $ & $ 0 $ 
 & $ -\sqrt{\frac{ 8 }{ 3 }} 
   $ 
\\
$ \Cascada \D $
 & $ \sqrt{\frac{ 3 }{ 2 }} 
   $ & $ \sqrt{\frac{ 1 }{ 2 
    }} $ & $ -2 $ 
 & $ -\sqrt{\frac{ 1 }{ 3 }} 
   $ & $ \sqrt{\frac{ 4 }{ 3 
    }} $ & $ \sqrt{\frac{ 9 
    }{ 2 }} $ 
 & $ -\sqrt{\frac{ 1 }{ 6 }} 
   $ & $ \sqrt{\frac{ 1 }{ 3 
    }} $ & $ \sqrt{\frac{ 4 
    }{ 3 }} $ 
 & $ \sqrt{\frac{ 32 }{ 3 }} 
   $ & $ -\sqrt{\frac{ 8 }{ 3 
     }} $ 
 & $ -\sqrt{\frac{ 2 }{ 3 }} 
   $ & $ 0 $ & $ 4 $ 
 & $ 0 $ 
\\
$ \Omegac \eta $& $ 0 $ 
 & $ \sqrt{ 6 } $ 
 & $ -\sqrt{\frac{ 1 }{ 3 }} 
   $ & $ 0 $ 
 & $ \frac{ 1 }{ 3  } $ 
 & $ -\sqrt{ 6 } $ 
 & $ \sqrt{ 8 } $ 
 & $ 0 $ & $ -2 $ 
 & $ \sqrt{\frac{ 8 }{ 9 }} 
   $ & $ 0 $ & $ 0 $ 
 & $ 0 $ 
 & $ -\sqrt{\frac{ 16 }{ 3 }} 
   $ & $ 0 $ 
\\
\end{tabular}
\end{ruledtabular}
\end{table*}

\newpage
\begin{table*}
\centering
\caption{
  $ C=1$, $ S=-2$, $ I=0$, $ J= 3/2$.
}
\label{tab:OmegacS}
\begin{ruledtabular}
\begin{tabular}{rrrrrrrrrrrrrrrrrrrrrrrrrrrrrrrrrrrrrrrrrrrrrrrrrr}
 &  $ \CascadacS \kaonb $ 
 &  $ \OmegacS \eta $ &  $ \Cascada \DS 
  $ &  $ \Cascadac \kaonSb $ 
 &  $ \CascadaS \D $ &  $ \Cascadacp 
  \kaonSb $ &  $ \Omegac \omega $ 
 &  $ \CascadacS \kaonSb $ 
 &  $ \CascadaS \DS $ &  $ \OmegacS 
  \omega $ &  $ \Omega \Ds $ 
 &  $ \Omegac \phi $ &  $ \OmegacS 
  \etap $ &  $ \Omega \DsS $ 
 &  $ \OmegacS \phi $ 
\\
$ \CascadacS \kaonb $& $ -1 $ 
 & $ \sqrt{ 6 } $ 
 & $ -\sqrt{\frac{ 2 }{ 3 }} 
   $ & $ -3 $ 
 & $ \sqrt{ 2 } $ 
 & $ -\sqrt{\frac{ 1 }{ 3 }} 
   $ & $ \sqrt{\frac{ 2 }{ 3 
    }} $ 
 & $ -\sqrt{\frac{ 5 }{ 3 }} 
   $ & $ \sqrt{\frac{ 10 }{ 3 
    }} $ & $ \sqrt{\frac{ 10 
    }{ 3 }} $ & $ 0 $ 
 & $ \sqrt{\frac{ 4 }{ 3 }} 
   $ & $ 0 $ & $ 0 $ 
 & $ \sqrt{\frac{ 20 }{ 3 }} 
   $ 
\\
$ \OmegacS \eta $
 & $ \sqrt{ 6 } $ 
 & $ 0 $ 
 & $ \frac{ 2 }{ 3  } $ 
 & $ \sqrt{ 6 } $ 
 & $ \sqrt{\frac{ 1 }{ 3 }} 
   $ & $ \sqrt{ 2 } $ 
 & $ 0 $ 
 & $ \sqrt{ 10 } $ 
 & $ \sqrt{\frac{ 5 }{ 9 }} 
   $ & $ 0 $ 
 & $ -\sqrt{ 2 } $ 
 & $ 0 $ & $ 0 $ 
 & $ -\sqrt{\frac{ 10 }{ 3 }} 
   $ & $ 0 $ 
\\
$ \Cascada \DS $
 & $ -\sqrt{\frac{ 2 }{ 3 }} 
   $ & $ \frac{ 2 }{ 3  } 
   $ & $ -\frac{ 8 }{ 3  } 
   $ & $ \sqrt{ 6 } $ 
 & $ -\sqrt{\frac{ 16 }{ 3 }} 
   $ & $ \sqrt{\frac{ 2 }{ 9 
    }} $ 
 & $ -\frac{ 2 }{ 3  } $ 
 & $ \sqrt{\frac{ 10 }{ 9 }} 
   $ & $ -\sqrt{\frac{ 80 }{ 9 
     }} $ 
 & $ -\sqrt{\frac{ 20 }{ 9 }} 
   $ & $ -\sqrt{ 8 } $ 
 & $ 0 $ 
 & $ \sqrt{\frac{ 8 }{ 9 }} 
   $ & $ -\sqrt{\frac{ 40 }{ 3 
     }} $ & $ 0 $ 
\\
$ \Cascadac \kaonSb $& $ -3 $ 
 & $ \sqrt{ 6 } $ 
 & $ \sqrt{ 6 } $ 
 & $ -1 $ & $ 0 $ 
 & $ -\sqrt{\frac{ 1 }{ 3 }} 
   $ & $ -\sqrt{\frac{ 2 }{ 3 
     }} $ 
 & $ -\sqrt{\frac{ 5 }{ 3 }} 
   $ & $ 0 $ 
 & $ -\sqrt{\frac{ 10 }{ 3 }} 
   $ & $ 0 $ 
 & $ \sqrt{\frac{ 4 }{ 3 }} 
   $ & $ 0 $ & $ 0 $ 
 & $ \sqrt{\frac{ 20 }{ 3 }} 
   $ 
\\
\end{tabular}
\end{ruledtabular}
\end{table*}

\newpage
\begin{sidewaystable*}
\vspace*{-6cm}
\centering
\caption{
  $ C=2$, $ S=0$, $ I=1/2$, $ J= 1/2$.
}
\label{tab:Xicc}
\begin{ruledtabular}
\begin{tabular}{rrrrrrrrrrrrrrrrrrrrrrrrrrrrrrrrrrrrrrrrrrrrrrrrrr}
 &  $ \Cascadacc \pion $ 
 &  $ \Omegacc \kaon $ &  $ \Cascadacc 
  \eta $ &  $ \Lambdac \D $ 
 &  $ \Cascadacc \rho $ &  $ \Lambdac 
  \DS $ &  $ \Cascadacc \omega $ 
 &  $ \Sigmac \D $ &  $ \Cascadac \Ds 
  $ &  $ \Omegacc \kaonS $ 
 &  $ \Sigmac \DS $ &  $ \Cascadacc 
  \etap $ &  $ \CascadaccS \rho $ 
 &  $ \SigmacS \DS $ &  $ \CascadaccS 
  \omega $ &  $ \Cascadacc \phi $ 
 &  $ \Cascadacp \Ds $ &  $ \Cascadac 
  \DsS $ &  $ \Cascadacp \DsS $ 
 &  $ \CascadacS \DsS $ &  $ \CascadaccS 
  \phi $ &  $ \OmegaccS \kaonS $ 
\\
$ \Cascadacc \pion $& $ -2 $ 
 & $ -\sqrt{\frac{ 3 }{ 2 }} 
   $ & $ 0 $ 
 & $ \frac{ 3 }{ 2  } $ 
 & $ \sqrt{\frac{ 4 }{ 3 }} 
   $ & $ \sqrt{\frac{ 3 }{ 4 
    }} $ & $ 0 $ 
 & $ \frac{ 1 }{ 2  } $ 
 & $ 0 $ 
 & $ \sqrt{\frac{ 1 }{ 2 }} 
   $ & $ \sqrt{\frac{ 25 }{ 12 
    }} $ & $ 0 $ 
 & $ \sqrt{\frac{ 32 }{ 3 }} 
   $ & $ \sqrt{\frac{ 2 }{ 3 
    }} $ & $ 0 $ & $ 0 $ 
 & $ 0 $ & $ 0 $ 
 & $ 0 $ & $ 0 $ 
 & $ 0 $ & $ 2 $ 
\\
$ \Omegacc \kaon $
 & $ -\sqrt{\frac{ 3 }{ 2 }} 
   $ & $ -1 $ 
 & $ -\sqrt{\frac{ 3 }{ 2 }} 
   $ & $ 0 $ 
 & $ \sqrt{\frac{ 1 }{ 2 }} 
   $ & $ 0 $ 
 & $ \sqrt{\frac{ 1 }{ 6 }} 
   $ & $ 0 $ 
 & $ -\sqrt{\frac{ 3 }{ 2 }} 
   $ & $ \sqrt{\frac{ 1 }{ 3 
    }} $ & $ 0 $ & $ 0 $ 
 & $ 2 $ & $ 0 $ 
 & $ \sqrt{\frac{ 4 }{ 3 }} 
   $ & $ \sqrt{\frac{ 1 }{ 3 
    }} $ & $ \sqrt{\frac{ 1 
    }{ 2 }} $ 
 & $ -\sqrt{\frac{ 1 }{ 2 }} 
   $ & $ \sqrt{\frac{ 25 }{ 6 
    }} $ & $ \sqrt{\frac{ 4 
    }{ 3 }} $ 
 & $ \sqrt{\frac{ 8 }{ 3 }} 
   $ & $ \sqrt{\frac{ 8 }{ 3 
    }} $ 
\\
\end{tabular}
\end{ruledtabular}
\end{sidewaystable*}

\newpage
\begin{sidewaystable*}
\vspace*{-6cm}
\centering
\caption{
  $ C=2$, $ S=0$, $ I=1/2$, $ J= 3/2$.
}
\label{tab:XiccS}
\begin{ruledtabular}
\begin{tabular}{rrrrrrrrrrrrrrrrrrrrrrrrrrrrrrrrrrrrrrrrrrrrrrrrrr}
 &  $ \CascadaccS \pion $ 
 &  $ \CascadaccS \eta $ &  $ \Cascadacc 
  \rho $ &  $ \Lambdac \DS $ 
 &  $ \Cascadacc \omega $ 
 &  $ \SigmacS \D $ &  $ \OmegaccS 
  \kaon $ &  $ \Omegacc \kaonS $ 
 &  $ \Sigmac \DS $ &  $ \CascadaccS 
  \rho $ &  $ \SigmacS \DS $ 
 &  $ \CascadaccS \omega $ 
 &  $ \Cascadacc \phi $ &  $ \Cascadac 
  \DsS $ &  $ \CascadacS \Ds $ 
 &  $ \Cascadacp \DsS $ &  $ \CascadaccS 
  \etap $ &  $ \CascadacS \DsS $ 
 &  $ \CascadaccS \phi $ 
 &  $ \OmegaccS \kaonS $ 
\\
$ \CascadaccS \pion $& $ -2 $ 
 & $ 0 $ 
 & $ -\sqrt{\frac{ 16 }{ 3 }} 
   $ & $ -\sqrt{ 3 } $ 
 & $ 0 $ & $ 1 $ 
 & $ -\sqrt{\frac{ 3 }{ 2 }} 
   $ & $ -\sqrt{ 2 } $ 
 & $ \sqrt{\frac{ 1 }{ 3 }} 
   $ & $ -\sqrt{\frac{ 20 }{ 3 
     }} $ & $ \sqrt{\frac{ 5 
    }{ 3 }} $ & $ 0 $ 
 & $ 0 $ & $ 0 $ 
 & $ 0 $ & $ 0 $ 
 & $ 0 $ & $ 0 $ 
 & $ 0 $ 
 & $ -\sqrt{\frac{ 5 }{ 2 }} 
   $ 
\\
$ \CascadaccS \eta $& $ 0 $ 
 & $ 0 $ & $ 0 $ 
 & $ \sqrt{\frac{ 1 }{ 3 }} 
   $ & $ 0 $ & $ 1 $ 
 & $ -\sqrt{\frac{ 3 }{ 2 }} 
   $ & $ -\sqrt{ 2 } $ 
 & $ \sqrt{\frac{ 1 }{ 3 }} 
   $ & $ 0 $ 
 & $ \sqrt{\frac{ 5 }{ 3 }} 
   $ & $ 0 $ & $ 0 $ 
 & $ \sqrt{\frac{ 4 }{ 3 }} 
   $ & $ -\sqrt{\frac{ 4 }{ 3 
     }} $ 
 & $ -\frac{ 2 }{ 3  } $ 
 & $ 0 $ 
 & $ -\sqrt{\frac{ 20 }{ 9 }} 
   $ & $ 0 $ 
 & $ -\sqrt{\frac{ 5 }{ 2 }} 
   $ 
\\
$ \Cascadacc \rho $
 & $ -\sqrt{\frac{ 16 }{ 3 }} 
   $ & $ 0 $ 
 & $ -\frac{ 7 }{ 3  } $ 
 & $ 2 $ 
 & $ -\sqrt{\frac{ 1 }{ 3 }} 
   $ & $ -\sqrt{\frac{ 1 }{ 3 
     }} $ & $ -\sqrt{ 2 } 
   $ & $ -\sqrt{\frac{ 8 }{ 3 
     }} $ & $ \frac{ 4 }{ 3 
     } $ 
 & $ -\sqrt{\frac{ 20 }{ 9 }} 
   $ & $ \sqrt{\frac{ 5 }{ 9 
    }} $ 
 & $ -\sqrt{\frac{ 20 }{ 3 }} 
   $ & $ 0 $ & $ 0 $ 
 & $ 0 $ & $ 0 $ 
 & $ 0 $ & $ 0 $ 
 & $ 0 $ 
 & $ -\sqrt{\frac{ 10 }{ 3 }} 
   $ 
\\
\end{tabular}
\end{ruledtabular}
\end{sidewaystable*}

\newpage
\begin{sidewaystable*}
\vspace*{-6cm}
\centering
\caption{
  $ C=2$, $ S=-1$, $ I=0$, $ J= 1/2$.
}
\label{tab:Omegacc}
\begin{ruledtabular}
\begin{tabular}{rrrrrrrrrrrrrrrrrrrrrrrrrrrrrrrrrrrrrrrrrrrrrrrrrr}
 &  $ \Cascadacc \kaonb $ 
 &  $ \Omegacc \eta $ &  $ \Omegacc 
  \omega $ &  $ \Cascadac \D $ 
 &  $ \Cascadacc \kaonSb $ 
 &  $ \Cascadacp \D $ &  $ \Cascadac 
  \DS $ &  $ \Omegacc \etap $ 
 &  $ \Omegacc \phi $ &  $ \Cascadacp 
  \DS $ &  $ \CascadaccS \kaonSb $ 
 &  $ \CascadacS \DS $ &  $ \Omegac \Ds 
  $ &  $ \OmegaccS \omega $ 
 &  $ \Omegac \DsS $ &  $ \OmegacS \DsS 
  $ &  $ \OmegaccS \phi $ 
\\
$ \Cascadacc \kaonb $& $ -2 $ 
 & $ \sqrt{ 3 } $ 
 & $ -\sqrt{\frac{ 1 }{ 3 }} 
   $ & $ \sqrt{\frac{ 3 }{ 2 
    }} $ & $ \sqrt{\frac{ 4 
    }{ 3 }} $ 
 & $ \sqrt{\frac{ 1 }{ 2 }} 
   $ & $ \sqrt{\frac{ 1 }{ 2 
    }} $ & $ 0 $ 
 & $ -\sqrt{\frac{ 2 }{ 3 }} 
   $ & $ \sqrt{\frac{ 25 }{ 6 
    }} $ & $ \sqrt{\frac{ 32 
    }{ 3 }} $ 
 & $ \sqrt{\frac{ 4 }{ 3 }} 
   $ & $ 0 $ 
 & $ -\sqrt{\frac{ 8 }{ 3 }} 
   $ & $ 0 $ & $ 0 $ 
 & $ -\sqrt{\frac{ 16 }{ 3 }} 
   $ 
 \\
$ \Omegacc \eta $
 & $ \sqrt{ 3 } $ 
 & $ 0 $ & $ 0 $ 
 & $ -\sqrt{\frac{ 1 }{ 2 }} 
   $ & $ -1 $ 
 & $ \sqrt{\frac{ 1 }{ 6 }} 
   $ & $ -\sqrt{\frac{ 1 }{ 6 
     }} $ & $ 0 $ 
 & $ 0 $ 
 & $ \sqrt{\frac{ 25 }{ 18 }} 
   $ & $ -\sqrt{ 8 } $ 
 & $ \frac{ 2 }{ 3  } $ 
 & $ -\sqrt{\frac{ 2 }{ 3 }} 
   $ & $ 0 $ 
 & $ -\sqrt{\frac{ 50 }{ 9 }} 
   $ & $ -\frac{ 4 }{ 3  } 
   $ & $ 0 $ 
\\
$ \Omegacc \omega $
 & $ -\sqrt{\frac{ 1 }{ 3 }} 
   $ & $ 0 $ & $ 0 $ 
 & $ -\sqrt{\frac{ 1 }{ 2 }} 
   $ & $ \frac{ 5 }{ 3  } 
   $ & $ \sqrt{\frac{ 25 }{ 6 
    }} $ 
 & $ -\sqrt{\frac{ 1 }{ 6 }} 
   $ & $ 0 $ & $ 0 $ 
 & $ -\sqrt{\frac{ 49 }{ 18 }} 
   $ & $ -\sqrt{\frac{ 8 }{ 9 
     }} $ & $ \frac{ 2 }{ 3 
     } $ & $ 0 $ & $ 0 $ 
 & $ 0 $ & $ 0 $ 
 & $ 0 $ 
\\
\end{tabular}
\end{ruledtabular}
\end{sidewaystable*}

\newpage
\begin{table*}
\centering
\caption{
  $ C=2$, $ S=-1$, $ I=0$, $ J= 3/2$.
}
\label{tab:OmegaccS}
\begin{ruledtabular}
\begin{tabular}{rrrrrrrrrrrrrrrrrrrrrrrrrrrrrrrrrrrrrrrrrrrrrrrrrr}
 &  $ \CascadaccS \kaonb $ 
 &  $ \Omegacc \omega $ &  $ \Cascadacc 
  \kaonSb $ &  $ \OmegaccS \eta $ 
 &  $ \Cascadac \DS $ &  $ \CascadacS 
  \D $ &  $ \Omegacc \phi $ 
 &  $ \Cascadacp \DS $ &  $ \CascadaccS 
  \kaonSb $ &  $ \CascadacS \DS $ 
 &  $ \OmegaccS \omega $ 
 &  $ \OmegacS \Ds $ &  $ \Omegac \DsS 
  $ &  $ \OmegaccS \etap $ 
 &  $ \OmegacS \DsS $ &  $ \OmegaccS 
  \phi $ 
\\
$ \CascadaccS \kaonb $& $ -2 $ 
 & $ \sqrt{\frac{ 4 }{ 3 }} 
   $ & $ -\sqrt{\frac{ 16 }{ 3 
     }} $ & $ \sqrt{ 3 } 
   $ & $ -\sqrt{ 2 } $ 
 & $ \sqrt{ 2 } $ 
 & $ \sqrt{\frac{ 8 }{ 3 }} 
   $ & $ \sqrt{\frac{ 2 }{ 3 
    }} $ 
 & $ -\sqrt{\frac{ 20 }{ 3 }} 
   $ & $ \sqrt{\frac{ 10 }{ 3 
    }} $ & $ \sqrt{\frac{ 5 
    }{ 3 }} $ & $ 0 $ 
 & $ 0 $ & $ 0 $ 
 & $ 0 $ 
 & $ \sqrt{\frac{ 10 }{ 3 }} 
   $ 
\\
$ \Omegacc \omega $
 & $ \sqrt{\frac{ 4 }{ 3 }} 
   $ & $ 0 $ 
 & $ \frac{ 2 }{ 3  } $ 
 & $ 0 $ 
 & $ -\sqrt{\frac{ 8 }{ 3 }} 
   $ & $ -\sqrt{\frac{ 2 }{ 3 
     }} $ & $ 0 $ 
 & $ \sqrt{\frac{ 32 }{ 9 }} 
   $ & $ -\sqrt{\frac{ 20 }{ 9 
     }} $ & $ \sqrt{\frac{ 10 
    }{ 9 }} $ & $ 0 $ 
 & $ 0 $ & $ 0 $ 
 & $ 0 $ & $ 0 $ 
 & $ 0 $ 
\\
$ \Cascadacc \kaonSb $
 & $ -\sqrt{\frac{ 16 }{ 3 }} 
   $ & $ \frac{ 2 }{ 3  } 
   $ & $ -\frac{ 8 }{ 3  } 
   $ & $ 2 $ 
 & $ \sqrt{\frac{ 8 }{ 3 }} 
   $ & $ -\sqrt{\frac{ 2 }{ 3 
     }} $ & $ \sqrt{\frac{ 32 
    }{ 9 }} $ 
 & $ \sqrt{\frac{ 32 }{ 9 }} 
   $ & $ -\sqrt{\frac{ 80 }{ 9 
     }} $ & $ \sqrt{\frac{ 10 
    }{ 9 }} $ 
 & $ -\sqrt{\frac{ 20 }{ 9 }} 
   $ & $ 0 $ & $ 0 $ 
 & $ 0 $ & $ 0 $ 
 & $ \sqrt{\frac{ 40 }{ 9 }} 
   $ 
\\
\end{tabular}
\end{ruledtabular}
\end{table*}

\newpage
\begin{table*}
\centering
\caption{
  $ C=3$, $ S=0$, $ I=0$, $ J= 1/2$.
}
\label{tab:Omegaccc}
\begin{ruledtabular}
\begin{tabular}{rrrrrrrrrrrrrrrrrrrrrrrrrrrrrrrrrrrrrrrrrrrrrrrrr}
 &  $ \Cascadacc \D $ 
 &  $ \Omegacc \Ds $ &  $ \Cascadacc 
  \DS $ &  $ \Omegacc \DsS $ 
 &  $ \CascadaccS \DS $ &  $ \Omegaccc 
  \omega $ &  $ \OmegaccS \DsS $ 
 &  $ \Omegaccc \phi $ \\
$ \Cascadacc \D $& $ 0 $ 
 & $ -\sqrt{ 2 } $ 
 & $ \sqrt{ 12 } $ 
 & $ \sqrt{\frac{ 2 }{ 3 }} 
   $ & $ \sqrt{ 24 } $ 
 & $ -\sqrt{ 8 } $ 
 & $ \sqrt{\frac{ 16 }{ 3 }} 
   $ & $ 0 $ \\
$ \Omegacc \Ds $
 & $ -\sqrt{ 2 } $ 
 & $ 1 $ 
 & $ \sqrt{\frac{ 2 }{ 3 }} 
   $ & $ \sqrt{\frac{ 25 }{ 3 
    }} $ & $ \sqrt{\frac{ 16 
    }{ 3 }} $ & $ 0 $ 
 & $ \sqrt{\frac{ 32 }{ 3 }} 
   $ & $ \sqrt{ 8 } $ 
 \\
$ \Cascadacc \DS $
 & $ \sqrt{ 12 } $ 
 & $ \sqrt{\frac{ 2 }{ 3 }} 
   $ & $ -\frac{ 4 }{ 3  } 
   $ & $ -\sqrt{\frac{ 2 }{ 9 
     }} $ 
 & $ -\sqrt{\frac{ 8 }{ 9 }} 
   $ & $ -\sqrt{\frac{ 8 }{ 3 
     }} $ 
 & $ -\frac{ 4 }{ 3  } $ 
 & $ 0 $ 
\\
$ \Omegacc \DsS $
 & $ \sqrt{\frac{ 2 }{ 3 }} 
   $ & $ \sqrt{\frac{ 25 }{ 3 
    }} $ 
 & $ -\sqrt{\frac{ 2 }{ 9 }} 
   $ & $ -1 $ 
 & $ -\frac{ 4 }{ 3  } $ 
 & $ 0 $ & $ 0 $ 
 & $ \sqrt{\frac{ 8 }{ 3 }} 
   $ \\
$ \CascadaccS \DS $
 & $ \sqrt{ 24 } $ 
 & $ \sqrt{\frac{ 16 }{ 3 }} 
   $ & $ -\sqrt{\frac{ 8 }{ 9 
     }} $ 
 & $ -\frac{ 4 }{ 3  } $ 
 & $ -\frac{ 20 }{ 3  } $ 
 & $ -\sqrt{\frac{ 4 }{ 3 }} 
   $ & $ -\sqrt{\frac{ 128 }{ 9 
     }} $ & $ 0 $ 
 \\
$ \Omegaccc \omega $
 & $ -\sqrt{ 8 } $ 
 & $ 0 $ 
 & $ -\sqrt{\frac{ 8 }{ 3 }} 
   $ & $ 0 $ 
 & $ -\sqrt{\frac{ 4 }{ 3 }} 
   $ & $ 0 $ & $ 0 $ 
 & $ 0 $  \\
$ \OmegaccS \DsS $
 & $ \sqrt{\frac{ 16 }{ 3 }} 
   $ & $ \sqrt{\frac{ 32 }{ 3 
    }} $ 
 & $ -\frac{ 4 }{ 3  } $ 
 & $ 0 $ 
 & $ -\sqrt{\frac{ 128 }{ 9 }} 
   $ & $ 0 $ & $ -4 $ 
 & $ \sqrt{\frac{ 4 }{ 3 }} 
   $ \\
$ \Omegaccc \phi $& $ 0 $ 
 & $ \sqrt{ 8 } $ 
 & $ 0 $ 
 & $ \sqrt{\frac{ 8 }{ 3 }} 
   $ & $ 0 $ & $ 0 $ 
 & $ \sqrt{\frac{ 4 }{ 3 }} 
   $ & $ 0 $  \\
\end{tabular}
\end{ruledtabular}
\end{table*}

\newpage
\begin{table*}
\centering
\caption{
  $ C=3$, $ S=0$, $ I=0$, $ J= 3/2$.
}
\label{tab:OmegacccS}
\begin{ruledtabular}
\begin{tabular}{rrrrrrrrrrrrrrrrrrrrrrrrrrrrrrrrrrrrrrrrrrrrrrrrrr}
 &  $ \Cascadacc \DS $ 
 &  $ \Omegaccc \eta $ &  $ \CascadaccS 
  \D $ &  $ \Omegacc \DsS $ 
 &  $ \CascadaccS \DS $ &  $ \Omegaccc 
  \omega $ &  $ \OmegaccS \Ds $ 
 &  $ \Omegaccc \etap $ &  $ \OmegaccS 
  \DsS $ &  $ \Omegaccc \phi $ 
\\
$ \Cascadacc \DS $
 & $ \frac{ 2 }{ 3  } $ 
 & $ \sqrt{\frac{ 4 }{ 3 }} 
   $ & $ -\sqrt{ 12 } $ 
 & $ -\sqrt{\frac{ 32 }{ 9 }} 
   $ & $ -\sqrt{\frac{ 20 }{ 9 
     }} $ 
 & $ -\sqrt{\frac{ 20 }{ 3 }} 
   $ & $ -\sqrt{\frac{ 8 }{ 3 
     }} $ & $ \sqrt{\frac{ 8 
    }{ 3 }} $ 
 & $ -\sqrt{\frac{ 40 }{ 9 }} 
   $ & $ 0 $ 
\\
$ \Omegaccc \eta $
 & $ \sqrt{\frac{ 4 }{ 3 }} 
   $ & $ 0 $ & $ 1 $ 
 & $ -\sqrt{\frac{ 8 }{ 3 }} 
   $ & $ \sqrt{\frac{ 5 }{ 3 
    }} $ & $ 0 $ 
 & $ -\sqrt{ 2 } $ 
 & $ 0 $ 
 & $ -\sqrt{\frac{ 10 }{ 3 }} 
   $ & $ 0 $ 
\\
$ \CascadaccS \D $
 & $ -\sqrt{ 12 } $ 
 & $ 1 $ & $ 0 $ 
 & $ -\sqrt{\frac{ 8 }{ 3 }} 
   $ & $ 0 $ 
 & $ \sqrt{ 5 } $ 
 & $ -\sqrt{ 2 } $ 
 & $ \sqrt{ 2 } $ 
 & $ -\sqrt{\frac{ 10 }{ 3 }} 
   $ & $ 0 $ 
\\
$ \Omegacc \DsS $
 & $ -\sqrt{\frac{ 32 }{ 9 }} 
   $ & $ -\sqrt{\frac{ 8 }{ 3 
     }} $ 
 & $ -\sqrt{\frac{ 8 }{ 3 }} 
   $ & $ 2 $ 
 & $ -\sqrt{\frac{ 40 }{ 9 }} 
   $ & $ 0 $ 
 & $ -\sqrt{\frac{ 16 }{ 3 }} 
   $ & $ \sqrt{\frac{ 4 }{ 3 
    }} $ & $ 0 $ 
 & $ \sqrt{\frac{ 20 }{ 3 }} 
   $ 
\\
$ \CascadaccS \DS $
 & $ -\sqrt{\frac{ 20 }{ 9 }} 
   $ & $ \sqrt{\frac{ 5 }{ 3 
    }} $ & $ 0 $ 
 & $ -\sqrt{\frac{ 40 }{ 9 }} 
   $ & $ -\frac{ 8 }{ 3  } 
   $ & $ \sqrt{\frac{ 1 }{ 3 
    }} $ 
 & $ -\sqrt{\frac{ 10 }{ 3 }} 
   $ & $ \sqrt{\frac{ 10 }{ 3 
    }} $ 
 & $ -\sqrt{\frac{ 50 }{ 9 }} 
   $ & $ 0 $ 
\\
$ \Omegaccc \omega $
 & $ -\sqrt{\frac{ 20 }{ 3 }} 
   $ & $ 0 $ 
 & $ \sqrt{ 5 } $ 
 & $ 0 $ 
 & $ \sqrt{\frac{ 1 }{ 3 }} 
   $ & $ 0 $ & $ 0 $ 
 & $ 0 $ & $ 0 $ 
 & $ 0 $ 
\\
\end{tabular}
\end{ruledtabular}
\end{table*}

\end{document}